\documentclass[11pt, fleqn]{article}

\usepackage{amsmath}
\usepackage{amssymb}
\usepackage{amsfonts}
\usepackage{amsthm}
\usepackage{dsfont}
\usepackage{mathrsfs}
\usepackage{ulem}

\usepackage{enumerate}

\usepackage[arrowdel]{physics}
\usepackage{tensor}

\usepackage{simplewick}
\usepackage{feynmf}

\usepackage{tabu}
\usepackage{physics}
\usepackage{mathdots}

\newcommand{\operator}[1]{\hat{#1}}

\newcommand{\R}{\mathbb{R}}
\newcommand{\C}{\mathbb{C}}
\newcommand{\Z}{\mathbb{Z}}

\newcommand{\RS}{\R^3\times\S^1_L}
\newcommand{\T}{\mathbb{T}}

\newcommand{\SU}[1]{SU(#1)}

\newcommand{\pauli}[1]{
    \ifnum#1=1
        \operator{\sigma}_{x}
    \else
        \ifnum#1=2
           \operator{\sigma}_{y}
        \else
            \ifnum#1=3
                \operator{\sigma}_{z}
            \else
                \errmessage{Incorrect number given to pauli}
            \fi
        \fi
    \fi
}

\usepackage{geometry}
\usepackage{comment}

\usepackage{soul}
\usepackage{subcaption}

\usepackage{jheppub}
\usepackage{amsmath,amssymb,amsthm}
\usepackage{bm}
\usepackage{slashed}
\usepackage{graphicx}
\usepackage[T1]{fontenc}
\usepackage{fix-cm}

\usepackage{tabu}
\usepackage{physics}

\makeatletter
\newcommand*\bigcdot{\mathpalette\bigcdot@{.5}}
\newcommand*\bigcdot@[2]{\mathbin{\vcenter{\hbox{\scalebox{#2}{$\m@th#1\bullet$}}}}}
\makeatother

\newlength{\dummysp}
\usepackage[font=scriptsize]{caption}
\usepackage{mwe}

\def\R{{\mathbb R}}
\def\S{{\mathbb S}}
\def\Z{{\mathbb Z}}
\def\T{{\mathbb T}}

\def\tr{\,{\rm tr}\,}

 \def\eps{\varepsilon}

\title{Gauge theory geography: charting a path between semiclassical islands
}
\author{Erich Poppitz, F. David Wandler}
\affiliation{Department of Physics, University of Toronto, Toronto, ON M5S 1A7, Canada}
\emailAdd{poppitz@physics.utoronto.ca}  \emailAdd{f.wandler@mail.utoronto.ca}

 \abstract{We study two semiclassical limits of $SU(2)$ Yang-Mills theory on a spatial torus with a 't Hooft twist: the ``femtouniverse,'' where all $\T^3$ directions are small, and deformed Yang-Mills theory on $\T^2 \times \S^1$, with small $\S^1$ and large or infinite $\T^2$. Carefully defining the symmetries, we show that the classical ground states, while different, have the same transformation properties under the 1-form center symmetry and parity. We argue that this is behind the identical multi-branch $\theta$-dependent vacuum structure of these theories. We then calculate the one-loop potential for the $\S^1$-holonomy in the presence of  twists on $\T^2$.  We use it to study the quantum stability of the semiclassical ground states in gauge theories with massive or massless adjoint fermions on spatial $\T^2 \times \S^1$, with a twist in the $\T^2$. The results point towards some interesting features worthy of further study. 
  }

\begin{document}

\maketitle

\section{Introduction}

The past 14 years have seen rapid development of new analytical tools for teasing nonperturbative information out of non-abelian gauge theories. The ones of interest to this paper fall in two main categories: new 't Hooft anomaly matching conditions and center-stabilized compactifications. 

The 't Hooft  anomaly matching    places strict limits on the symmetry structure of a gauge theory  in the IR \cite{Hooft1980}. Within the last decade, a new class of 't Hooft anomalies were discovered, initiated in the remarkable set of papers \cite{Aharony:2013hda,Kapustin:2014gua,Gaiotto:2014kfa,Gaiotto:2017yup,Gaiotto:2017tne}, involving generalized higher-form symmetries. These symmetries act on extended objects, such as Wilson loops, instead of local fields and they do not allow for the usual definitions of current and charge operators but require their generalization. Nevertheless, it was shown that the generalized symmetries can be involved in 't Hooft anomalies and can provide constraints beyond those found from 0-form continuous 't Hooft anomalies. One such symmetry is the 1-form center symmetry, whose breaking dictates whether the theory is in a confining or deconfining phase \cite{Wilson:1974sk,Polyakov:1975rs,Polyakov:1978vu}.

Center-stabilizing compactifications involve studying a Yang-Mills-like theory on a space-time with some or all directions compactified, with some feature added to ensure the center symmetry does not break as the volume becomes smaller than the inverse of the strong coupling scale $\Lambda$. This feature can take the form of adjoint representation fermions (i.e.~supersymmetric Yang-Mills (SYM) or QCD(adj)) \cite{Unsal2008adj,Unsal2009bions} or a non-local double trace deformation potential (i.e.~deformed Yang-Mills (dYM)) \cite{Unsal:2008ch,Shifman:2008ja}. Recently, it has been shown \cite{Tanizaki:2022ngt} that compactification on a 2-torus with an 't Hooft flux\footnote{Throughout this paper we use the phrase ``'t Hooft flux'' to refer to the twist  (given by equation (\ref{cocycle})) on the boundary conditions. Technically, the use of the term ``flux'' could be misleading since the boundary conditions do not guarantee a non-zero gauge field strength (which is the traditional definition of  ``flux''). However, ``'t Hooft flux'' is now a standard term in the community (understood to mean the flux of the topological two-form $\Z_N$ gauge field through noncontractible two-surfaces, which can be used to represent the twisted boundary conditions, as in \cite{Kapustin:2014gua}),  so we hope there is no confusion.} also provides the correct stabilization properties (see \cite{Gonzalez-Arroyo:1998hjb,Montero:1999by,GarciaPerez:1999hs,Montero:2000pb} for earlier relevant remarks). 

These methods have proven very useful and thus there is a large literature surrounding them, more than we can reasonably cite here.\footnote{See \cite{Poppitz:2021cxe} for a pedagogical review of circle compactifications.} Having an unbroken center symmetry guarantees that the symmetry breaking structure is identical to the predicted structure on $\R^4$, and hence the phases are conjectured to be continuously connected to their infinite volume limits. Moreover, the combination of compactification and center-stabilization conspire to prevent the coupling from running to large values at low energies. This weak-coupling setup allows the use of semiclassical and perturbative methods to understand the IR physics in these models.

The authors and A. Cox  recently studied the explicit matching of the new 't Hooft anomalies in certain center-stabilized compactified theories within the Hamiltonian formalism  \cite{Cox:2021vsa}. In particular, we showed that anomalies involving the center symmetry determine the ground state degeneracies for YM and SYM on $\R\times \T^2 \times \S^1$ with an 't Hooft flux through the $\T^2$. It was noticed that the vacuum structure of YM on $\R\times \T^2 \times \S^1$ (with 't Hooft flux, for arbitrary sizes of $\T^2 \times \S^1$)  matched the vacuum structure found in dYM on $\R^3 \times \S^1_L$\footnote{Hereafter, we use $\S^1_L$ to denote the spatial circle of size $L$.}.  In particular the $\theta$-angle dependent multi-branch structure, studied in  dYM in \cite{Unsal:2008ch,Thomas:2011ee,Unsal:2012zj,Bhoonah:2014gpa,Anber:2017rch,Aitken:2018kky,Aitken:2018mbb}, was the same as in the small-$\T^3$ theory \cite{vanBaal:2000zc}.   This stood out as peculiar since the argument in \cite{Cox:2021vsa} relied heavily on the 't Hooft flux, but such a flux cannot be defined for dYM on $\R^3 \times \S^1_L$.

The present paper explores this peculiarity by compactifying dYM on $\R \times \T^2\times \S^1_L$ with an 't Hooft flux and studying the ground states in the large (ultimately, infinite) volume limit: $\T^2\rightarrow \R^2$. We focus on just $\SU{2}$ gauge theory for the sake of clarity and simplicity of notation, but the main ideas are expected to generalize to other gauge groups. Given the center-stabilization property of the deformation, as we take the $\T^2$ large, we should retain the validity of semiclassical calculations for any volume of the $\T^2$. Hence, here we focus on  studying the classical vacua of the theory, on their perturbative stability, and on the action of the zero- and one-form symmetries.\footnote{The details of the nonperturbative semiclassical  dynamics at finite $\T^2$ with a twist, suggested by \" Unsal  \cite{Unsal:2020yeh} to also smoothly connect to the infinite-$\T^2$ limit,  are of great interest as well, but have not yet been fleshed out  and  are left for future study. See Section \ref{sec:discussion} for discussion of some interesting questions that arise.} 

We begin by conjecturing ground states for dYM on $\T^2\times \S^1_L$ with an 't Hooft flux. In the infinite $\T^2$ limit, these states agree with the classical vacua of dYM on $\RS$.\footnote{These classical ground states were first considered by P\'erez, Gonz\'alez-Arroyo, and Okawa for three-dimensional YM  theory \cite{GarciaPerez:2013idu}, though they dismissed them as ground states due to the presence of a tachyonic mode. However, with the deformation potential,  as already noted in \cite{Unsal:2020yeh}, the ground state is made stable, or at least meta-stable, at a point in parameter space where there is no tachyon,  in the context of both the Polyakov model and dYM.} We study the symmetries' action on these ground states at  finite $\T^2$ and show that  it alone implies a two-fold degeneracy of these states at $\theta=\pi$, in accordance with the 't Hooft anomaly  \cite{Cox:2021vsa}. We also show that the dYM  symmetry realization in the space of classical ground states is identical to the one in the ``femtouniverse'' with twists (see \cite{vanBaal:2000zc} for a review), a set-up which offers another semiclassical  limit. We argue that this fact, along with some reasonable assumptions about the semiclassical nonperturbative effects, explains the identical expressions for the $\theta$-dependence of the vacuum energy obtained in these two limits.

We next turn to a study of the stability of these ground states in the framework of a local asymptotically-free  ultraviolet (UV) completion of dYM theory. We use this completion to  calculate the Gross-Pisarski-Yaffe (GPY) potential for the $\S^1_L$ holonomy, similar to \cite{GPY1981}, but now including an extra finite $\T^2$ with 
't Hooft twists.
The UV completion of dYM is Yang-Mills theory with a number of adjoint fermions of mass $M\sim 1/L$, where $L$ is the period of the $\S^1_L$. The calculation involves determining the spectra of gauge boson and fermion excitations around the ground state  and summing up the vacuum energy contributions of each mode. Our conclusion is that the conjectured $\T^2 \times \S^1_L$ ground states are perturbatively stable and smoothly evolve into the known $\R^3 \times \S^1_L$ ground states of dYM.

The calculation of the GPY potential for the $S^1_L$ holonomy with the ``finite-$\T^2$ with  twist'' effects included is one of the main technical results of this work, which may be of use in future studies of other QCD-like theories. Our other technical result is the explicit study of the relation between the implementations of 't Hooft twists  via different sets of transition functions. Understanding this relation in detail enables us to study the symmetry realization and perform dynamical calculations in a most convenient manner.\footnote{This is because the symmetries are realized most simply with one choice of transition functions, while the conjectured ground states and the study of the dynamics, including the determination of the spectra, are simpler with a different choice.}

Recently \cite{Tanizaki:2022ngt}, Tanizaki and \"Unsal proposed that compactifications of various YM and QCD-like \cite{Tanizaki:2022plm} theories on   the manifold $\R^2 \times \T^2$ with an 't Hooft flux should be continuously connected to their infinite volume limits on $\R^4$. We note that the present paper looks at the circle compactified and deformed version of their set up: we take dYM on $\R \times \T^2 \times \S^1$ with 't Hooft flux and explicitly show continuity with dYM on $\RS$. Hence, this work can be seen as a proof of concept for the use of $\T^2$ compactifications with flux for studying uncompactified theories.

{\flushleft{The paper is organized as follows.}}

\smallskip
\begin{itemize}
\item[-] Section \ref{sec:notation} gives a full description of our notation and conventions. It also specifies two useful gauges for the transition functions and explicitly constructs the transformation between them. 

 \item[-] Section \ref{sec:dYM} describes YM and dYM on  $\R \times \T^2\times \S^1$ and constructs the conjectured ground states. 
  We begin (Section \ref{sec:gammagauge}) with a review of the Hamiltonian formalism and symmetry algebra of \cite{Cox:2021vsa}, recalling how it implies  the double degeneracy of all states at $\theta= \pi$ at finite  $\T^2\times \S^1$. To compare  the symmetry realization in various semiclassical limits, we first study (Section \ref{sec:femto}) the classical ground states and symmetries of pure YM  on  $\R \times \T^2\times \S^1$ with a 't Hooft flux on $\T^2$. Here, both $\T^2$ and $\S^1$ are small compared to the inverse strong coupling scale, i.e.~this is the ``femtouniverse'' limit, whose studies date back to \cite{Bjorken:1979hv,Luscher:1982ma}.  Then (Section \ref{sec:dym}), we study the classical ground states of dYM on $\R \times \T^2 \times \S^1_L$ with 't Hooft flux. We find,  following \cite{Unsal:2020yeh}, that the infinite-$\T^2$ limit of the classical dYM ground states smoothly connects to the ones on $\R^3 \times \S^1_L$.
We also show that the symmetries' action in the space of classical ground states is identical in the femtouniverse and dYM. We  argue (Section \ref{sec:theta}) that this fact, along with some assumptions about the semiclassical dynamics, explains the identical $\theta$-dependence of the vacuum energy obtained in these two semiclassical limits.

\item[-] Section \ref{sec:GPYpotential} is devoted to a study of the stability of the dYM classical ground states in a UV complete framework. Here, we describe the calculation of the GPY potential for the $\S^1_L$ holonomy due to gauge fields and massless or massive adjoint fermions with 't Hooft twist effects included. We investigate the stability of the conjectured ground states in the UV complete dYM and in a variety of other theories.

\item[-] Section \ref{sec:discussion}  discusses the implications of the work presented and looks to future directions. 

We have also provided two appendices, showing  details that may be helpful in the future. Appendix \ref{appx:transform} details the construction of a gauge transformation between our two useful gauges (as well as some others, likely useful in generalizations of this work). Appendix \ref{appx:spectra} details the calculation of the spectra and GPY potential in the background of the conjectured ground states.
\end{itemize}

\section{Gauges and notations}
\label{sec:notation}

In this Section, we present our notation, the various gauges for the transition functions we use and the transformation between them, derived in Appendix 
\ref{appx:transform}. This will be useful in the following discussion.

\subsection{'t Hooft flux on a spatial torus}

Motivated by the study of the anomaly \cite{Cox:2021vsa}, we consider  $G=SU(N)$ YM theory in the background of a unit 't Hooft flux $\vec{m} = (0,0,1)$ in a spatial $\T^3$. We consider the spatial $\T^3$: $x^1 \sim x^1 + L^1$, $x^2 \sim x^2 + L^2$, $x^3 \sim x^3 + L^3$, where, to agree with the $\R^3 \times \S^1_L$ literature, we shall often denote $L^3 \equiv L$. We are   interested in two calculable limits that we define below:\footnote{There are other limits of interest, left for future study, see e.g. \cite{Yamazaki:2017ulc}.}
\begin{enumerate} 
\item The {\bf ``small-circle, or $\R^3 \times \S^1_L$, limit''} is that of an asymmetric $\T^3$: we take $L^{1,2}\Lambda \gg 1$ with the size of the third circle, $L \Lambda \ll \pi$  kept fixed and small. Ultimately, our interest is in the $L^{1,2} \rightarrow \infty$ limit, with $\T^3$ approaching $\R^2 \times \S^1_L$. This limit is of interest because in large classes of theories the dynamics abelianizes and the theory is weakly coupled provided $L$ is kept small.
\item  The {\bf ``femtouniverse limit''} is the one where the entire $\T^3$ is small, so that $L^{1,2}  \Lambda \ll 1$ as well as $L \Lambda \ll 1$; usually the equal side torus $L^1=L^2=L$ has been considered. Here, the coupling is small due to asymptotic freedom and the small $\T^3$-size.
\end{enumerate}
We   view the spatial $\T^3$ as $\R^3/\Z^3$. 
We denote by $x \in \R^3$ a point in space, while $\hat e_j$ is a unit vector in the $j$-th direction, $j,k=1,2,3$. We shall study the theory quantized in the $A_0=0$ gauge. The 1-form gauge potential $A = A_k dx^k = A_k^a T^a dx^k$ is hermitean and the hermitean generators obey $\tr T^a T^b = \delta^{ab}/2$. Sometimes we use form notation for the field strength as well, $F = d A + i A \wedge A$.

All fields are defined on the entire $\R^3$ covering space and obey periodicity conditions, following from the requirement that local gauge invariants are periodic functions of $x^j$ with period $L^j$. For example, gauge fields obey:
\begin{eqnarray}
\label{fields}
A(x + L^j \hat e_j) &=& \Omega_j(x) \circ A(x),\; \text{where} \; \Omega_j: \R^3 \rightarrow G \; \text{and} \\
 g(x) \circ A(x) &\equiv& g( A- i d) g^{-1},  \; \nonumber
\end{eqnarray}
where no sum over $j$ is implied.
All functions appearing above are smooth functions defined on $\R^3$. Adjoint matter fields obey similar conditions but the ``$\circ$'' operation used to denote gauge transformations is defined without the nonhomogeneous  derivative term. Clearly, (\ref{fields}) ensures that the field strength as well as any other local gauge invariants (involving e.g. adjoint matter fields) are   periodic functions on $\R^3$ of periods $L^j$. In contrast, the periodicities of gauge invariant noncontractible Wilson loops, such as the one winding in the $j$-th direction, 
\begin{equation}\label{wilsonloopdef}
W_j(x) = \tr[ {\cal{P}} \exp({i \int\limits_{x}^{x+ L^j \hat e_j} A})\; \Omega_j],
\end{equation} can   depend on the twists $n_{ij}$  from (\ref{cocycle}).\footnote{This is because shifting $x$ in the argument of $W_j(x)$ by a torus lattice vector $\hat e_i L_i$ with $i \ne j$ makes the Wilson loop ``sweep 't Hooft flux'' and changes it by a center transformation $\sim e^{i {2 \pi \over N}n_{ij}}$, with $n_{ij}$ defined in (\ref{cocycle}) below.}

The transition functions $\Omega_j(x): \R^3 \rightarrow G$ are smooth  functions obeying cocycle conditions that we now describe.
These cocycle conditions  ensure that the gauge and adjoint fields are single valued on any chosen unit cell of $\R^3/\Z^3$. For any $i$-$  j$ two-plane in $\R^3$, we have 
\begin{eqnarray}
\label{cocycle}
\Omega_i(x + L^j \hat{e}_j)\; \Omega_j(x) &=& e^{i {2\pi \over N} n_{ij}} \; \Omega_j(x+ L^i \hat{e}_i) \;\Omega_i(x) ~, ~ \forall \; i,j=1,2,3, \; \forall x \in \R^3.
\end{eqnarray}
Here, $n_{ij} = - n_{ji}$ are integer (mod $N$)  't Hooft twists, defined in every 2-plane of $\T^3$. Nontrivial, i.e.~nonzero (mod $N$), twists corresponds to turning on nondynamical topological background gauge fields for the $Z_N^{(1)}$ 1-form symmetry in the  spatial 2-planes. 

Under arbitrary gauge transformations $g(x): \R^3 \rightarrow G$, $A$ and $\Omega_j$ transform as 
\begin{eqnarray}
\label{transforms}
A(x) \rightarrow A^g(x) &=& g(x) \circ A(x), \nonumber \\
\Omega_j(x)\rightarrow \Omega_j^g(x) &=& g(x+ L^j \hat{e}_j)\; \Omega_j(x) \;g^{-1}(x)~.
\end{eqnarray}
It is easy to check that the twists $n_{ij}$ are invariant under  (\ref{transforms}).

As (\ref{transforms}) shows, the transition functions $\Omega_j(x)$ are not invariant under general gauge transformations.  In the framework of studying  gauge bundles over the torus, it has been argued that any two sets of transition functions obeying  cocycle conditions (\ref{cocycle}) with the same $n_{ij}$ can be mapped to each other using a smooth gauge transformation. We shall explicitly construct and use such a map, smooth and defined on all of $\R^3$, for the two sets of transition functions we use below. 

For further use, we note that canonical quantization of  the theory  in the $A_0=0$ gauge proceeds by choosing a fixed set of transition functions $\Omega_j $ obeying (\ref{cocycle}) with the chosen $n_{ij}$ background. The choice of fixed $\Omega_j$ amounts to partial gauge fixing. Further, one constructs a Hilbert space of eigenstates of the field operator $\hat A_k$ obeying (\ref{fields}) with the chosen $\Omega_j$ and  then demands invariance of physical states under  small (with nontrivial $\pi_3(G)$) gauge transformations preserving $\Omega_j$. {We call the  gauge transformations preserving the $\Omega$ transition functions {\it $\Omega$-periodic}. The set of $\Omega$-periodic gauge transformations is defined as 
\begin{eqnarray}
\label{gaugeomega}
 \{g_\Omega: \R^3 \rightarrow G,  g_\Omega (x + L^j \hat{e}_j) = \Omega_j(x)\; g_\Omega (x)\; \Omega_j^{-1}(x)\}. 
\end{eqnarray}
We note that the $\Omega$-periodic gauge transformations also include gauge transformations with nontrivial $\pi_3(G)$, under which physical states transform by $\theta$-angle dependent phases. 

In what follows, we largely concentrate on $G=SU(2)$, thus setting $N=2$. We shall also take $n_{12}= m_3 = 1$ as the only nonzero twist.

\subsection{Some useful gauges and the transforms between them}
\label{sec:gauges1}

In order to study  dYM, QCD(adj), and other theories of interest with semiclassically calculable dynamics, we shall employ two different gauges for the nontrivial transition functions responsible for the $\vec{m} = (0,0,1)$ background. As we shall see, different aspects of the theory are analyzed more conveniently in one or the other gauge. {Hence, the ability to switch from one to the other is useful, so we construct an explicit map between the two.}

{\flushleft{\bf ``$\mathbf{\Gamma}$-gauge:''}} The first gauge employs the often-used   constant transition functions, $\Omega_i$, denoted by $\Gamma_i$. These are constant group elements obeying $\Gamma_i \Gamma_j = e^{i {2 \pi \over N} n_{ij}} \Gamma_j \Gamma_i$ (for $SU(N)$):
\begin{eqnarray}
\label{gammagauge}
\text{``$\Gamma$-gauge:''} ~\Gamma_1 &=& i \sigma_1, \nonumber\\
  \Gamma_2 &=& i \sigma_3, \\
  \Gamma_3 &=& \bf{1},\nonumber
\end{eqnarray}
where the only nontrivial commutator is $\Gamma_1 \Gamma_2 = - \Gamma_2 \Gamma_1$, showing that the  ``$\Gamma$-gauge'' transition functions defined above obey (\ref{cocycle}) with $N=2$, $n_{12}=1$. The $\Gamma$-gauge has been  used to study the spectrum of the theory in the femtouniverse, where all $\T^3$ directions are small \cite{GonzalezArroyo:1987ycm}, as well as in the original calculations of the Witten index \cite{Witten:1982df}.  The analysis of the mixed zero-form/1-form anomaly of \cite{Cox:2021vsa} is also most conveniently done in this ``nice''   gauge. We shall, therefore, be interested in the gauge transforms between this and other gauges discussed below.

{\flushleft{\bf ``$\mathbf{\Omega}$-gauge:''}} In this gauge,  the transition functions are given by 
\begin{eqnarray}
\label{omegagauge}
\text{``$\Omega$-gauge:''}~ \Omega_1 &=& \bf{1},\nonumber \\
\Omega_2 &=&e^{i \pi {x^1 \over L^1} \sigma_3 },\\
\Omega_3 &=& \bf{1}.\nonumber
\end{eqnarray}
Owing to $\Omega_2(x^1 + L^1) = e^{i \pi} \; \Omega_2(x^1)$, and $\Omega_1=\Omega_3={\bf 1}$, these transition functions obey the same cocycle condition as (\ref{gammagauge}) with $n_{12}=1$. The transition functions (\ref{omegagauge}) are abelian and it is not surprising that they are useful to study the dynamics of   theories that abelianize in the small-circle limit defined above.

{\flushleft{\bf ``$\mathbf{\Omega_{(k)}}$-gauge:''}} Other $\Omega$-gauges with abelian transition functions are possible, most notably the identical to (\ref{omegagauge}) but with a different $\Omega_2$. These gauges are characterized by an integer, $k$. For a given $k$, we define the {\bf ``$\mathbf{\Omega_{(k)}}$-gauge''} to be
\begin{eqnarray}
\label{omegakgauge}
\text{``$\Omega_{(k)}$-gauge:''} ~\Omega_{1, (k)} &=& \bf{1},\nonumber \\
\Omega_{2, (k)} &=&e^{i \pi (1+2 k) {x^1 \over L^1} \sigma_3 }, \; k \in \Z,\\
\Omega_{3, (k)} &=& \bf{1}.\nonumber
\end{eqnarray}
which has the same $x^1$ periodicity for all $k$. There is a simple constant flux abelian background\footnote{Running ahead, we note that this generalizes the background  studied later, eqn.~(\ref{a1}), which is the $k=0$ case of (\ref{kflux}).} obeying the boundary conditions in this gauge  
\begin{eqnarray}
\label{kflux}
A  &=& - {2\pi (1+2k) x^2 dx^1 \over L^1 L^2} {\sigma_3 \over 2}, \; \text{with} \; A(x^2+L^2)=A(x^2) - i \Omega_{2,(k)} d \Omega_{2,(k)}^{-1}, \nonumber \\
F_{12, (k)} &=& {2 \pi (1+2 k) \over L^1 L^2} {\sigma_3 \over 2},
\end{eqnarray}  
The magnetic energy of this flux is thus lowest for $k=0, -1$. {We stress that all backgrounds can be mapped to backgrounds with the same energy and flux in any gauge, just with a generally less simple form. In particular, }these abelian flux configurations are not the lowest energy ones with the given transition functions: there exist configurations of zero classical energy and no flux which are most easily exhibited in the $\Gamma$-gauge \cite{Witten:1982df}. These abelian constant flux configurations can also be exhibited in the gauge (\ref{omegagauge}).\footnote{One can find the gauge transformation $g_{(k)}$, relating the $\Omega_{(k)}$ to $\Gamma$ transition functions, similar to  Appendix \ref{appx:omega0} for $k=0$, by convolution with the $\Omega$ to $\Omega_{(k \ne 0)}$ transformation described in Appendix \ref{omegaktoomegazero}.} We shall mostly focus on the lowest fluxes in what follows.\footnote{We note, however, that understanding nonperturbative effects and the dual photon picture will have to involve higher fluxes as well. The role of higher fluxes in the Polyakov model and dYM has been discussed in \cite{Banks:2014twn,Anber:2011gn,Unsal:2020yeh}. See also our comments in Section \ref{future}.}

Let us note that the transformation relating the transition functions $\Omega_{i, (-1)}$ to $\Omega_i$ (here $i=1,2,3$)\footnote{More generally, $i \sigma_2$ can be seen to relate $\Omega_{(k)}$ to $\Omega_{(-k-1)}$,   for which the energy of the abelian fluxes (\ref{kflux}) are the same. We stress, however, that the transformation is so simple only between gauges {where their natural abelian flux backgrounds have the same energy.}} of (\ref{omegagauge}) is especially simple\begin{equation}
\label{omega0minus1}
\Omega_i = (i\sigma_2) \;\Omega_{i, (-1)} (-i \sigma_2) ~.
\end{equation}
For further use, note that the $SU(2)$ Weyl reflection $i \sigma_2$ is neither $\Omega$-, nor $\Omega_{(-1)}$-  or $\Gamma$-periodic. 
Both $\Omega_i$ and $\Omega_{i, (-1)}$ gauges will be useful for us since the constant abelian fluxes exhibited in these two respective gauges have the same $|F_{12}^{3}|$ and thus the same magnetic field energy. 

There exists a smooth (infinitely differentiable) map $g$:  $\R^3 \rightarrow SU(2)$ between the $\Omega$- and $\Gamma$-gauges, which can be taken $x^3$-independent. It maps the transition functions in the $\Omega$-gauge to the ones in the $\Gamma$-gauge, as in (\ref{transforms}):\footnote{To avoid overcrowding, in the formulae below, as well as in Appendix \ref{appx:transform} where they are derived, we set $L^1=L^2=L = 1$. It is trivial to restore dimensions in the end.}
\begin{eqnarray}
\label{mapgauge1}
\Gamma_1 &=& g(x^1+1, x^2) \; \Omega_1 \;g^{-1}(x^1,x^2),
\nonumber \\
\Gamma_2 &=& g(x^1 , x^2+1) \; \Omega_2 \;g^{-1}(x^1,x^2),\\
\Gamma_3 &=&   \Omega_3 = \bf{1}, \nonumber 
 \end{eqnarray} where the $\Omega_i$ are from 
  (\ref{omegagauge}) and $\Gamma_i$ from  (\ref{gammagauge}).
  
The usefulness of having the map between $\Omega$-gauge and $\Gamma$-gauge is that any  field configuration $A^\Omega(x^1,x^2)$ obeying the periodicity condition with the transition functions (\ref{omegagauge}) can be mapped to a field configuration, denoted by $A^\Gamma(x^1, x^2)$, obeying the $\Gamma$-gauge periodicity conditions (\ref{gammagauge}). The explicit form of the map is:
\begin{equation}\label{mapgauge}
A^\Gamma(x^1,x^2) = g(x^1,x^2)\circ A^\Omega(x^1,x^2)~.
\end{equation}
The local gauge invariants, such as the various contributions to the energy densities, of the $A^\Gamma$ and $A^\Omega$ field configurations are identical. Adjoint fields are mapped analogously to (\ref{mapgauge}) by omitting the nonhomogeneous term in the gauge transformation.

Similar to (\ref{mapgauge}), we can also map a configuration in the $\Omega_{(-1)}$ gauge to one in the $\Gamma$-gauge, using also (\ref{omega0minus1}):\begin{equation}\label{mapgauge2}
A^\Gamma(x^1,x^2) = (g(x^1,x^2) (i \sigma_2)) \circ A^{\Omega_{(-1)}}(x^1,x^2)~,
\end{equation}
and, of course, from (\ref{omega0minus1}) we also have
\begin{equation}\label{mapgauge3}
A^\Omega(x^1,x^2)(x^1,x^2) = (i \sigma_2)  \circ A^{\Omega_{(-1)}}(x^1,x^2)~.
\end{equation}

 \begin{figure}[h]
 \centerline{
\includegraphics[width=11cm]{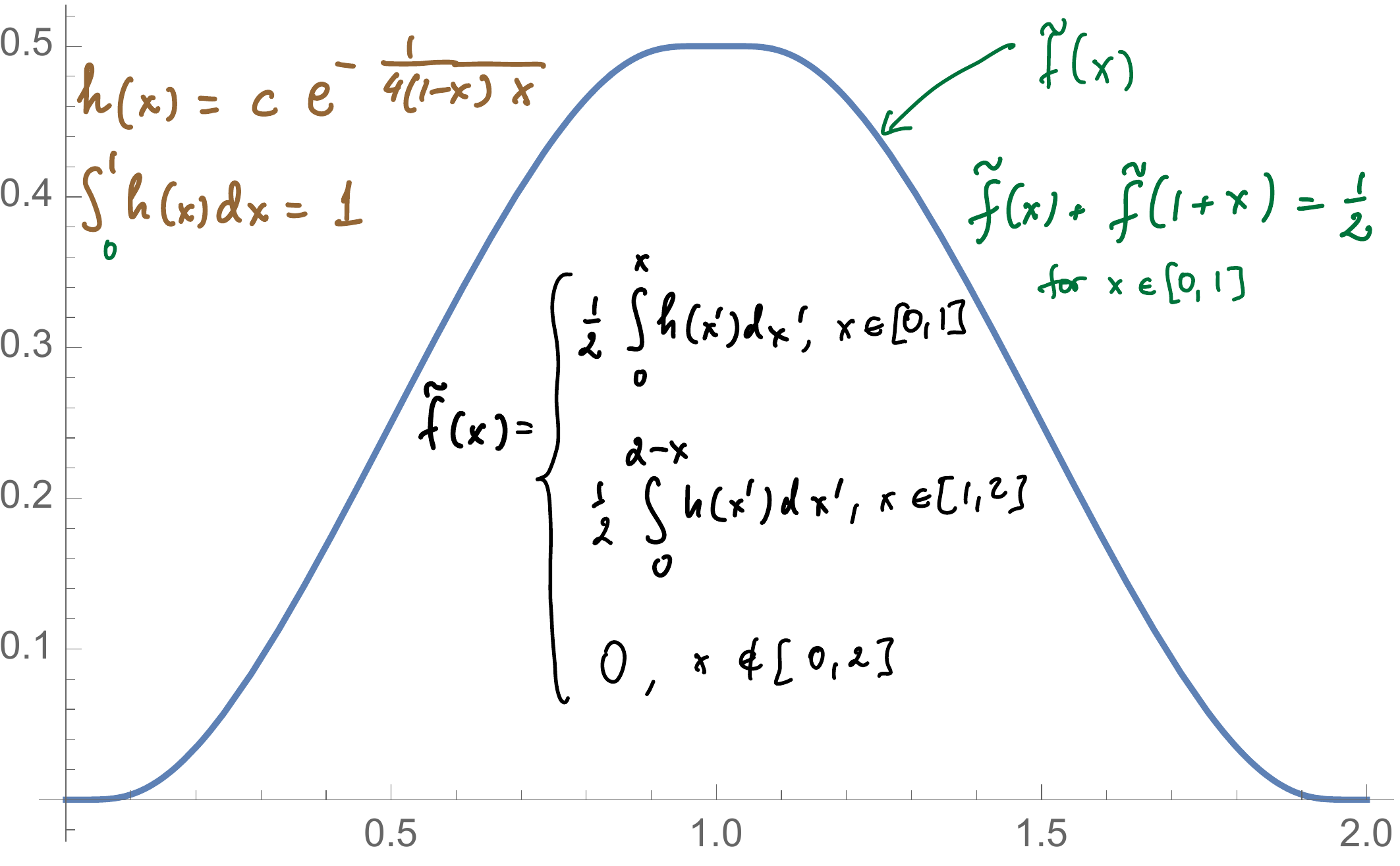}}
\caption{A plot of the infinitely differentiable function $\tilde f(x)$, vanishing outside  $x \in [0,2]$ and obeying  (\ref{tildef1}). Its construction using a ``bump function'' $h(x) \sim e^{- {1 \over 4 x (1-x)}}$, only nonzero for $0<x < 1$, is illustrated on the plot. The function $f(x)$ entering (\ref{gunitsquare1})   is the square root of $\tilde f(x)$. \label{fig1}}
\end{figure}

To end this Section, we describe the explicit form, derived in Appendix \ref{appx:transform}, of the gauge transformation\footnote{A different form of these transformations, given in terms of $\theta$-functions and unknown to us at the time we obtained (\ref{gunitsquare1}), has been derived earlier in \cite{GarciaPerez:2013idu}. Furthermore, this reference also considered the abelian flux background (in the $2+1$ framework) that we study later in this paper. We thank A. Gonz\'alez-Arroyo for pointing this out to us.}
  $g$: $\R^2 \rightarrow SU(2)$ obeying (\ref{mapgauge1}). For all $x^1 \in \R$ and for $0 \le x^2 \le 1$ it is given by
\begin{eqnarray}\label{gunitsquare1}
 &&g(x^1,x^2)\big\vert_{\forall x^1\in \R\; {\text{and}} \; 0 \le x^2 \le 1} =\\
 &&\qquad \qquad \left(\begin{array}{cc} e^{i {\pi x^1 \over 2}}[f(x^2) -i e^{i \pi x^1} f(x^2+1)] &  -e^{-i {\pi x^1 \over 2}} [f(x^2) - ie^{-i \pi x^1} f(x^2 + 1)] \cr     e^{i {\pi x^1 \over 2}} [f(x^2) + ie^{i \pi x^1} f(x^2 + 1)]&e^{-i {\pi x^1 \over 2}}[f(x^2) +i e^{-i \pi x^1} f(x^2+1)]  \end{array} \right)~. \nonumber
\end{eqnarray}
Here $f(x)$: $\R \rightarrow \R$ is an infinitely differentiable ``bump function'' which is  nonzero only for $x  \in (0, 2)$. In fact, $f(x)$ equals the square root of the function $\tilde{f}(x)$ shown on Figure \ref{fig1} and obeying
\begin{equation}\label{tildef1}
\tilde{f}(x) = \left\{ \begin{array}{ccc} 0 & \text{for} &x \notin [0, 2], \cr
{1 \over 2} & \text{for}& x = 1, \cr
{1 \over 2} - \tilde{f}(1+x)& {\text{for}} & x \in [0,1]~. \end{array} \right.
\end{equation} 
Thus, $f(x) = \sqrt{\tilde{f}(x)}$  obeys $f(1)= {1\over\sqrt{2}}$ and  $f^2(x)+f^2(1+x)= {1\over 2} $ for $0 \le x \le 1$. 
 
It is easy to check that the properties of $f(x^2)$ given above guarantee the unitarity of (\ref{gunitsquare1}). The validity of the condition from the first line of (\ref{mapgauge1}) can also be easily verified from the form given in (\ref{gunitsquare1}). 
Note, however, that verifying that (\ref{gunitsquare1}) obeys the condition on the second line in (\ref{mapgauge1}) requires knowledge of the extension of (\ref{gunitsquare1}) outside $0 \le x^2 \le 1$. 
The  extension of (\ref{gunitsquare1})  over the entire $\R^2$-plane and the construction of the smooth map $g(x)$: $\R^2 \rightarrow SU(2)$ for all $x \in \R^2$ are described in detail in  Appendix \ref{appx:transform}, in particular (\ref{gauge2x}). 

In the following Sections, we make extensive use of the transition function change affected by $g(x^1, x^2)$ to define the action of center symmetry and parity in the different gauges of interest.

\section{dYM vs femtouniverse with $\mathbf{n_{12}=1}$: classical vacua and symmetries}
\label{sec:dYM}

The classical Minkowski-space action density of (deformed) Yang-Mills theory with gauge group  $SU(2)$ on $\R^t \times \T^2 \times \S^1$ is
\begin{eqnarray}
\label{action}
{\cal{L}} &=& {1 \over 2 g^2} F_{0 i}^a F^{a}_{0 i}  + {1 \over 2 g^2} F_{0 3}^a F^{a}_{03}- {1 \over 2 g^2} F_{12}^a F^a_{12} - {1 \over 2 g^2} F_{3 i}^a F^a_{3 i} - {c \over L^4} | \tr W_3|^2~,
\end{eqnarray}
where, for future convenience, we have separated the spatial components into ones along $\T^2$ ($x^{i}$, $i=1,2$) and $\S^1$ ($x^3$) with periodicities given earlier. We denote time by $x^0=t$, summation over the values of the repeated index $i=1,2$ is understood, and all Minkowski metric factors have already been accounted for.  
The  generators are $T^a = {\sigma^a \over 2}$, and the field strength is $F_{ij}^a = \partial_i A_j^a - \partial_j A_i^a - \epsilon^{abc} A_i^b A_j^c$ (here, $i,j$ can  take values $1,2,3$).

The first four terms in $L$ constitute the usual pure YM action. The last term is the ``double-trace'' deformation. When $c=0$, the action (\ref{action}) reduces to that of pure YM theory. The {quantity} $W_3$, whose trace enters the double-trace deformation, is the fundamental Wilson loop operator, eq.~(\ref{wilsonloopdef}) without the trace (recalling that in either gauge $\Omega_3=\bf{1}$), winding in the $\S^1$ direction:
\begin{eqnarray}\label{w3}
  W_3(t, x^i, x^3)&\equiv& {\cal{P}} e^{i \int_{x^3}^{x^3+L } A_3^a(t, x^i, y^3) T^a d y^3} ~.
\end{eqnarray}
The deformation term in (\ref{action}) is a nonlocal term, leading to a nonrenormalizable theory. Here, we only note that its effect has been shown to be produced by integrating out massive adjoint fermions on $\R^3 \times \S^1$. The question of its origin here will not   concern  us for now: at first, when studying dYM theory, we shall treat $c$ as a free parameter dialing it sufficiently large to impose $\tr W_3=0$ on the classical ground state. We relegate the study the UV completion of dYM to Section \ref{sec:GPYpotential}.

In what follows, we shall study (d)YM theory in the $n_{12}$ 't Hooft flux background in the $A_0=0$ gauge.
Thus, all (d)YM fields are taken to obey periodicity conditions (\ref{fields}) with transition functions on $\T^3 = \T^2 \times \S^1$ taken to be either the $\Gamma$- or $\Omega$-gauge ones, (\ref{gammagauge}) or (\ref{omegagauge}), respectively. We note that since $\Omega_3=\bf{1}$ in both gauges, the trace of the fundamental Wilson loop $W_3$ in (\ref{action}) is  invariant under the respective $\Omega$- or $\Gamma$- periodic gauge transformations. In addition, $\tr W_3$ is invariant under $x^3$ translations by $L$ and is thus  $x^3$-independent. It is also invariant under $x^1$, $x^2$ translations by $L^1$ and $L^2$, as follows from the vanishing of the $n_{13}, n_{23}$ twists in (\ref{cocycle}). 

Thus, the entire action density ${\cal{L}}$ is invariant under the $\Z^3$ translations defining $\T^3$ and can be used to define a theory on  $\T^3$ by integrating ${\cal{L}}$ over a unit cell in $\R^3$. In what follows, we take the action and energies to be integrated over the unit cell:
\begin{equation}
\label{unitcell}
\T^3 = \left\{ (x^1,x^2,x^3) \in \R^3 \; : \; 0 \le x^1 \le L^1,  0 \le x^2 \le L^2,  0 \le x^3 \le L\right\}.
\end{equation}
In the $A_0=0$ gauge, the energy density ${\cal{E}}$ following from (\ref{action}) is 
\begin{eqnarray}
\label{energydensity}
{\cal{E}} &=& {g^2 \over 2} (\Pi_i^a \Pi_i^{a}  + \Pi_3^a \Pi_3^a) + {1 \over 2 g^2} F_{12}^a F^a_{12} + {1 \over 2 g^2} F_{3 i}^a F^a_{3 i} + {c \over L^4} | \tr W_3|^2~,
\end{eqnarray}
where the conjugate momenta are $\Pi_i^a$ and Gauss' law constraint $G^a =0$ should be imposed:
\begin{eqnarray}\label{gausslaw}
g^2 \Pi_{i (3)}^a &=& \partial_t A_{i (3)}^a, ~
G^a = \partial_i \Pi_i^a + \partial_3 \Pi_3^a - \epsilon^{abc} (A_i^b \Pi_i^c + A_3^b \Pi_3^c).
\end{eqnarray}
Finally, the classical energy functional is
\begin{eqnarray} \label{energy}
E &=& \int_{\T^3} d^3 x\; {\cal{E}}~. \end{eqnarray}
We note that  the separate terms appearing in the energy density (\ref{energydensity}) are invariant under time-independent gauge transformations and are identical for fields obeying either the $\Gamma$- or $\Omega$-gauge boundary conditions. 

We now study the equations determining the static ($\Pi_{1,2,3}=0, G^a=0$) extrema of the classical energy functional. This is most straightforward in the $\Gamma$-gauge, where the values of the gauge field and its fluctuation on the opposite sides of $\T^3$ are related by a homogeneous transformation via the constant twist matrices $\Gamma_i$: $A^\Gamma(x^i = L^i) = \Gamma_i A^\Gamma(x^i=0) \Gamma_i^{-1}$ (here $i=1,2,3$). There are thus no boundary terms and the equations determining the extrema of the classical energy obtained by varying $A_{1}$ and $A_2$ are the usual ones \begin{eqnarray}
\label{energyextrema}
{\delta{E}  \over \delta A_{1}} = 0&=&  \partial_2 F_{12} + i [A_2, F_{12}] + \partial_3 F_{13} + i [A_3, F_{13}] = D_2 F_{12} + D_3 F_{13} ~,\nonumber \\ 
{\delta{{E}}  \over \delta A_{2}} = 0&=&  \partial_1 F_{21} + i [A_1, F_{21}] + \partial_3 F_{23} + i [A_3, F_{23}] = D_1 F_{21} + D_3 F_{23} ~,
\end{eqnarray}
where, we denoted by $D_1 F_{21}$ the covariant derivative of $F_{21}$ (implicitly defined above and similar  for the other terms). The variation with respect to $A_3$ yields a more complicated expression due to the deformation term
\begin{eqnarray}\label{energyextrema2}
{\delta {{E}} \over \delta A_3} =0&=&  \partial_1 F_{31} + i [A_1, F_{31}] + \partial_2 F_{32} + i [A_2, F_{32}]+ {g^2 c \over L^3} \left[ i\; W_3 \; (\tr W_3^*) - i \; W_3^*\; (\tr W_3) \right]\nonumber \\
&=& D_1 F_{31} + D_2 F_{32} + {g^2 c \over L^3} \left[ i\; W_3 \; (\tr W_3^*) - i \; W_3^*\; (\tr W_3) \right]
\end{eqnarray}
where $W_3$ was given in (\ref{w3}).\footnote{To obtain the variation of the deformation potential we used  $\delta \left({\cal{P}} e^{i \int_0^1 dy C(y)}\right) = \int_0^1 dt \left({\cal{P}} e^{i \int_0^t dy C(y)}\right) i \delta C(t)  \left({\cal{P}} e^{i \int_t^1 dy C(y)}\right)$, which follows most naturally from the lattice definition of the Wilson line, see e.g.~\cite{Polyakov:1987hqn}.}
As is clear from the second form of (\ref{energyextrema}, \ref{energyextrema2}) shown above, the equations determining the extrema of the energy functional transform homogeneously under gauge transformations and  thus hold when expressed via either the $\Gamma$- or $\Omega$-gauge fields.

\subsection{$\Gamma$-gauge quantization and symmetries} 
\label{sec:gammagauge}

Quantization and the symmetries are exhibited  most straightforwardly in the $\Gamma$-gauge. In the rest of this Section we review the results of \cite{Cox:2021vsa}. Borrowing notation from 't Hooft \cite{tHooft:1979rtg,tHooft:1981sps}, one studies the $A_0=0$ gauge and constructs a  Hilbert space of $\hat A$ field operator eigenstates that satisfy the boundary conditions (\ref{fields}) with constant transition functions (\ref{gammagauge}). This results in the ``large'' Hilbert space: \begin{eqnarray}\label{bc}
&& \mathcal{H} = \left\{ \ket{A}, | A(x + \hat e_1 L^1)  = \Gamma_1 \circ A(x), A(x + \hat{e}_2 L^2)  = \Gamma_2 \circ A(x), A(x+ \hat e_3 L^3)  = \Gamma_3 \circ A(x)  \right\},\nonumber\\
&& \, 
\end{eqnarray} where $\ket A$ stands for an eigenvector of the ``position'' operator $\hat A(x)\ket{A} = \ket{A} A(x)$.
Consider the set of gauge transformations preserving the boundary conditions (\ref{bc}), i.e.~the ``$\Gamma$-periodic'' gauge transformations, which we denote by $U$, defined earlier in (\ref{gaugeomega}), with $\Omega_j \rightarrow \Gamma_j$. A gauge transformation $U$ uniquely determines an operator on the large Hilbert space by the relation
\begin{equation}
\hat{U} \ket{A} = \ket{U \circ A}~.
\end{equation} 
Gauss' law requires that the physical states $\ket\psi \in \mathcal{H}$ obey $\hat U \ket\psi = \ket\psi$, i.e. are invariant under gauge transformations $U$ which are $\Gamma$-periodic and are homotopic to the identity.
In addition to gauge transformations homotopic to the identity, {$\Gamma$-periodic maps from $\T^3$ to $G$ can have non-zero instanton number $\nu \in \Z$}, associated with $\pi_3(G)$. These ``large''
gauge transformations do not leave physical states invariant but act as 
\begin{equation}\label{Htheta}
{\cal{H}}^{phys.}_\theta = \left\{\ket{\psi} \in \mathcal{H} \,:\, \hat{U} \ket{\psi} = e^{-i\theta \nu} \ket{\psi}, \forall U\right\}
\end{equation}
where $\nu$ is the instanton number associated with the transformation $U$ ($\nu$ vanishes for the ``small'' gauge transformations). ${\cal{H}}^{phys.}_\theta $ defines the physical Hilbert space, where all vectors have  definite theta angle. 

  In terms of the position, $\hat A_i^a$,  and momentum, $\hat \Pi_i^a(\vec{x}) = - i {\delta \over \delta A_i^a(\vec{x})}$ (where for brevity now we take $i=1,2,3$ to include both $\T^2$ and $\S^1$ directions in the sum), operators, the Hamiltonian in the physical Hilbert space is 
 \begin{eqnarray} \label{hamiltonian}
 \hat{H} &=& \int\limits_{\T^2 \times \S^1} d^3 x \left( {g^2} \;\sum\limits_{i=1}^3 \tr \hat \Pi_i  \hat \Pi_i + {1 \over  g^2}\;\tr \;\sum\limits_{i=1}^3 \hat B_{i} \hat B_{i} + {c \over L^4} |\tr W_3|^2\right), \\
 &&[ \hat\Pi_i^a(\vec{x}), \hat A_j^b(\vec{y})] = -i \delta^{ab} \delta_{ij} \delta^{(3)}(\vec{x}-\vec{y}).
 \end{eqnarray}
Here, as discussed above, the integral is over the unit cell (\ref{unitcell}),  $\hat{B}_i = {1 \over 2} \eps_{ijk} \hat F_{jk}$, and  the operators $\hat \Pi_i (\vec{x})$ and $\hat{A}_i (\vec{x})$ obey the  boundary conditions (\ref{bc}) twisted by  $\Gamma_j$ of (\ref{gammagauge}). $W_3$ is as given in (\ref{w3}) and is $x^3$ independent. 

The 1-form center symmetry generators $\hat T_i$, $i=1,2,3$ are defined, up to small gauge transformations,  by their action on the large Hilbert space:
\begin{eqnarray}
\Z_2^{(1)}: \hat T_1 \ket{A} &\equiv& \ket{\Gamma_2 \circ A},\nonumber \\ 
\hat T_2 \ket{A} &\equiv& \ket{\Gamma_1 \circ A}, \\
\hat T_3 \ket{A} &\equiv& \ket{T_3(x) \circ A},\nonumber 
\end{eqnarray}
where 
the non-constant transformation $T_3(x)$ generating center symmetry in the $\S^1$ direction can be taken to be\footnote{It should be clear that this form is not unique. A different form of $T_3(x)$ was given in \cite{Selivanov:2000kg}, which is, however, not well-suited for our dynamical calculations. Yet another expression, given in \cite{Cox:2021vsa} (see eq.~(3.28) there), is similar in appearance to (\ref{center3}), obeys the right  boundary conditions (\ref{tboundaryconditions}) and has the winding number (\ref{windingt3}), but the corresponding function $g$  does not smoothly extend outside the unit cell (\ref{unitcell}), while the above  (\ref{center3}) does, by virtue of its construction. A smooth extension  beyond the unit cell of $\T^3$ is needed to properly define a bundle over the torus. The lack of a smooth extension can be seen to lead to pathologies once one attempts explicit dynamical calculations making use of the map $g$. } \begin{equation}\label{center3}
T_3(x) = g({x^1 \over L^1}, {x^2 \over L^2}) \; e^{-i \pi {x^3 \over L} \sigma_3} \; g^{-1}({x^1 \over L^1}, {x^2 \over L^2})~. 
\end{equation}
Here, $g$ is the map constructed in Appendix \ref{appx:transform}, already given in (\ref{gunitsquare1}) (see also (\ref{gauge2x})).
We note that the gauge transformations\footnote{{Note that center symmetry transformations are not gauge transformations. While the maps $T_j = (\Gamma_2, \Gamma_1, T_3)$ can be used to define a gauge transformations, they are here used as center symmetry transformations. The distinction comes from whether the transition functions are changed according to (\ref{transforms}). A center symmetry transformation applies the map $T_j$ to the fields but does not make the corresponding change the transition functions.}} defining $\hat T_j$,   $T_j = (\Gamma_2, \Gamma_1, T_3)$, are easily seen to obey 
\begin{equation}
\label{tboundaryconditions}
T_j(x + \hat e_k L^k) =e^{i \pi \delta_{jk}}\; \Gamma_k \; T_j(x)\; \Gamma_k^{-1},
\end{equation}
 as appropriate for non-$\Gamma$-periodic  transformations generating center symmetry. The relation obeyed by $T_3$ follows from the property (\ref{mapgauge1}) of $g$.
 
Another important property of $T_3$ (\ref{center3}) is that it has a half-integer winding number $\T^3 \rightarrow SU(2)$, as argued  by 't Hooft long ago \cite{tHooft:1981sps} (or see \cite{Cox:2021vsa}). This can also be seen explicitly, using the form of $g(x^1,x^2)$ in the unit cell (\ref{unitcell}) given in eqn.~(\ref{gunitsquare1}). It is straightforward  (we note that using computer algebra can help) to explicitly calculate the $T_3(x)$ winding number
\begin{equation}
\label{windingt3}
{1 \over 24 \pi^2} \int_{\T^3} \tr (T_3 d T_3^{-1})^3 = \prod\limits_{i=1}^3 \left(\int\limits_{0}^1 dx^i\right)\;(1 - \cos (2 \pi x^3)){d\over d x^2} f^2(x^2) = f^2(1)- f^2(0) = {1\over 2},
\end{equation}
where we used the properties of the ``bump'' function $f(x)$ mentioned after (\ref{gunitsquare1}).

Further, recall  that because the $\hat T_{i}$ generate a $\Z_2$ symmetry, their eigenvalues are $e^{i \pi e_k} = \pm 1$ ($e_k = 0,1$), with the exception of $\hat T_3$ whose eigenvalues in ${\cal{H}}_\theta$  are $e^{i \pi e_3} e^{ - i {\theta \over 2}}$. The (mod $2$) integer $e_k$ represents a $\Z_2$ electric flux in the $k$-th direction of $\T^3$.

The nontrivial winding of $T_3$ implies that 
$\hat T_3$ does not commute with the operator performing a $2\pi$ shift of the theta angle. The latter operator is defined by its action on $\hat A$ eigenstates:
\begin{eqnarray}
\label{2pishiftoperator}
\hat V_{2 \pi} \ket{A}&=& \ket{A} e^{ i 2 \pi S_{CS}[A]}, ~\text{where}  ~ S_{CS}  = {1 \over 8 \pi^2} \int\limits_{\T^2 \times \S^1}  \tr( A\wedge F -{i \over 3} A \wedge A \wedge A)~.
\end{eqnarray}
Using the well-known transformation law of the Chern-Simons action\footnote{We note that here we correct   the sign of the second term given  in \cite{Cox:2021vsa}.}
\begin{eqnarray} \label{cstransform}
  S_{CS}[g \circ A] - S_{CS}[A]&=& {1 \over 24 \pi^2} \int_{\T^3} \tr (g d g^{-1})^3 - {1 \over 8 \pi^2} \int_{\T^3} d \tr( i A d g^{-1} g),
  \end{eqnarray}
  recalling that $\hat T_3$ is defined via the improper gauge transformation (\ref{center3}) with half-integer winding number (\ref{windingt3}), and noting that the boundary term in (\ref{cstransform}) vanishes due to the $\Gamma$-gauge boundary conditions, we find the commutation relation 
  \begin{eqnarray}\label{anomaly1}
 \hat T_3 \hat V_{2 \pi} &=& - \hat V_{2 \pi} \hat T_3.
\end{eqnarray}
The nontrivial group commutator between $\hat T_3$ and $\hat V_{2 \pi}$ reflects the mixed anomaly between the center symmetry and parity, as we discuss below.

The parity operation acts on $A$ in a way familiar from $\R^3$, except for the fact that consistency with the $\Gamma$-periodicity, eqn.~(\ref{bc}),  on $\R^3/\Z^3$ requires the inclusion of the matrix $\Gamma_P$ in the transformation:
\begin{equation}\label{parity1}
P: \; A_k(x^1,x^2,x^3)  \rightarrow A_k^P(x^1,x^2,x^3) =  -\Gamma_P A_k( -x^1, -x^2, -x^3) \Gamma_P^{-1}~,~ x \in \R^3.
\end{equation}
Demanding that  $A_k^P(x)$ also satisfy the $\Gamma$-periodic boundary conditions \eqref{bc}, it can be easily checked that the matrix $\Gamma_P \in SU(2)$, $\Gamma_P^2=\pm 1$, should obey 
\begin{equation}\label{gammap}
\Gamma_i \Gamma_P \Gamma_i = e^{ i \phi_i} \Gamma_P , 
\end{equation}
with $e^{i \phi_i}$ an arbitrary $\Z_2$ phase. For our choice of $\Gamma_i$ from  (\ref{gammagauge}), we  take $\Gamma_P = i \sigma_1$. 
We also recall that parity is only a symmetry of (d)YM at $\theta=0$ and $\theta=\pi$.

We begin with   $\theta = 0$.
Let $\hat P_0$ denote the operator that implements the transformation (\ref{parity1}) on our large Hilbert space. The subscript denotes that this is the correct parity symmetry operator for $\theta = 0$. Notice that also $\hat P_0^2 = 1$ as required.
By considering the above action of $\hat P_0$  and $\hat T_i$ on an arbitrary eigenstate of $A$ in the large Hilbert space, it follows that $\hat P_0 \hat T_i \hat P_0$ acts as a center symmetry transformation $\hat{T}_i' $ with
\begin{equation}\label{tprime}
{T}_i' (x^1,x^2,x^3) = \Gamma_P T_i( -x^1, -x^2, -x^3) \Gamma_P.
\end{equation} 
Now recall that $T_i$ obeys the boundary conditions \eqref{tboundaryconditions}. From   \eqref{tprime}, making use of (\ref{gammap}), we find $\hat{T}_i'(x + \hat e_k L^k) = \Gamma_P T_i( - x - \hat e_k L^k) \Gamma_P = e^{i \pi \delta_{ik}}\Gamma_P \Gamma_k^{-1} T_i (- x) \Gamma_k \Gamma_P = e^{i \pi \delta_{ik}} \Gamma_k \Gamma_P T_i(-x) \Gamma_P \Gamma_k^{-1} =  e^{i \pi \delta_{ik}} \Gamma_k   T_i'( x)  \Gamma_k^{-1}$. Thus, $T_i'$ obeys the same\footnote{Note that this is so only for  $SU(2)$ only, for $SU(N)$ $T_i'$ obeys the boundary conditions of $T_i^{-1}$.} boundary conditions as $T_i$. Therefore, on the space of physical states, we find that  parity and center-symmetry  commute at $\theta =0$:
\begin{equation}
\label{eqn:P0_algebra}
\hat P_0 \; \hat T_i\; \hat P_0 = \hat T_i.
\end{equation}
Hence, $\hat P_0$ leaves the eigenstates of $\hat T_i$, the $\Z_2$ electric flux states, invariant $\hat P_0: \ket{\vec{e}} \rightarrow\ket{\vec{e}}$. 
Note also that $\hat P_0$ does not change the sign of the magnetic field, $\hat P_0 \hat B_i(x,y,z) \hat P_0 = \Gamma_P \hat B_i( -x, -y, -z) \Gamma_P$, but changes the sign of $\hat \Pi_i$, the electric field. 

 For $\theta =\pi$ (the other value where $P$ is a symmetry), it is convenient to work in 
the $\theta=0$ Hilbert space ${\cal{H}}_{\theta = 0}$ and put the $\theta$-dependence in the Hamiltonian:
\begin{equation}\label{hamiltoniantheta} \hat H_\theta = \int\limits_{\T^3} d^3 x \left[ \sum\limits_{i=1}^3 \left( {g^2\over 2} ( \hat \Pi_i^a -{\theta \over 8 \pi^2} \hat B_i^a)(\hat \Pi_i^a -{\theta \over 8 \pi^2} \hat B_i^a)+ {1 \over 2 g^2}\; \hat B_{i}^a \hat B_{i}^a\right) + {c \over L^4} |\tr W_3|^2 \right] .\end{equation} Since parity reverses the sign of the electric field but not the magnetic field, this form makes it clear that at $\theta=\pi$ parity involves a $2\pi$ shift of the theta-angle. Explicitly,  
using the commutation relation $\hat V_{2 \pi} \hat \Pi_i^a \hat V_{2\pi}^{-1} = \hat \Pi_i^a - {1 \over 4 \pi} \hat B_i^a$, we find
\begin{equation}\label{hamiltoniantheta2}
\hat V_{2\pi} \hat P_0  \hat{H}_{\theta=\pi} \hat P_0 \hat V_{2\pi}^{-1} = \hat{H}_{\theta=\pi}.
\end{equation} In other words,  parity  at $\theta=\pi$ is generated by the operator\begin{equation} \label{paritypi}
 \hat P_\pi = \hat V_{2\pi} \hat P_0~.
\end{equation} Notice that $\hat P_0 \hat V_{2\pi}  \hat P_0 = \hat V_{2\pi}^{-1}$, so $\hat P_\pi^2 = 1$ as required for a parity symmetry.
From (\ref{anomaly1}) and the fact that $\hat T_1$ and $\hat T_2$ commute with $\hat V_{2\pi}$, we then find that the $\theta=\pi$ center-symmetry and parity algebra is
\begin{equation}
\label{eqn:Ppi_algebra}
\hat P_\pi \; \hat T_i\; \hat P_\pi = e^{i \pi \delta_{i3}} \hat T_i.
\end{equation}
Thus, $\hat P_\pi$ acts on $\Z_2$ electric flux\footnote{These are the eigenstates of $\hat T_i$: $\hat T_i |\vec{e}\rangle = e^{i \pi e_i} |\vec{e} \rangle$. Since $T_i$ commute with $H$, the energy eigenstates are also flux eigenstates.} states as $\hat P_\pi: \ket{e_1, e_2, e_3} \rightarrow \ket{ e_1, e_2,  1+e_3 (\text{mod} 2)}$, where $e_i = \{0,1\}$.
Since $\hat P_\pi$ and $\hat T_i$ are both symmetries,  eq.~(\ref{eqn:Ppi_algebra}), representing the Hilbert space incarnation of the mixed center-parity anomaly, implies that all energy eigenstates with $e_3 =0$ and $e_3=1$ are exactly degenerate. 

We note that the above discussion of symmetries and their realization is valid on arbitrary-size $\T^3$ irrespective of whether the theory is weakly coupled. However, performing dynamical calculations is only possible in  various semiclassical limits. The perturbative spectrum of YM in the femtouniverse with a twist has been studied in \cite{GonzalezArroyo:1987ycm}. The semiclassical configurations  responsible for nonperturbative effects on $\T^2 \times \S^1$ in the different semiclassical limits are, in most cases, not explicitly known. A notable exception are 't Hooft's constant flux instanton solutions on a ``symmetric'' torus \cite{tHooft:1981nnx}, studied further in \cite{vanBaal:1984ar,GarciaPerez:2000aiw,Gonzalez-Arroyo:2019wpu,Anber:2022qsz}. Improving the understanding of instanton configurations on the torus is important for further progress.  In this paper, however, we shall not consider details of nonperturbative calculations; instead, our focus is on symmetries and perturbative stability.

\subsection{Femtouniverse: pure Yang-Mills ($c=0$)}\label{sec:femtogamma}
\label{sec:femto}

Here, we study the ground sates of pure YM theory (whose Lagrangian is (\ref{action}) with $c=0$) in the framework of femtouniverse with a twist. This is an old subject, reviewed in \cite{vanBaal:2000zc}, but we include the discussion in order to facilitate comparison with dYM. In addition, the understanding \cite{Cox:2021vsa} of the mixed parity/center-symmetry anomaly in this framework is new. 

\subsubsection{$\Gamma$-gauge vacua,  anomaly and degeneracy} 

The theory is weakly coupled due to the small size of $\T^3$. Minimizing the classical energy (\ref{energydensity}) with $c=0$, one finds that there are two classical field configurations, denoted by $A^\pm$,\footnote{
A proof that these are the only   $F=0$ configurations is given in Sect. 7 of  \cite{Witten:1982df}.}
 with $F=0$ and thus classically zero energy (other states where $A$ has an $x$-dependent variation have higher energy):
\begin{equation}\label{femtogamma}
A^{+, \Gamma}=0, ~ A^{-, \Gamma} = T_3 (A^{+, \Gamma} - i d) T_3^{-1} =  - i T_3 d T_3^{-1}~. 
\end{equation}  
These are not gauge transformations of each other, since $T_3$ is not $\Gamma$-periodic and is thus not a gauge transformation. Instead, the zero-energy configurations $A^{\pm, \Gamma}$ are related by the action of center symmetry in the $x^3$ direction. 
Let us denote by $\ket{+}$ and $\ket{-}$ the  quantum states in the physical Hilbert space built by gauge averaging  the eigenstates of $\hat{A}$ with eigenvalues $A^{\pm, \Gamma}$. From eq.~(\ref{femtogamma}) above, it is clear that $\hat T_3: \ket{+} \leftrightarrow \ket{-}$. Thus, the $\hat T_3$ eigenstates, the $e_3$-flux states are $\ket{e_3=0} = {1 \over \sqrt{2}} (\ket{+} + \ket{-})$ and $\ket{e_3=1} =  {1 \over \sqrt{2}} (\ket{+}-\ket{-})$.  The two  classically degenerate ground states $\ket{\pm}$ both have vanishing flux in $x^{1,2}$, $e_1=e_2=0$. To summarize, the   two classical ground states in the femtouniverse with a unit twist in the $\T^2$ are
\begin{eqnarray}
\vert + \rangle, \vert -\rangle &=& \hat T_3 \vert + \rangle~, ~{\text{or}}~~ 
\vert e_3 \rangle =  {1 \over \sqrt{2}}(\vert + \rangle + (-1)^{e_3} \vert - \rangle)~, ~ e_3 = \left\{0, 1\right\}~. \label{femtoground}
\end{eqnarray}

Consider parity at $\theta =0$ and study the action of $P$ of (\ref{parity1}) on the classical configurations $A^{\pm, \Gamma}$ (\ref{femtogamma}). Clearly, we have that $\hat P_0 \ket{+} = \ket{+}$, since $A^{+, \Gamma}=0$ is invariant under (\ref{parity1}). In addition from (\ref{eqn:P0_algebra}),  $\hat P_0$ commutes with $T_3$, so the $A^{-, \Gamma}$ state is also parity invariant; this is consistent with the already noted  fact that $\hat P_0$  acts trivially on the $\ket{e_3=\{0,1\}}$ eigenstates. Taken together, the previous two relations imply that  $\hat P_0 \ket{-} = \ket{-}$. Parity commutes with the Hamiltonian and so the $\ket{e_3=0,1}$ states can be taken to be energy eigenstates; however, there is no symmetry reason that the states $\ket{e_3=0}$ and $\ket{e_3=1}$ be degenerate.

 Consider now $\theta=\pi$. We have that  $\hat V_{2\pi} \ket{+} = \ket{+}$ and $\hat V_{2\pi} \ket{-} = - \ket{-}$. The latter equation follows from the commutation relation between $\hat T_3$ and $\hat V_{2\pi}$ or, equivalently, simply by noting that $V_{2\pi}[A^{-,\Gamma}] = -1$ due to the half-integer winding number of $T_3$, while $V_{2 \pi}[A^{+,\Gamma}]=1$. Since by (\ref{paritypi}) $\hat P_\pi = \hat V_{2\pi} \hat P_0 $ we have that $\hat P_\pi: \ket{e_3=0} \leftrightarrow \ket{e_3=1}$, i.e. these states are exactly degenerate, in accordance with the anomaly (\ref{eqn:Ppi_algebra}). No effect can lift the degeneracy since the commutation relation is an exact property of the theory.
 
 \subsubsection{$\Omega$-gauge vacua}
 
 In this gauge, the boundary conditions (\ref{omegagauge}) are now $x^1$-dependent, hence finding the classical configuration of zero energy is less trivial. However, the map (\ref{mapgauge}) comes to rescue. The two classical backgrounds with $F=0$ easily seen in the $\Gamma$-gauge  are mapped to classical backgrounds with $F=0$ in $\Omega$ gauge. These are (setting $L^1=L^2=L=1$ for brevity)
 \begin{eqnarray}\label{zerofluxomega}
 A^{\pm, \Omega} &=& g^{-1}({x^1}, {x^2})\circ A^{\pm, \Gamma}(x), \nonumber \\
 A^{+, \Omega} &=& - i (g^{-1} d g)(x^1,x^2),   \\
 A^{-,\Omega} &=& - i g^{-1} T_3 d ( T_3^{-1} g) = \pi \sigma_3 dx^3 - i e^{- i \pi x^3 \sigma_3} (g^{-1} d g)(x^1,x^2) e^{i \pi x^3 \sigma_3},\nonumber
 \end{eqnarray}
 where explicit expressions for $(g^{-1} dg)(x^1,x^2)$ can be found in the Appendix, see (\ref{gderivatives}).
 By construction, these $\Omega$-gauge backgrounds have $F=0$, obey the right boundary conditions (\ref{omegagauge}), and are distinguished by the expectation value of the Wilson line in the $x^3$-direction   parallel to the magnetic flux, $W_3[A^{\pm, \Omega}] = \pm {\bf 1}$. This corresponds to the breaking of the $\Z_2$ center symmetry in the $\vec{m}$ direction by the classical ground states. On the other hand, the  Wilson lines orthogonal to the magnetic flux, $W_{1,2}[A]$, are at   center symmetric points when evaluated at the solutions $A^{\pm, \Omega}$.\footnote{The properties of $T_3$ and $g$, and the definition of the gauge invariant Wilson line  can be used to find explicit expressions.}
We stress that there is no continuous classical vacuum degeneracy in the femtouniverse with boundary conditions twisted by $n_{12}=1$. This was, in fact, noted and used for the calculation of the Witten index \cite{Witten:1982df}.

Our final remark is that it is possible to study the pure Yang-Mills femtouniverse symmetries in the $\Omega$-gauge as well, but it is more straightforward to do so in the $\Gamma$ gauge and appeal to the equivalence between the two.

\subsection{dYM: small-circle ($c \ne 0$)}

\label{sec:dym}

We now revert to the study of dYM, with Lagrangian (\ref{action}) with $c\ne 0$, focusing on the classical vacua and the action of symmetries. Our goal is to contrast the findings with those of the femtouniverse from the previous Section \ref{sec:femto}.

\subsubsection{$\Omega$-gauge dYM} 

In contrast to the femtouniverse limit, here the minimization of the classical energy is more straightforward in $\Omega$-gauge. The static energy given in (\ref{energydensity}) and reproduced below  has  contributions due to the various magnetic field energies and the double-trace deformation: \begin{eqnarray}
\label{energydensity2}
{\cal{E}} &=& {g^2 \over 2} (\Pi_i^a \Pi_i^{a}  + \Pi_3^a \Pi_3^a) + {1 \over 2 g^2} F_{12}^a F^a_{12} + {1 \over 2 g^2} F_{3 i}^a F^a_{3 i} + {c \over L^4} | \tr W_3|^2~,
\end{eqnarray}

We begin by taking the deformation term to dominate and 
thus require that the classical vacuum   minimize  the deformation energy. Being a nonnegative  quantity, it is minimized for $\tr W_3=0$. The center-symmetric holonomy  
 \begin{equation}\label{centerholonomy}
 A^{\Omega_3, \pm} = \mp {\pi \over L} {\sigma_3 \over 2}
 \end{equation}
  makes the deformation energy vanish. To obtain (\ref{centerholonomy}), one uses an $\Omega$-periodic gauge transformation to put the center symmetric $A_3$ into the third isospin direction; on the distinction between the two signs, see discussion after (\ref{centerstabilitydym}) below.
 
  Next, we consider the  contributions to the energy due to the $F_{i3}$-terms in (\ref{energydensity2}). Since the center-symmetric holonomy (\ref{centerholonomy}) is $ \sim {\sigma_3 \over L}$, at small $L$, any nonzero classical field $A_{i=1,2}$ which does not commute with $\sigma_3$ will have an energy cost growing at small $L$ due to the commutator terms in $F_{i3}$. Thus, in the small-$L$ limit of interest to us, the lowest energy configurations  should commute with $\sigma_3$. Likewise, at small $L$, the lowest energy configurations should be $x^3$-independent, as an $x^3$ dependence would also lead to $1/L$ contributions to $F_{i 3}$.

Thus, we  focus on the most obvious possibility: we let $A_{1,2}$ point in the 3\textsuperscript{rd} isospin direction and be $x^3$-independent. Recall that the $\Omega$-gauge boundary condition can not be satisfied by $A_{1,2}$ identically vanishing  or constant, because of the space-dependence of $\Omega_2(x^1)$ in (\ref{omegagauge}), which implies the appearance of an inhomogeneous term in the relation (\ref{fields}) between $A_1(x^2=L^2)$ and $A_1(x^2=0)$. Thus, it   precludes an $x$-independent $A_1$. We are thus forced, in this abelian sector, to consider the following configuration\footnote{The need to consider the abelian sector is to preclude $1/L$ terms in the energy density due to the nonzero $\S^1$ holonomy (\ref{centerholonomy}) imposed by minimizing the deformation energy. Likewise, $x^3$-dependence would also incur $1/L$ terms. Thus, we note that the most general $x^{1,2}$-dependent Cartan-subalgebra field configuration obeying the $\Omega$-gauge boundary conditions is the one given in (\ref{a1}), with the addition of arbitrary periodic functions of $x^1$ and $x^2$. The periodic functions lead to extra energy cost (and are constrained by the energy extremum equations), leaving us with (\ref{a1}) as the lowest energy abelian configuration.} as a ground-state candidate
\begin{eqnarray}
\label{a1}
A_1(x^2) &=& \left(- {2 \pi x^2 \over L_1 L_2}  + {a_1 \over L_1}\right){\sigma^3 \over 2},~\text{obeying} \; A_1^3(x^2+L_2) = \Omega_2 (A_1^3(x^2) - i  \partial_1) \Omega_2^{-1}, 
\end{eqnarray}
\begin{eqnarray}
A_2 &=& {a_2 \over L_2} {\sigma^3 \over 2}, ~A_3  =  \pm {\pi \over L} {\sigma^3 \over 2}, ~\text{where} ~ A_{2 (3)} = \Omega_2 A_{2 (3)} \Omega_2^{-1},  \nonumber~
\end{eqnarray}
where, for brevity, we dropped the superscript $A^\Omega$, indicating that the fields are in the $\Omega$-gauge. We also showed that  this configuration obeys the boundary condition (\ref{fields}) with $\Omega_2$ from (\ref{omegagauge}). The equations for the extremization of the energy (\ref{energyextrema}, \ref{energyextrema2}) are also obeyed, as the field strength of (\ref{a1}) is constant and all commutator terms there vanish. 

 We have also included  constant terms to $A_1^3$ and $A_2^3$,  $a_1$ and $a_2$, which are allowed both by the $\Omega$-gauge boundary conditions and  the classical energy minimization   (\ref{energyextrema}, \ref{energyextrema2}). The values of  $a_1$ and $a_2$ are shifted by $4 \pi$ by the $\Omega$-periodic gauge transformations  $G_{k_1, k_2} (x^1, x^2) = e^{i 4 \pi \left(k_1 {x^1 \over L_1} + k_2 {x^2 \over L_2}\right) {\sigma^3 \over 2} }$, so we have $a_{i=1,2} \simeq a_{i=1,2} + 4 \pi \Z$.    The traces of the Wilson loop operators winding in the $x^{1,2}$ directions in the abelian configurations (\ref{a1}) are  
\begin{eqnarray}
\label{wilson12}
W_1(x^2) &=&\tr e^{i ({a_1 } - {2 \pi x^2 \over L_2  }) {\sigma_3\over 2} }= 2 \cos({a_1 \over 2} - {  \pi x^2 \over   L_2})~,\\
W_2(x^1) &=&\tr e^{i  a_2 {\sigma_3 \over 2}} \Omega_2(x^1) =2   \cos({a_2 \over 2 } + {  \pi x^1 \over L_1    }) ~, \nonumber
\end{eqnarray}while   $\tr W_3 =0$ to minimize the deformation energy.
Notice that for fixed $a_{1,2}$, $W_1(x^2+1) = - W_1(x^2)$, and likewise for $W_2(x^1)$, as expected in the $n_{12}=1$ background.   The nonvanishing values of $W_{1,2}$ in the background (\ref{a1}) mean that center symmetry in the $x^{1,2}$-directions is broken in this background in an $x^{1,2}$-dependent manner (for any $a_1, a_2$).
  
While the arbitrary $a_1, a_2$ appearing in (\ref{a1}) might appear as a continuous vacuum degeneracy, we shall argue that they correspond to a choice of origin of coordinates (as one can already infer from (\ref{wilson12})) and no physical gauge invariant quantities depend on them.  Physically, this is because the background (\ref{a1}) represents a constant homogeneous magnetic field in the translationally invariant $\T^2$ of the $1-2$ plane where there is no preferred point. We shall discuss this in detail in the next Section, as  the $\Gamma$-gauge action of translations is more transparent.
  
In summary,  we found an  abelian configuration that obeys $\Omega$-gauge boundary conditions in the $\T^2$ directions and minimizes the deformation energy:\footnote{In the $\Omega_{(-1)}$ gauge, the constant abelian flux configurations are 
$A^{\pm, \Omega_{(-1)}}(x) = \left(\left( {2 \pi x^2 \over L^1 L^2}- {a_1 \over L^1}\right) dx^1 - {a_2  \over L^2} dx^2    \mp {\pi \over  L} dx^3\right) {\sigma_3 \over 2}$; transformed to $\Omega$-gauge via (\ref{omega0minus1}) these map to $A^{\pm, \Omega}(x)$. As noted earlier near (\ref{kflux}), there are fluxes where $|F_{12}^3|$ is larger but these require studying $\Omega_{(k)}$ gauges with $|2k+1|>1$. In other words,  (\ref{omegavacuumdym})  contains the two degenerate lowest flux configurations   of \cite{Unsal:2020yeh}.} \begin{equation}\label{omegavacuumdym}
A^{\pm, \Omega}(x) = \left(\left(- {2 \pi x^2 \over L^1 L^2} + {a_1 \over L^1}\right) dx^1 + {a_2 \over L^2} dx^2    \mp {\pi \over  L} dx^3\right) {\sigma_3 \over 2}~. 
\end{equation}
We shall henceforth call the classical configurations (\ref{omegavacuumdym}) ``classical vacua'' (in addition, we shall put $a_{1,2}=0$ most of the time). 

{\flushleft{W}}e now summarize  the  properties of (\ref{omegavacuumdym}):
\begin{enumerate}
\item $A^{\pm, \Omega}(x)$ minimize the ``deformation'' energy, setting it to zero:
\begin{equation}\label{centerstabilitydym}
W_3\big\vert_{\text{evaluated for} \; A^{\pm, \Omega}}= \mp i \sigma_3, \; \text{hence} \; \tr W_3 = 0.
\end{equation}
The two center-symmetric points above are distinct and are not related by an $\Omega$-periodic  gauge-group Weyl transform (since $\sigma_2$ is not $\Omega$-periodic: as per (\ref{omega0minus1}), it changes the transition functions and maps to a different Hilbert space). One can also consider the effect of the   transformations $G_k (x^3) = e^{i 2 \pi k {x^3 \over L} {\sigma^3 \over 2}}$, $k \in \Z$. For even-$k$, $G_{k=2p}(x^3)$ is an $\Omega$-periodic gauge transformation, but only shifts (\ref{centerholonomy}) by 
${4 \pi p\over L} {\sigma^3 \over 2}$. Thus, it can not map between the two values of the holonomy (\ref{centerholonomy}). 

As we discuss further below, the two configurations in (\ref{omegavacuumdym}), while not distinguished by the value of the gauge invariant $\tr W_3$  are distinguished by a different gauge invariant operator, see  (\ref{centerorderdym}) below.

\item $A^{\pm, \Omega}(x)$ obey $A^{\pm, \Omega}(x+ \hat e_2 L^2) = \Omega_2(x)(A^{\pm, \Omega}(x )- i d)\Omega_2^{-1}(x)$, with $\Omega_2(x)$ from (\ref{omegagauge}). At the same time, they are  periodic in $x^1$ and $x^3$, all in accordance with the $\Omega$-gauge boundary conditions (\ref{omegagauge}).
\item The two field configurations $A^{\pm, \Omega}(x)$ have the same field strength $F^{ \Omega} = {2 \pi dx^1 \wedge dx^2 \over L^1 L^2}  {\sigma_3 \over 2}$, or $F_{12}^{ \Omega, 3} =  {2 \pi \over L^1 L^2}$, while $F_{13}=0$ due to the constancy of $A_3^{\pm, \Omega}$ and the abelian nature of (\ref{omegavacuumdym}). Thus, the classical   energy (\ref{energydensity}, \ref{energy}) of the field configuration (\ref{omegavacuumdym}) is 
\begin{equation}
\label{dymenergy}
E_{class.} = {1 \over 2 g^2} {4 \pi^2 L \over L^1 L^2}.
\end{equation} 

\item While we have not proven that (\ref{omegavacuumdym}) is the lowest energy configuration in dYM with $n_{12}=1$, this appears very plausible, at least in the fixed-$L$, large-$L^{1,2}$ limit where the classical energy (\ref{dymenergy}) can be made vanishingly small. 
 
\item The $SU(2)$-gauge invariant expression for the field-strength of (\ref{omegavacuumdym}) is
\begin{equation}\label{centerorderdym}
\tr (F_{12} W_3)\big\vert_{\text{evaluated for} \; A^{\pm, \Omega}}= \mp i {2 \pi \over L^1 L^2}.
\end{equation}
The order parameter (\ref{centerorderdym}) is winding in the $x^3$ direction  hence its nonvanishing in the classical vacuum (\ref{omegavacuumdym})  indicates  that, classically, the $\S^1$-center symmetry is ``broken'' in dYM in the 't Hooft flux background\footnote{{The center symmetry does not break in the quantum theory where the vacuum states are given by flux states, see (\ref{fluxstatedef}). It is straightforward to show that the expectation value $\langle e_3| \tr (F_{12} W_3) | e_3 \rangle$ vanishes identically, thus restoring center symmetry. This is similar to the situation in the femtouniverse (see Section \ref{sec:discussion} for a further discussion).}}.

This ``breaking'' occurs despite the center stabilizing deformation and the associated vanishing of $W_3$ (\ref{centerstabilitydym}). Here, instead, the order parameter for the center-breaking (in $\S^1$) is the non-Lorentz invariant (under 3d Lorentz transforms), $\tr (F_{12} W_3)$. This breaking, however, vanishes in the infinite $L^{1,2}$, or $\R^3 \times \S^1$, limit. 
\end{enumerate}
Thus, the picture that emerges is that the classical ground states of dYM in $n_{12}=1$ background, in the large-$L^1, L^2$ limit are the two configurations (\ref{omegavacuumdym}), which are mapped into each other by the action of center symmetry in the $\S^1$ direction, similar to (\ref{femtoground}). That this is so is further corroborated by the study of the classical background (\ref{omegavacuumdym}) in the $\Gamma$-gauge, to which we now turn. 

\subsubsection{$\Gamma$-gauge dYM: symmetries,   degeneracy, and anomaly}

Here, we shall use (\ref{mapgauge}) to map (\ref{omegavacuumdym}) to $\Gamma$-gauge and study the symmetry properties of the two classically degenerate ground states found above.
The reason the $\Gamma$ gauge is somewhat preferred is that the action of the symmetries is most straightforward (the definition of the transformations under the $2\pi$ shift of the $\theta$-angle in $\Omega$-gauge would have to include various boundary terms absent in the $\Gamma$-gauge). 

The $\Gamma$-gauge backgrounds corresponding to $A^{\pm, \Omega}$ from (\ref{omegavacuumdym}) are the two configurations, setting $L_1=L_2=L=1$ for brevity  \begin{eqnarray}\label{gammavacuumdym}
A^{\pm, \Gamma}(x) &=& g(x^1, x^2) \left[A^{\pm, \Omega}(x^1,x^2) - i d \right]g^{-1} (x^1, x^2)\\
 &=&g(x^1, x^2) \left[((a_1- {2 \pi x^2}) dx^1 + a_2 dx^2 \mp {\pi } dx^3) {\sigma_3 \over 2} - i d \right]g^{-1} (x^1, x^2)~.\nonumber
\end{eqnarray}
By our discussion above, $A^{\pm, \Gamma}$ obey $\Gamma$-periodicity and have the same vacuum energy (\ref{dymenergy}) and the same nonzero value of the gauge invariant order parameter  $\tr (F_{12} W_3)$  (\ref{centerorderdym}).

{\flushleft \bf{Translations:}} The transition functions in $\Gamma$-gauge are constant and hence translationally invariant. The action of translations is especially simple, we simply have 
\begin{eqnarray}\label{translation1}
  x^i \rightarrow x^i + \epsilon^i: ~
A^\Gamma(x)  \rightarrow  A^\Gamma(x + \epsilon^i).  \end{eqnarray}
It is illuminating to apply this to our vacuum configuration (\ref{gammavacuumdym}). We use $g(x+ \epsilon^i) = g(x)[1 +\epsilon^i g^{-1}(x) \partial_i g(x)]$ and obtain:
\begin{eqnarray}\label{shiftedA}
A^{\pm, \Gamma}(x+ \epsilon^i) &=& g(x)\left\{A^{\pm, \Omega}(x) + \epsilon^i \partial_i A^{\pm,\Omega}(x) + d \omega_\epsilon + i [A^{\pm, \Omega}, \omega_\epsilon] - i d \right\} g^{-1}(x)~\nonumber \\
&=&g(x)\left\{A^{\pm, \Omega}(x^i + \epsilon^i) + d \omega_\epsilon + i [A^{\pm, \Omega}, \omega_\epsilon] - i d \right\} g^{-1}(x) \\
&=& g(x) \left\{A^{\pm, \Omega}(x^i + \epsilon^i) + D[A^{\pm, \Omega}(x)] \omega_\epsilon(x) - i d \right\} g^{-1}(x). \nonumber
\end{eqnarray}
Here, $\omega_\epsilon$ is an infinitesimal gauge transformation, whose components $g^{-1}(x) \partial_j g(x)$  are found in eqn.~(\ref{gderivatives}) of the Appendix. In the $0 \le x^2 \le 1$ strip of the $x^{1,2}$-plane,   $\omega_\epsilon$ is given by:
\begin{eqnarray}\label{omegacompensate}
\omega_\epsilon &=&   \epsilon^1 \omega_1 + \epsilon^2 \omega_2, ~\text{with} \\
\omega_1 &=&  {\sigma^3 \over 2} \pi (  4 f(x^2)^2 -3) + 2 \pi f(x^2) f(x^2+1) (\sigma^1 \sin 2\pi x^1 - \sigma^2 \cos 2 \pi x^1), \nonumber\\
\omega_2 &=& (\sigma^1 \cos 2 \pi x^1 + \sigma^2 \sin 2 \pi x^1) (f'(x^2) f(x^2+1) - f(x^2) f'(x^2+1))~, \nonumber 
\end{eqnarray}
where $f(x^2)$ is the ``bump'' function entering the definition of $g(x^1,x^2)$.
Before discussing the consequences of (\ref{shiftedA}), we note that it is of the same form as (\ref{gammavacuumdym}), but with the $x^i$ argument of $A^{\pm, \Omega}$ shifted by $\epsilon^i$ and an additional compensating gauge transformation $\omega_\epsilon$ defined in (\ref{omegacompensate}).  

The most interesting properties of $\omega_\epsilon$ are its periodicity properties. First, clearly, it is periodic in $x^1$ with periodicity $1$ (recall we set $L_1=1$, etc.). In addition, all terms in $\omega_\epsilon$ are periodic in $x^2$, owing to the properties of $f(x^2),$\footnote{Recall that $f(x^2)$ is the square root of the bump function $\tilde f$ from  Fig.~\ref{fig1}. Its properties  make  it easy to see that both $f(x^2) f(x^2+1)$ and $f'(x^2) f(x^2+1) - f(x^2) f'(x^2+1)$ take the same values at $x^2=0$ and $x^2=1$, ensuring periodicity of the non-Cartan parts of $\omega_\epsilon$.}  except for the first term in $\omega_1$, proportional to $\sigma^3$. Thus, this is the only term that can affect the Wilson loops (\ref{wilson12}) and  $a_{1,2}$.
From (\ref{omegacompensate}) we find $\omega_1(x^2=1)- \omega_1(x^2=0) = {\sigma^3 \over 2} 4 \pi (f^2(1)-f^2(0)) =  {\sigma^3 \over 2} 2 \pi$. Thus, the change of $\omega_1$ upon crossing the torus in the  $x^2$ direction generates precisely a $2 \pi \epsilon^1$ shift of the $a_2$ coefficient in (\ref{gammavacuumdym}), as  already surmised from (\ref{wilson12}).\footnote{To see this, it is best to consider gauge invariant observables depending on $a_2$, i.e. $x^2$-winding Wilson loops. They are affected by the non-periodic gauge transformation $\omega_\epsilon$ in a manner equivalent to shifting $a_2$. {This shift can also be seen directly in the $\Omega$-gauge, where a shift of $x^1$ requires a compensating gauge transformation to restore the $\Omega$-gauge transition functions. It is straightforward to show that this compensating gauge transformation produces the necessary $2\pi\epsilon^1$ shift of $a_2$.}}  The shift of $a_1$, on the other hand, is simply due to the change of the $x^2$ argument of $A^{\pm, \Omega}$ from the $\epsilon^2 \partial_2 A^{\pm,\Omega}(x)$ term in (\ref{shiftedA}), and is also as implied by (\ref{wilson12}). 

The moral of the above discussion is that a proper definition of translations confirms the intuition that $a_{1,2}$ correspond to a choice of origin on the translationally invariant homogeneous magnetic background in the $1-2$ plane. Thus, in our further discussion, we shall set them to zero.

{\flushleft \bf{Center symmetries:}}  Next, we note that $A^{\pm, \Gamma}$ are mapped to each other by $T_3$, the center symmetry transform in the $\S^1$ direction. Explicitly, one easily verifies that the two configurations (\ref{gammavacuumdym}) obey
\begin{eqnarray}
\label{a3center}
 A^{-, \Gamma}(x) = T_3 (A^{+, \Gamma} - i d) T_3^{-1}. 
\end{eqnarray}
This follows after substituting the  form of $T_3$ from (\ref{center3}), where it is expressed in terms of $g$.
 Recalling the femtouniverse relation, eqn.~(\ref{femtogamma}) and the subsequent discussion, we can not fail to notice the parallel, as far as the $T_3$ action on the states $\ket{\pm}$  of the femtouniverse  is concerned. 
 
The action of $\hat T_3$ on  $\tr (F_{12} W_3)$ can also be found in the $\Gamma$-gauge. Since the fields are not abelian, we revert to the proper definition of $W_3$, recalling that $\Gamma_3 =1$, $\tr F_{12} W_3 (x^1,x^2,x^3) = \tr   F_{12}\; {\cal P} e^{i \int\limits_{x^3  }^{x^3 + L} d x^3 A_3^{\pm, \Gamma}(x^1,x^2,x^3)}$. Then the action of $\hat T_3$ is represented by $T_3$  from (\ref{tboundaryconditions}):
\begin{eqnarray}\label{ordertransform}
\hat T_3: \tr F_{12} W_3 (x^1,x^2,x^3) &\rightarrow &\tr T_3(x^1,x^2,x^3) F_{12}   {\cal P} e^{i \int\limits_{x^3  }^{x^3 + L} d x^3 A_3^{\pm, \Gamma}(x^1,x^2,x^3) }   {T_3^{-1}}(x^1,x^2,x^3+L), \nonumber \end{eqnarray}
thus, $\hat T_3: \tr F_{12} W_3 (x^1,x^2,x^3)  \rightarrow   - \tr F_{12}W_3(x^1,x^2,x^3)$, owing to the antiperiodicity of $T_3$ in $x^3$. Thus, the two nonzero values of (\ref{centerorderdym}) are indeed related by center symmetry transforms in $x^3$.

 Notice also that the $\hat T_{1,2}$ action is also interesting. We already determined that in the $\Gamma$-gauge, $\hat T_1$ is represented by the constant gauge transformation $\Gamma_2$ and $\hat T_2$ is represented by $\Gamma_1$. Thus, because of the boundary condition on $A$ in $\Gamma$-gauge, we have that 
 \begin{eqnarray} \label{actionoft1}
 \hat T_1:&& A^{\pm, \Gamma}(x) \rightarrow \Gamma_2 A^{\pm, \Gamma}(x) \Gamma_2^{-1} = A^{\pm, \Gamma}(x + \hat e_2 L_2), \nonumber \\
 \hat T_2:&& A^{\pm, \Gamma}(x) \rightarrow  \Gamma_1 A^{\pm, \Gamma}(x) \Gamma_1^{-1}= A^{\pm, \Gamma}(x + \hat e_1 L_1),
 \end{eqnarray}
 showing that in this gauge center symmetry along $x^{1,2}$ acts as a lattice translation. 
Along with (\ref{a3center}), this equation completes the $\hat T_i$ actions on $A^{\pm, \Gamma}$. All gauge invariant local operators are center-symmetry invariant. 
 
 To find the transformation of the winding  Wilson loops, as in (\ref{ordertransform}), we go back to their path-ordered definition. Consider $\tr   W_1 (x^1,x^2,x^3)  \equiv \tr   {\cal P} e^{i \int\limits_{x^1  }^{x^1 + L_1} d x^1 A_1^{\pm, \Gamma}(x^1,x^2,x^3)} \Gamma^1 $. Thus, recalling the property (\ref{tboundaryconditions}) obeyed by the gauge transformations representing $\hat T_{1,2,3}$ (given by $\Gamma_2, \Gamma_1, T_3(x)$, respectively) and the fact that they also act on transition functions, we have, using (\ref{actionoft1}) in the first two lines and (\ref{tboundaryconditions}) in the last line below:
 \begin{eqnarray}\label{tracew1gammagaue}
\hat T_1: \tr   W_1 (x^1,x^2,x^3) &\rightarrow&\tr  {\cal P} e^{i \int\limits_{x^1  }^{x^1 + L_1} d x^1 \Gamma_2  A_1^{\pm, \Gamma}(x^1,x^2,x^3)  \Gamma_2^{-1}} \;   \Gamma_1   \nonumber \\
 &=&\tr \Gamma_2  {\cal P} e^{i \int\limits_{x^1  }^{x^1 + L_1} d x^1 A_1^{\pm, \Gamma}(x^1,x^2,x^3)} \Gamma_2^{-1}   \Gamma_1   = - \tr W_1 (x^1,x^2,x^3)\nonumber \\
\hat T_2: \tr   W_1 (x^1,x^2,x^3) &\rightarrow&\tr \Gamma_1  {\cal P} e^{i \int\limits_{x^1  }^{x^1 + L_1} d x^1 A_1^{\pm, \Gamma}(x^1,x^2,x^3)} \Gamma_1^{-1} \Gamma_1 = \tr   W_1 (x^1,x^2,x^3)\nonumber \\
\hat T_3: \tr   W_1 (x^1,x^2,x^3) &\rightarrow&\tr T_3(x^1,x^2,x^3)  {\cal P} e^{i \int\limits_{x^1  }^{x^1 + L_1} d x^1 A_1^{\pm, \Gamma}(x^1,x^2,x^3)} (\Gamma_1 T_3(x^1,x^2,x^3) \Gamma_1^{-1})^{-1} \Gamma^1\nonumber \\
&& = \tr W_1(x^1,x^2,x^3)
\end{eqnarray}
Thus, as expected, $\tr W_1$ is invariant under $\hat T_{2,3}$ and changes sign under $\hat T_1$.
Similarly, we find that $\tr W_2$ is invariant under $\hat T_{1,3}$ and changes sign under $\hat T_2$, as quite naturally expected. The values of the Wilson loops were already given in (\ref{wilson12}). 
 
  {\bf\flushleft{Parity at $\theta=0$}}: We could use (\ref{parity1}) and follow the transformation properties of the classical dYM vacuum configurations  (\ref{gammavacuumdym}). However, a shortcut allowing us to argue that both $A^{\pm, \Gamma}$ vacua are parity invariant is to study the transformation properties of the gauge invariants characterizing the classical background. These are the Wilson loops  winding in the $\T^2$ directions, given in (\ref{wilson12}) with $a_1=a_2=0$ as discussed above, the Wilson loop winding in $x^3$ (whose trace vanishes), and the order parameter measuring the background flux of (\ref{centerorderdym}). All of these gauge invariants do not change upon $x^i \rightarrow - x^i$, showing that  parity is respected by the dYM classical ground states, similar to the situation in the femtouniverse.

{\bf\flushleft{$2\pi$ shifts of $\theta$ and mixed anomaly}}: As discussed above, there are  two minimum energy states, $A^{\pm,\Gamma}$. Then, as in the femtouniverse, 
we can build two quantum states $\ket{+}$ and $\ket{-}$ around the  minimum energy classical field configurations (\ref{gammavacuumdym}), by averaging over gauge transformations, 
and study their symmetry transforms. From (\ref{a3center}) it follows that $\hat T_3$ interchanges the two, $\hat T_3: \ket{+} \leftrightarrow \ket{-}$. Then, we conclude, similar to the femtouniverse, that the two classical ground states are
\begin{eqnarray}
\label{fluxstatedef}
\vert + \rangle, \vert -\rangle &=& \hat T_3 \vert + \rangle~, ~{\text{or}}~~ 
\vert e_3 \rangle =  {1 \over \sqrt{2}}(\vert + \rangle + (-1)^{e_3} \vert - \rangle)~, ~ e_3 = \left\{0, 1\right\}~, \label{dymground}
\end{eqnarray}
noticing that this is the same as (\ref{femtoground}). 
 Also as in that discussion, at $\theta=0$, we have that each of these states is an eigenstate of $\hat P_0$, so no degeneracy is expected. However, at $\theta=\pi$, the algebra (\ref{eqn:Ppi_algebra}) implies that $\hat P_\pi: \ket{e_3=0} \leftrightarrow \ket{e_3 = 1}$, implying double degeneracy. 
The point that we want to stress is that the symmetries' action on the  classical vacua are  exactly as in the femtouniverse:  {namely that $T_3$, the center symmetry in the $x^3$ direction, exchanges $\vert + \rangle \leftrightarrow \vert - \rangle$  in dYM with $n_{12}=1$, despite $\tr W_3=0$ in both states, at any finite $L^{1,2}$.}

{\flushleft{To}} end this Section, let us study more explicitly  the mixed anomaly at $\theta = \pi$. To this end, we need to find the action of $V_{2 \pi}$, the operator performing $2\pi$ shifts of the $\theta$-angle (which, along with $P_0$, is part of the definition of parity at $\theta = \pi$) acting on the $\ket{\pm}$ states. Since these physical states are built from the eigenstates of the field operator $\hat A$ with eigenvalues $A^{\pm, \Gamma}$ (the classical field configurations (\ref{gammavacuumdym})) by gauge averaging, it suffices to calculate the value of the classical functional $V_{2\pi}[A]$ on the classical field configurations.
 
  \begin{enumerate}
  \item Recall from (\ref{2pishiftoperator}) that $V_{2 \pi}[A] = e^{i 2 \pi S_{CS}[A]}$ and consider, recalling (\ref{a3center})
 \begin{eqnarray}\label{v2pionvacua}
  V_{2\pi}[A^{-,\Gamma}] &=& V_{2\pi}[A^{+,\Gamma}] e^{i 2 \pi \left[ S_{CS}[T_3 \circ A^{+,\Gamma}] - S_{CS}[A^{+,\Gamma}]\right]}
  \nonumber \\
  &=&V_{2\pi}[A^{+,\Gamma}]e^{ i 2\pi\left[{1 \over 24\pi^2} \int_{\T^3} \tr (T_3 d T_3^{-1})^3 - {1 \over 8 \pi^2} \int_{\T^3} d \tr( i A^{+,\Gamma} \wedge d T_3^{-1} T_3)\right]}   \nonumber\\
  &=& - V_{2 \pi}[A^{+,\Gamma}], \end{eqnarray}
where we used (\ref{windingt3}). The vanishing of the boundary term  in the second line above  is due to  $A^{+, \Gamma}(x+\hat e_i) = \Gamma_i A^{+, \Gamma}(x) \Gamma_i^{-1}$ and  (\ref{tboundaryconditions}),  $T_3(x+ \hat e_i) = \Gamma_i T_3(x) \Gamma_i^{-1} e^{i \pi \delta_{i3}}$, which, for constant $\Gamma_i$, implies that $d T_3^{-1} T_3$ is also $\Gamma_i$-periodic. Thus, the boundary term in the variation of $S_{CS}$ vanishes, showing that the last line in (\ref{v2pionvacua}) is correct. Thus, the action of $V_{2\pi}$ on the dYM vacua is the same as in the femtouniverse, as per the discussion of Section \ref{sec:femtogamma}. 
\item But what about the value of $V_{2\pi}[A^{-, \Gamma}]$ itself? As a sanity cross check, let us compute it. We use the CS term transformation  (\ref{cstransform}) with $g$ from (\ref{gunitsquare1})
\begin{equation}
  S_{CS}[g \circ A^{\pm, \Omega}] - S_{CS}[A^{\pm, \Omega}]= {1 \over 24 \pi^2} \int_{\T^3} \tr (g d g^{-1})^3 - {1 \over 8 \pi^2} \int_{\T^3} d \tr( i A^{\pm, \Omega} d g^{-1} g),
\end{equation}
and notice that  
  $(g d g^{-1})^3=0$, as follows upon inspection.  Hence we have  (replacing $dg^{-1} g = - g^{-1} d g$) that 
  \begin{eqnarray}
  \label{vapm}
 V_{2 \pi}[A^{\pm, \Gamma}] &=& e^{i 2 \pi( S_{CS}[g \circ A^{\pm, \Omega}] - S_{CS}[A^{\pm, \Omega}])}  \; e^{i 2 \pi S_{CS}[A^{\pm, \Omega}]}\nonumber \\
 &=&   e^{i {1 \over 4 \pi} \int_{\T^3} d \tr( i A^{\pm, \Omega}  g^{-1} d g)} \; e^{i 2 \pi S_{CS}[A^{\pm, \Omega}]}.~\nonumber \\
  \end{eqnarray}
Next, we evaluate the two factors appearing in (\ref{vapm}).
Using $A^{\pm, \Omega}$ from (\ref{omegavacuumdym}) and $g^{-1} d g$ from 
 (\ref{gderivatives}) we obtain (setting $L_1$$=$$L_2$$=$$L$$=1$ again)
\begin{eqnarray}
\label{v2pivacua}
e^{ i {1 \over 4 \pi} \int_{\T^3} d \tr( i A^{\pm, \Omega}  g^{-1} d g)}&=&e^{ -{1 \over 4 \pi} \int_{\T^3} d \tr[ \mp \pi dx^3 {\sigma^3 \over 2} (g^{-1} \partial_1 g\; dx^1 + g^{-1} \partial_2 g \;dx^2)]} \nonumber \\
&=& e^{\pm {i \pi \over 8} \int_{\T^3} d (-4 f^2(x^2)  d x^3 \wedge dx^1)} = e^{ \mp {i \pi \over 2} (f^2(1)-f^2(0))} = e^{ \mp {i \pi \over 4}}, \nonumber \\
 e^{i 2 \pi S_{CS}[A^{\pm, \Omega}]} &=& e^{{i \over 4 \pi}  \int_{\T^3}  \tr[ A^{\pm, \Omega} \wedge d A^{\pm, \Omega}] }  
= e^{\mp {i \pi \over 4 }  \int_{\T^3}  dx^3 \wedge  dx^1 \wedge dx^2 } = e^{ \mp {i \pi \over 4}}.
\end{eqnarray}

Inserting (\ref{v2pivacua}) in (\ref{vapm}), we find that all is in agreement, with $V_{2\pi}$ giving $i$ or $-i$ on the two states $\ket{\pm}$ described above:\begin{eqnarray} \label{v2ponpm}
\hat V_{2\pi} |\pm \rangle = |\pm \rangle e^{\mp  2 i {\pi \over 4}} = \mp i | \pm \rangle~.  \end{eqnarray}
\item To complete the study of the exact degeneracy at $\theta=\pi$, we recall that $\hat P_\pi =\hat V_{2\pi} \hat P_0$.
Combined with the discussion of $\hat P_0$ earlier in this section, where we argued that $| \pm\rangle$ are $\hat P_0$ invariant, (\ref{v2ponpm}) implies that $\hat P_\pi| \pm\rangle = \mp i \eta_P | \pm \rangle$, where $\eta_P$ is an arbitrary phase associated with the parity action on the quantum states. This in turn shows that $\hat P_\pi$ maps the flux states $|e_3 =0\rangle$ and $|e_3 = 1\rangle$ to each other, as required by the anomaly and implying the double degeneracy.
\end{enumerate}
We further discuss the implications of the above findings and the similarities between the classical ground states in dYM (\ref{dymground}) and the femtouniverse (\ref{femtoground}), in Section \ref{sec:theta}.

\section{Spectra and ``GPY'' potential due to gauge bosons and fermions}
\label{sec:GPYpotential}
In this Section, we turn to a
 study of the stability of the gauge field background (\ref{omegavacuumdym}), within a UV completion 
of dYM theory obtained by adding massive adjoint fermions. We also consider the theory with massless fermions. This does not describe dYM, but does provide an interesting comparison and may also be relevant for future studies.
On $\R^3 \times \S^1$, massless fermions have been known to ensure center stability since \cite{Unsal2008adj,Unsal2009bions,Unsal:2008ch,Shifman:2008ja}.

Here, we find the potential governing the Wilson loop in the $x^3$ direction, evaluating the gauge boson and fermion contributions, similarly to the familiar Gross-Pisarski-Yaffe (GPY) potential on $\R^3\times \S^1$ \cite{GPY1981}, but including the effect of the finite $\T^2$ with twist.

\subsection{Calculating the potential}
\label{sec:calculating}

The first step in finding the potential is to find the spectra of excitations around the general background:
\begin{equation}
\label{genbackground}
A^{W, \Omega}(x) = \left(\left(- {2 \pi x^2 \over L^1 L^2} + {\alpha_1 \over L^1}\right) dx^1 + {\alpha_2 \over L^2} dx^2   + {W \over  L} dx^3\right) {\sigma_3 \over 2}~. 
\end{equation}
Notice that this background is identical to the background (\ref{omegavacuumdym}), but has an arbitrary dimensionless constant, $W$, instead of $\pm \pi$ for the coefficient of the Cartan component. The points $W = \pm \pi$, considered in the previous section, are the ones ensuring $\tr W_3 = 0$. The vacuum energy associated with the spectrum of modes in the background (\ref{genbackground})  gives the potential governing $W$.

The deformation potential makes it difficult to  study the quantum corrections and stability beyond the classical level, due to its non-local and ultimately non-renormalizable nature, so we instead consider the UV completion in terms of local fields. The UV completion replaces the deformation potential with $n_f$ flavours of adjoint Weyl fermions with Majorana mass $M\sim 1/L$. We find that $n_f=2$ tends to be sufficient to provide center-stabilization, as we will demonstrate with the final potential. For completeness, we state the Lagrangian for the fermion fields $\psi$:
\begin{equation}\label{fermionlagrangian}
{\cal{L}}_{ferm} = \Tr\left[i\psi^\dag\bar{\sigma}^\mu\partial_\mu\psi - \psi^\dag\bar{\sigma}^\mu\left[A_\mu,\psi\right] + \frac{M}{2}\left(\psi\psi+\psi^\dag\psi^\dag\right)\right]~.
\end{equation}

In this Section, we merely state the spectra and leave most of the details of the calculation for Appendix \ref{appx:spectra}. 
We begin by defining a dimensionless parameter
\begin{equation}\label{epsilon}
\epsilon = \frac{L^2}{L_1L_2}~,
\end{equation}
which determines the ratio of the  size of the $\T^2$ with respect to the small $S^1$ circle.  In the case where we take the $\T^2$ large to approach the infinite volume limit, $\epsilon$ is a small parameter and we will treat it as such.

The derivation of the gauge-boson energy levels   is in  Appendix \ref{appx:spectra}. Here, we only list the energies and degeneracies of the non-Cartan gauge bosons (which are the only ones depending on $W$). They are given by 
\begin{equation}
\label{bosonspectrum}
E_{k,n}^b = \sqrt{\frac{2\pi}{L_1L_2}(2n+1) + \frac{1}{L^2}\left(2\pi k + W\right)^2}~,~ \text{for} ~ k \in \Z, ~n=-1,0,1,2,...
\end{equation}
For $n\geq 1$, these levels correspond to 4 degenerate modes, whereas for the $n=0$ and $n=-1$ cases, the degeneracy is only 2. Here it is important to note that for $\frac{W}{2\pi}$ within $\sqrt{\frac{L^2}{2\pi L_1L_2}} = \sqrt{\epsilon/{2\pi}}$ of an integer, the $n=-1$ levels contain a tachyon mode, usually attributed to Nielsen and Olesen \cite{Nielsen:1978rm}. This indicates that backgrounds of the form (\ref{genbackground}) cannot be the true ground state in this regime. 

Similarly, the energy levels of the non-Cartan components of a Weyl fermion with Majorana mass $M$ are
\begin{equation}
\label{fermionspectrum}
E_{k,n}^f = \sqrt{\frac{4\pi}{L_1L_2}n + \frac{1}{L^2}\left(2\pi k + W\right)^2 + M^2}~,~ \text{for} ~ k \in \Z, ~n=0,1,2,...
\end{equation}
As in the bosonic case, all the $n\geq 1$ levels have fourfold degeneracy and the $n=0$ levels are doubly degenerate. 

Below, we evaluate the vacuum energy densities $\rho_{vac} = \pm {1 \over L_1 L_2 L} \sum\limits_E {E \over 2}$ of bosons ($+$) and fermions ($-$), where the sum over $E$ includes the appropriate degeneracy factors. For evaluating the sums of the corresponding zero-point energies, eqns.~(\ref{bosonrho}) and (\ref{fermionrho}) below, we use zeta-function regularization, keeping only the $W$-dependence. 

In the rest of this Section, we describe our procedure, which allows the sums over the zero point energies  of bosons and fermions to be cast in the form of convergent series. This can then be evaluated numerically to any desired precision. 
The reader interested in the results can proceed to Section \ref{sec:infinite}, where we show that the 
vacuum energies approach the well-known GPY result in the $\epsilon \rightarrow 0$ limit (infinite-$\T^2$), and to Section \ref{sec:stability}, where the sums are numerically evaluated for different theories and for various values of $\epsilon < 1$ and the stability of the various backgrounds is discussed.

We begin with the gauge bosons, whose spectrum (\ref{bosonspectrum}) leads to the following expression for its vacuum energy density, written using (\ref{epsilon}):
\begin{eqnarray}
\label{bosonrho}
\rho_{vac}^b(W)
& = & \frac{2\pi \epsilon}{L^4} \sum_{k\in\Z} \left[ \sqrt{\left(k + \frac{W}{2\pi}\right)^2 - \frac{\epsilon}{2\pi}} +  \sqrt{\left(k + \frac{W}{2\pi}\right)^2 + \frac{\epsilon}{2\pi}} \right.\nonumber\\
&& \left. + 2 \sum_{n=1}^{\infty}  \sqrt{\left(k + \frac{W}{2\pi}\right)^2 + \frac{\epsilon (2n+1) }{2\pi}} \right]~. 
\end{eqnarray}
The vacuum energy density contribution from the fermions is likewise given by 
\begin{eqnarray}
\label{fermionrho}
\rho_{vac}^f(W,M)
& = &-\frac{2\pi \epsilon}{L^4} \sum_{k\in\Z} \left[  \sqrt{\left(k + \frac{W}{2\pi}\right)^2 + \left(\frac{LM}{2\pi}\right)^2} + 2 \sum_{n=1}^{\infty}  \sqrt{\left(k + \frac{W}{2\pi}\right)^2 + \frac{\epsilon n}{\pi} + \left(\frac{LM}{2\pi}\right)^2} \right]~.\nonumber 
\\
\end{eqnarray}

Next, we observe that the $n\geq 1$ sums are similar between the boson and fermion contributions, as the boson sum is equivalent to the fermion sum with $LM = \sqrt{2\pi \epsilon}$, so we only show the regularization process for the $n\geq 1$ sum for the fermion case. Consider
\begin{equation}
\rho_{vac}^{f,n\geq1}(W,M) =  \lim_{s\rightarrow -1/2} -\frac{4\pi \epsilon}{L^4} \sum_{n=1}^{\infty}  \sum_{k\in\Z}  \left(\left(k + \frac{W}{2\pi}\right)^2 + \frac{\epsilon n}{\pi} + \left(\frac{LM}{2\pi}\right)^2\right)^{-s}~.
\end{equation}
Using the well-known formula
\begin{eqnarray}
\label{Fidentity}
F(s;a,c) &=& \sum_{k\in\Z} \frac{1}{((k+a)^2+c^2)^s} \\
&=& \frac{\sqrt{\pi}}{\Gamma(s)}\left|c\right|^{1-2s}\left[\Gamma\left(s-\frac12\right) + 4 \sum_{p=1}^{\infty}\left(\pi p \left|c\right|\right)^{s-1/2}\cos(2\pi pa)K_{s-1/2}\left(2\pi p\left|c\right|\right)\right],\nonumber
\end{eqnarray}
we can evaluate the sum over $k$. It leaves us with 
\begin{equation}
\begin{split}
&\rho_{vac}^{f,n\geq1}(W,M) = \\
& \lim_{s\rightarrow -1/2} -\frac{4\pi \epsilon}{L^4} \sum_{n=1}^{\infty} \frac{\sqrt{\pi}}{\Gamma(s)}\sqrt{\frac{\epsilon n}{\pi}+\left(\frac{LM}{2\pi}\right)^2}^{1-2s} \\ 
& \times \left[\Gamma\left(s-\frac12\right) + 4 \sum_{p=1}^{\infty}\left( \pi p \sqrt{\frac{\epsilon n}{\pi}+\left(\frac{LM}{2\pi}\right)^2}\right)^{s-1/2}\cos(pW)K_{s-1/2}\left(2\pi p \sqrt{\frac{\epsilon n}{\pi}+\left(\frac{LM}{2\pi}\right)^2}\right)\right]~.
\end{split}
\end{equation}
The term with $\Gamma\left(s-\frac12\right)$ is divergent, but also does not depend on $W$, so it does not contribute to the potential. Therefore, we ignore this term and are left with a double sum over $n$ and $p$. This sum in now convergent in the limit $s\rightarrow -\frac12$, so we apply the limit to get
\begin{equation}
\label{fermiondoublesum}
\rho_{vac}^{f,n\geq1}(W,M) = \frac{8 \epsilon}{L^4} \sum_{n=1}^{\infty} \sum_{p=1}^{\infty} \frac{\cos(pW)}{p} \sqrt{\frac{\epsilon n}{\pi}+\left(\frac{LM}{2\pi}\right)^2}K_{-1}\left(2\pi p \sqrt{\frac{\epsilon n}{\pi}+\left(\frac{LM}{2\pi}\right)^2}\right)~.
\end{equation} 
In the massless fermion case, the sum (\ref{fermiondoublesum})  is a bit simpler:
\begin{equation}\label{masslessfdoublesum}
\rho_{vac}^{f,n\geq1}(W,0) = \frac{8 \epsilon}{L^4} \sum_{n=1}^{\infty} \sum_{p=1}^{\infty} \frac{\cos(pW)}{p} \sqrt{\frac{\epsilon n}{\pi}}K_{-1}\left(2\pi p \sqrt{\frac{\epsilon n}{\pi}}\right)~.
\end{equation}

The bosonic version of (\ref{fermiondoublesum}) can be obtained by making the replacement, $LM\rightarrow\sqrt{2\pi\epsilon}$ and flipping the overall sign:
\begin{equation}
\label{bosondoublesum}
\rho_{vac}^{b,n\geq1}(W,M) = -\frac{8 \epsilon}{L^4} \sum_{n=1}^{\infty} \sum_{p=1}^{\infty} \frac{\cos(pW)}{p} \sqrt{\frac{\epsilon}{\pi}\left(n+\frac12\right)}K_{-1}\left(2\pi p \sqrt{\frac{\epsilon}{\pi}\left(n+\frac12\right)}\right)~.
\end{equation} 
As already noted, the  double sums (\ref{fermiondoublesum},\ref{bosondoublesum},\ref{masslessfdoublesum}) are convergent and may be evaluated numerically.

We now turn to the  remaining terms ($n<1$) of the boson and fermion vacuum energy densities, which can also be evaluated using zeta-function regularization. In the fermion case, we have 
\begin{equation}
\rho_{vac}^{f,n=0}(W,M) = \lim_{s\rightarrow-\frac12}-\frac{2\pi \epsilon}{L^4} \sum_{k\in\Z} \left(\left(k + \frac{W}{2\pi}\right)^2 + \left(\frac{LM}{2\pi}\right)^2\right)^{-s}~.
\end{equation}
In the massless case, this can be evaluated with the Hurwitz zeta function:
\begin{equation}
\rho_{vac}^{f,n=0}(W,0) = -\frac{2\pi \epsilon}{L^4}\left(\zeta\left(-1,\frac{W}{2\pi}\right) + \zeta\left(-1,1-\frac{W}{2\pi}\right)\right)~.
\end{equation}
In the massive case, this can be evaluated using (\ref{Fidentity}) and throwing away the infinite constant term:
\begin{equation}
\rho_{vac}^{f,n=0}(W,M) = \frac{4\epsilon}{L^4} \frac{LM}{2\pi} \sum_{p=1}^{\infty} \frac{\cos(pW)}{p} K_{-1}(pLM)~. 
\end{equation}
This again converges and can be evaluated numerically to arbitrary precision. 

For the $n=-1$ and $n=0$ terms in the boson case, it is cleanest to expand in a Taylor series around $\epsilon=0$. The $n=-1$ term cannot be evaluated using (\ref{Fidentity}) because of the negative sign between the terms in the square root. The $n=0$ term does not have this issue and may be evaluated using (\ref{Fidentity}); however, expanding both as a Taylor series in $\epsilon$ allows for a nice cancellation of the infinite constants. The Taylor expansion gives 
\begin{equation}
\label{tachyonseries}
\rho_{vac}^{b, n<1}(W)  = \frac{4\pi\epsilon}{L^4}\sum_{m=0}^{\infty} \left(\frac{\epsilon}{2\pi}\right)^{2m} \frac{\Gamma\left(2m-\frac12\right)}{\Gamma\left(-\frac12\right)\Gamma\left(2m+1\right)} \left(\zeta\left(4m-1,\frac{W}{2\pi}\right) + \zeta\left(4m-1,1-\frac{W}{2\pi}\right)\right)~.
\end{equation} 
For values of $W$ that are not within the interval $[2\pi n - \sqrt{2\pi\epsilon}, 2\pi n + \sqrt{2\pi\epsilon}]$ for any integer $n$, this sum converges and can be numerically evaluated to arbitrary precision. For $W\in[2\pi n - \sqrt{2\pi\epsilon}, 2\pi n + \sqrt{2\pi\epsilon}]$ for some integer $n$, a tachyon is present in the system. Hence, this sum still contains the information of the potential tachyon. 

See Figure \ref{fig:B} for a plot of the bosonic potential and Figure \ref{fig:fermcomp} for the massive fermion contribution, obtained by numerically evaluating the converging sums obtained in this Section. The same plot, but for massless Weyl fermions is given in Figure \ref{fig:masslessfermcomp}.

\begin{figure}[h]
\includegraphics[width=0.8\textwidth]{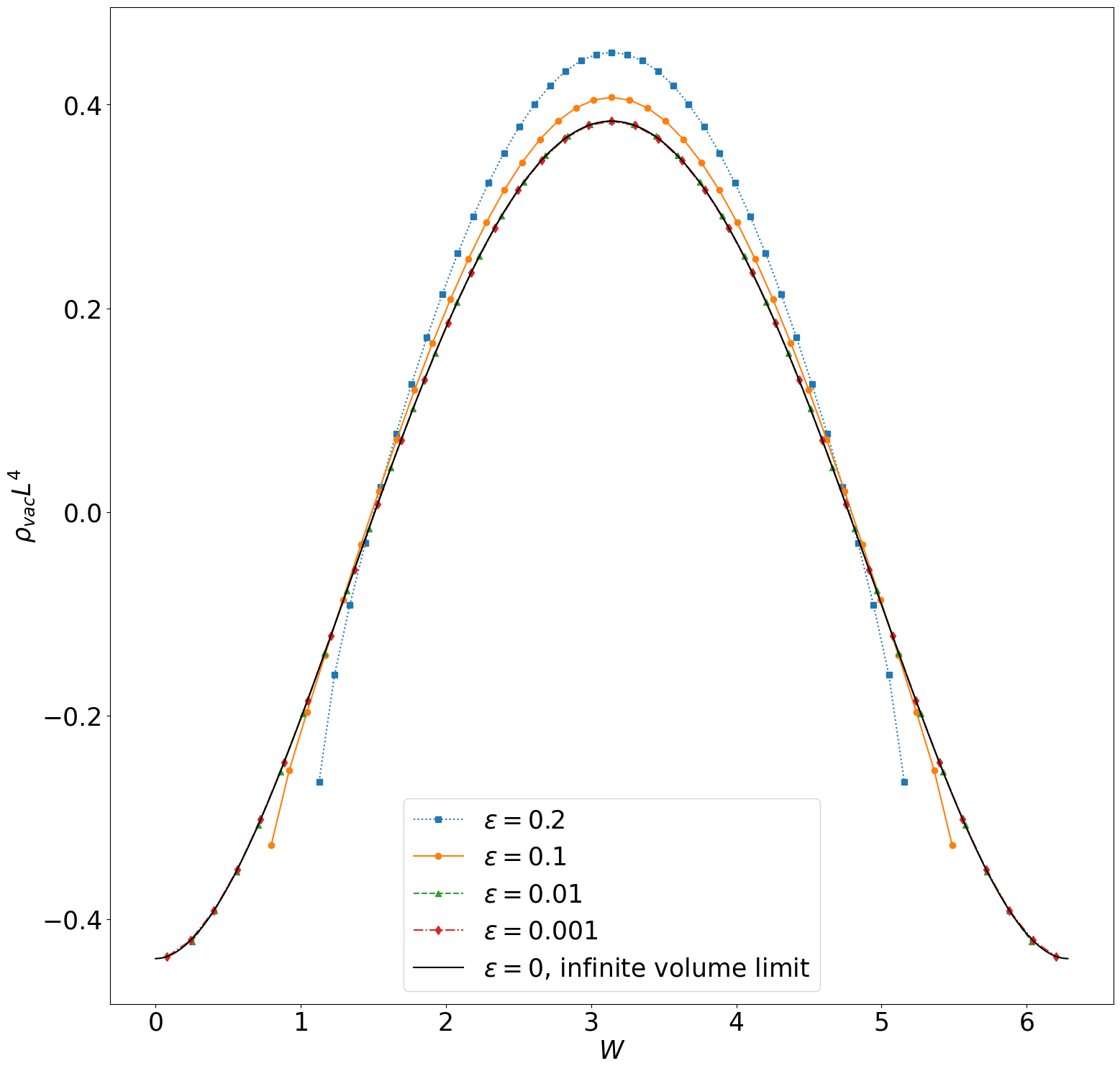}
\centering
\caption{The purely bosonic contribution  to the GPY potential. The numerical evaluation used $N_n=100000$ and $N_p=500$ as the upper limit for the $n$- and $p$-sums in Equation  (\ref{bosondoublesum}). The upper limit for the Taylor series in Equation (\ref{tachyonseries}) was taken to be $N_m=500$. The numerical reliability breaks down near the boundary of the region without a tachyon, so the plot   only covers the values in the interval, $\left[\sqrt{\frac{2\pi \epsilon}{0.99}},2\pi - \sqrt{\frac{2\pi \epsilon}{0.99}}\right]$. The value diverges down to negative infinity as $W$ approaches the values $W=\sqrt{2\pi \epsilon}$ and $W=2\pi - \sqrt{2\pi \epsilon}$. The infinite volume limit follows from Equation (\ref{infinitevolbosons}).\label{fig:B}}
\end{figure}

\begin{figure}[h]
\includegraphics[width=0.8\textwidth]{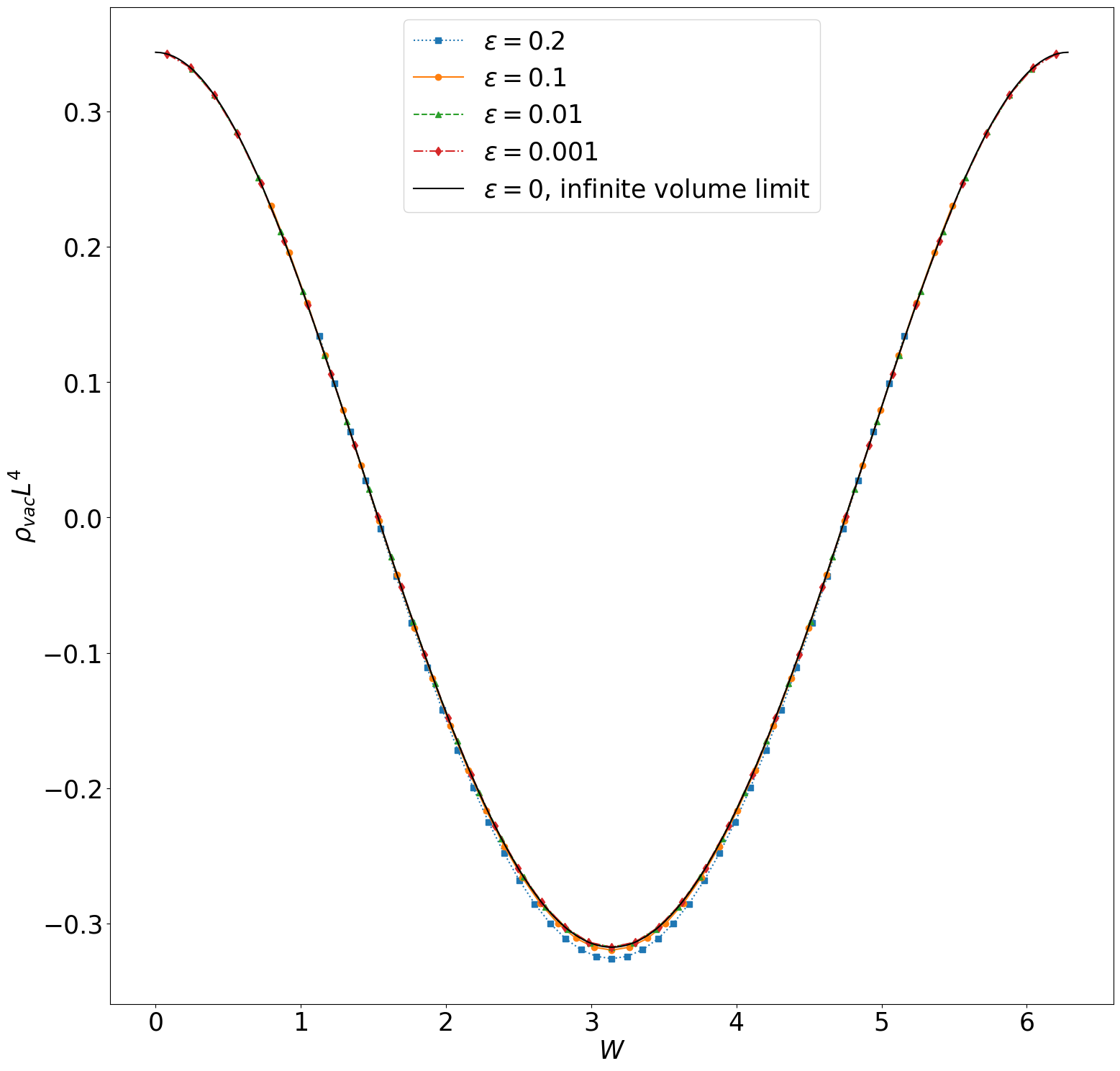}
\centering
\caption{The contributions of a single massive fermion (with mass $M\sim 1/L$)  to the GPY potential. The numerical evaluation used $N_n=100000$ and $N_p=500$ as the upper limit for the $n$- and $p$-sums in Equation (\ref{fermiondoublesum}). Here there are no issues with convergence, as for the Taylor series required for the $n=-1$ modes of the boson. The infinite volume limit was calculated from Equation (\ref{massivefermIVL}) with upper limit $N_m=500$.\label{fig:fermcomp}
}
\end{figure}

\subsection{The infinite volume limit}
\label{sec:infinite}

In this Section, we consider taking the partial infinite volume limit where the $\T^2$ is taken arbitrarily large while the $S^1$ remains small and finite. This is equivalent to the limit $\epsilon\rightarrow 0$, since we take $L_1,L_2\rightarrow\infty$, but keep $L$ fixed.  For the massless fermions and the gauge bosons, we should reproduce the familiar GPY potential \cite{GPY1981} in the infinite volume limit. We confirm this explicitly with the following calculation, which also presents a check on our calculation of the spectrum.

 \begin{figure}[h]
\includegraphics[width=0.8\textwidth]{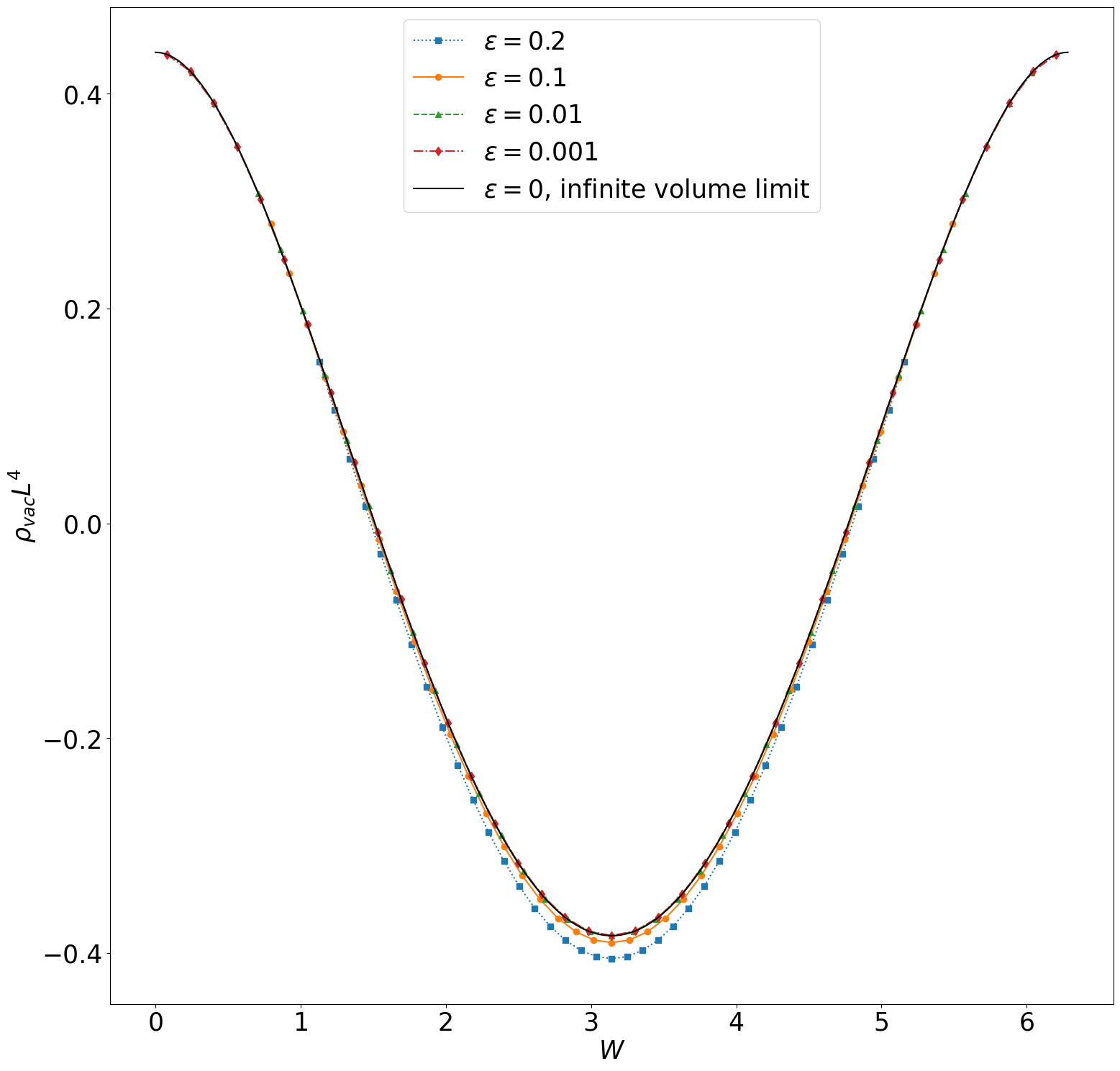}
\centering
\caption{The contributions of a single massless fermion to the GPY potential. The numerical evaluation used $N_n=100000$ and $N_p=500$ as the upper limit for the $n$- and $p$-sums in Equation (\ref{fermiondoublesum}). Here there are no issues with convergence, as for the Taylor series required for the $n=-1$ modes of the boson. The infinite volume limit was calculated from Equation (\ref{fermioncontinuum}). \label{fig:masslessfermcomp}}
\end{figure}

In this limit, the $n=0$ and $n=-1$ modes become insignificant due to the overall factor of $\epsilon$ in (\ref{bosonrho}, \ref{fermionrho}). Turning to the sums over $n\geq1$, we see that they take the form
\begin{equation}
\sum_{p=1}^\infty \epsilon \sum_{n=1}^\infty f(\epsilon n) \rightarrow  \sum_{p=1}^\infty \int_0^\infty dn f(n)
\end{equation} 
where we used the definition of Riemann integrals  in the infinite volume ($\epsilon \rightarrow 0$) limit to obtain the r.h.s.
Thus converting the sums to integrals, we find that for the massless fermion, the result takes the form
\begin{eqnarray}
\rho_{vac}^{f}(\epsilon=0) &= & \frac{8}{L^4} \sum_{p=1}^{\infty} \frac{\cos(pW)}{p} \int_0^\infty dn \sqrt{\frac{n}{\pi}}K_{-1}\left(2\pi p \sqrt{\frac{n}{\pi}}\right) \nonumber
\end{eqnarray}
\begin{eqnarray}
&= & \frac{8}{L^4} \sum_{p=1}^{\infty} \frac{\cos(pW)}{2\pi^2 p^4} = {2 \over \pi^2 L^4} \sum_{p=1}^\infty {e^{i p W} + e^{- i p W} \over p^4} = \frac{2}{\pi^2 L^4} \left(\text{Li}_4\left(e^{iW} \right) + \text{Li}_4\left(e^{-iW}\right)\right) \nonumber \\
&= & \frac{8\pi^2}{3L^4} \left(\zeta\left(-3,\frac{W}{2\pi}\right) + \zeta\left(-3,1-\frac{W}{2\pi}\right)\right), \label{fermioncontinuum}
\end{eqnarray}
where $\text{Li}_4$ denotes the polylogarithm of order 4. 
 The final result agrees with the usual GPY potential for massless Weyl fermions (equal to minus the GPY result for gauge bosons \cite{GPY1981}) and we have given several equivalent expressions for it, see e.g.~\cite{Shifman:2008ja}.

The infinite volume limit for the gauge bosons can be similarly calculated. In fact, the only difference (other than an overall sign) is that the $n$'s in the sum are shifted by $1/2$. In terms of the Riemann sum, this is just a different choice of points within the partitioned intervals, so it leads to an identical integral. Hence the bosons give
\begin{equation}\label{infinitevolbosons}
\rho_{vac}^{b}(\epsilon=0) = -\frac{8\pi^2}{3L^4} \left(\zeta\left(-3,\frac{W}{2\pi}\right) + \zeta\left(-3,1-\frac{W}{2\pi}\right)\right) =  - \frac{2}{\pi^2 L^4} \left(\text{Li}_4\left(e^{iW} \right) + \text{Li}_4\left(e^{-iW}\right)\right) ~.
\end{equation}
Again, this agrees with the GPY result. We also see directly the restoration of supersymmetry in the infinite volume limit where the boson and fermion energy densities exactly cancel out.

For the case of massive fermions, the integral is slightly more complicated, but we find that the $\epsilon\rightarrow 0$ limit can be written as 
\begin{equation}
\label{massivefermIVL}
\begin{split}
\rho_{vac}^{f,M>0}(\epsilon=0) = & \frac{8\pi}{L^4} \sum_{p=1}^{\infty} \frac{\cos(pW)}{p} \int_{\left(\frac{LM}{2\pi}\right)^2}^\infty dx \sqrt{x}K_{-1}\left(2\pi p \sqrt{x}\right)\\
= & \frac{2}{\pi^2L^4}  \sum_{p=1}^{\infty} \frac{\cos(pW)}{p^4} (pLM)^2K_2(pLM)~.
\end{split}
\end{equation}
This agrees with the usual infinite volume result for massive Weyl fermions, see e.g.~\cite{Poppitz:2021cxe}.

\begin{figure}[h]
\includegraphics[width=0.8\textwidth]{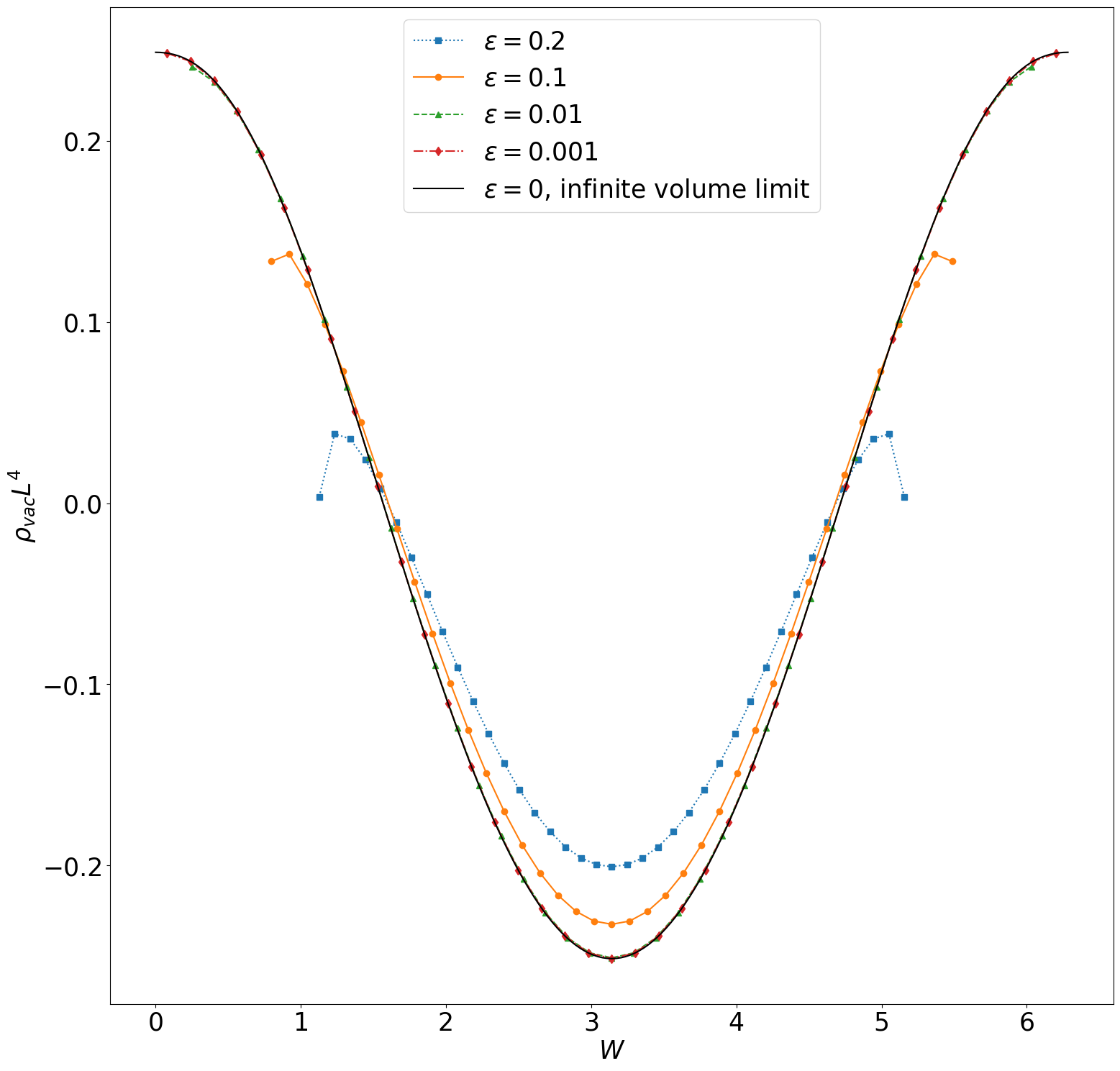}
\centering
\caption{The dimensionless vacuum energy as a function of the dimensionless parameter, $W$, for dYM with $n_f=2$ flavours of fermions with mass $LM = 1.00$. The numerical evaluation used $N_n=100000$ and $N_p=500$ as the upper limit for the $n$- and $p$-sums in Equations (\ref{fermiondoublesum}) and (\ref{bosondoublesum}). The upper limit for the Taylor series in Equation (\ref{tachyonseries}) was taken to be $N_m=500$. The numerical reliability breaks down near the boundary of the region without a tachyon, so the plot is only covers the values in the interval, $\left[\sqrt{\frac{2\pi \epsilon}{0.99}},2\pi - \sqrt{\frac{2\pi \epsilon}{0.99}}\right]$. The value diverges down to negative infinity as $W$ approaches the values $W=\sqrt{2\pi \epsilon}$ and $W=2\pi - \sqrt{2\pi \epsilon}$ making the $W=\pi$ vacuum metastable. The infinite volume limit was calculated from Equation (\ref{massivefermIVL}) using the upper bound $N_p=500$.\label{fig:epscomp}}
\end{figure}

\subsection{Stability of the semiclassical vacua}
\label{sec:stability}

 Here, we numerically evaluate the potential of interest for the UV completion of dYM. We sum the boson contribution shown on Figure \ref{fig:B} and the contributions of two Weyl flavours of fermions of mass $1/L$ of Figure \ref{fig:fermcomp}. The results at different volumes, parameterized by the dimensionless $\epsilon = \frac{L^2}{L_1L_2}$, are given in Figure  \ref{fig:epscomp}. The infinite volume limit is included on the plot (labelled $\epsilon=0$) to demonstrate how quickly the limit converges. The conclusion for dYM is that the same UV completion as the one often invoked on $\R^3 \times \S^1_L$ ensures stability of the $W = \pm \pi$ vacua\footnote{We stress that the GPY potential has the same shape near $W = -\pi$, a region we chose not to plot.} also at finite $\T^2$ with a twist, for values of $\epsilon$ as large as $0.2$ ($\sqrt{\epsilon} \sim .45$). For these values of $\epsilon$, the largest we've studied, the $\T^2$ and $\S^1_L$ are of comparable size as $L \sim \sqrt{L_1 L_2}/2$. Thus, both $\Lambda L \ll \pi$ and $\sqrt{L_1 L_2} \Lambda \ll \pi$ can be obeyed, meaning that there is an overlap between the dYM semiclassical regime and the one of the femtouniverse; thus, as we further mention in Section \ref{future}, it might be of interest to study the possible transitions between the corresponding semiclassical vacua. 

To broaden the class of theories, we also compare the GPY potentials for dYM, QCD(adj) with $n_f=2$ massless Weyl flavors, and SYM ($n_f=1$ massless Weyl fermion), on Figure \ref{fig:theorycomp}. For brevity, we only chose to show the results for a single value of $\epsilon=0.01$. The conclusion, in the case of dYM and massless $n_f \ge 2$ QCD(adj) is that already for not-so-large $\T^2$ with twists, the center-symmetric value of the $\S^1_L$ holonomy is perturbatively stabilized, as in the $\R^3 \times \S^1_L$ case. Thus the study of the finite $\T^2 \times \S^1_L$ (with twist) case using the backgrounds (\ref{omegavacuumdym}) as  classical vacua is self-consistent, just as it is on $\R^3 \times \S^1$. 

The SYM case, on the other hand, presents us with  an exception---as the value $W=\pm \pi$ is a local maximum of the GPY potential, albeit with a negative mass squared that vanishes as $\epsilon \rightarrow 0$ (this becomes even clearer from Figure \ref{fig:symcomp})---and we shall discuss some related puzzles in the  Section \ref{sec:discussion}.  

Finally, we note that our treatment of the $n <1$ terms in the bosonic sum (\ref{bosonrho}) is only consistent for $W$ sufficiently far away from the edges of the Weyl chamber. Clearly, this is because of the tachyonic term in (\ref{bosonspectrum}). Evaluating a potential in this regime using solely the quadratic fluctuations and without taking into account further nonlinearities may be  possible---see the calculation for the case of $\R^3$ of the original paper \cite{Nielsen:1978rm}---but the physical interpretation of the imaginary result that one obtains is not obvious. For values of $W$ close to the tachyon, our series converges too slowly for our numerical approach.

\begin{figure}[h]
\includegraphics[width=0.8\textwidth]{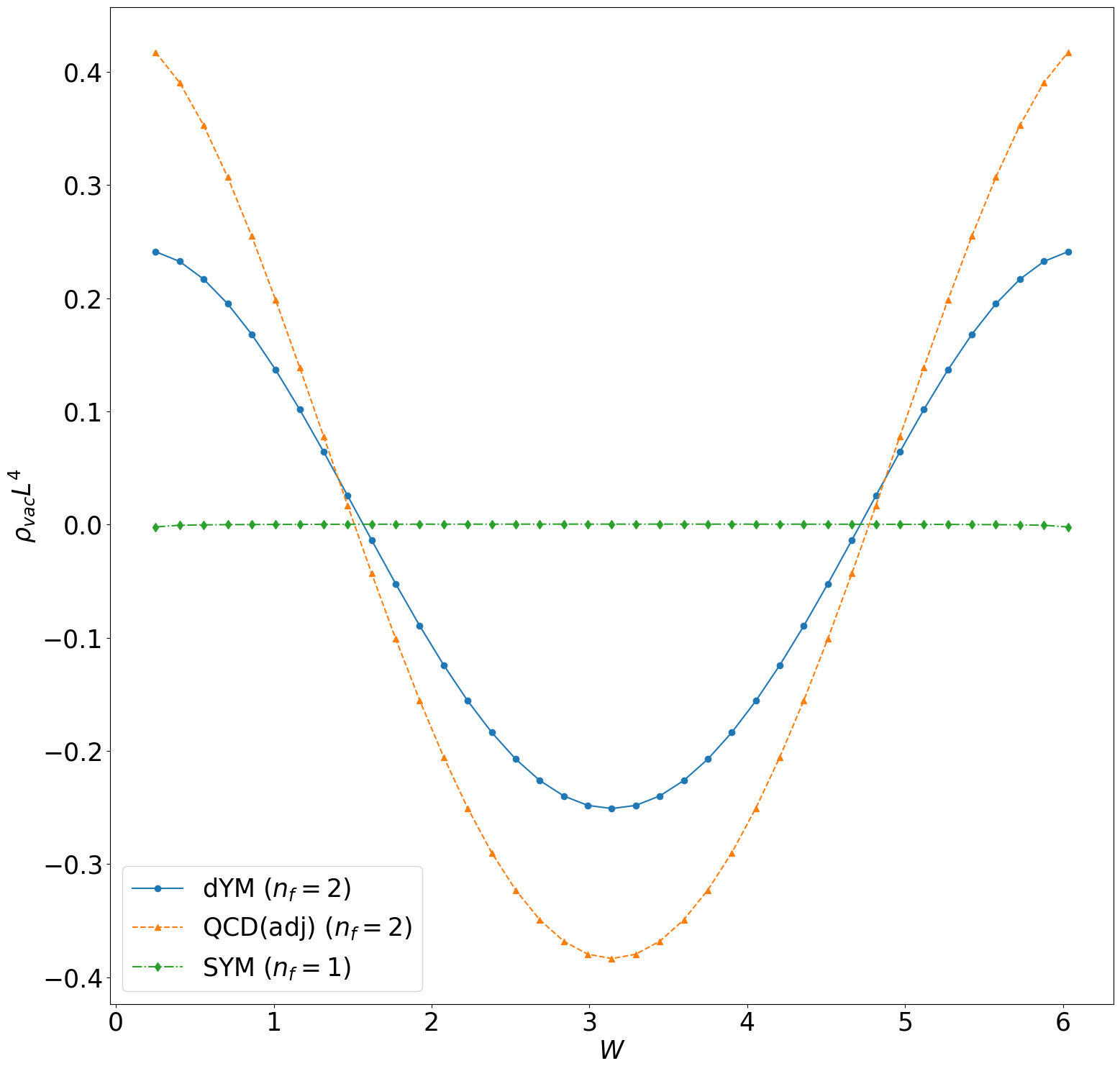}
\centering
\caption{The dimensionless vacuum energy as a function of the dimensionless parameter, $W$, for various theories. The dYM fermions are given mass $LM=1.0$, and the QCD(adj) and SYM are massless. The volume parameter is set to be $\epsilon = 0.01$. The numerical reliability breaks down near the boundary of the region without a tachyon, so the plot is only covers the values in the interval, $\left[\sqrt{\frac{2\pi \epsilon}{0.99}},2\pi - \sqrt{\frac{2\pi \epsilon}{0.99}}\right]$. The value diverges down to negative infinity as $W$ approaches the values $W=\sqrt{2\pi \epsilon}$ and $W=2\pi - \sqrt{2\pi \epsilon}$ making the $W=\pi$ vacuum metastable or slightly unstable for the case of SYM.\label{fig:theorycomp}}
\end{figure}

\section{Discussion and future directions}
\label{sec:discussion}
 
Let us now summarize our findings and discuss various puzzles and interesting directions for future studies. 
 
 \subsection{dYM vs. femtouniverse}
 \label{sec:theta}
 
 In   Sections \ref{sec:femto} and \ref{sec:dym}, we determined the classical vacua of YM in the femtouniverse and in dYM on $\T^2 \times S_L^1$, both with a twist in $\T^2$. These theories represent two weak-coupling limits of YM theory, where perturbative and semiclassical nonperturbative calculations should be possible, at least in principle. We found that while the classical field configurations which minimize the energy are different in the two limits, the action of parity and center symmetry on the corresponding semiclassical states are identical.

\subsubsection{Implications for  $\mathbb{\theta}$-dependence}
We now argue that the similarity in the action of the $T_3$ center symmetry explains why the $\theta$-dependence of the vacuum energy found via instanton calculations---performed roughly two decades apart, see below---in these two limits is identical.

To begin, recall that we found in Sections \ref{sec:femto} and \ref{sec:dym} that in each case  there are two gegenerate classical ground states, $|+\rangle$ and $|-\rangle$, and that $|-\rangle = \hat T_3 |+\rangle$, as per (\ref{femtoground}, \ref{dymground}). To all orders of perturbation theory, the $|\pm\rangle$ vacua remain degenerate \cite{Luscher:1982ma,GonzalezArroyo:1987ycm},  but they can mix due to tunnelling effects. 
A quick derivation of the $\theta$-dependence of the vacuum energy for $SU(2)$ follows below, along with a discussion of the assumptions about the nature of the relevant semiclassical objects, whose explicit form is known, in each case, to a different extent.

Since the Hamiltonian commutes with $\hat T_3$, the eigenstates of $\hat H$ can be taken to be the states with defined electric flux 
\begin{equation}\label{flux12}
|e_3 \rangle = {1\over \sqrt{2}}(|+\rangle + (-1)^{e^3} |-\rangle).
\end{equation} Let $E_{e_3 = 0,1}$ be the energies of the  lowest eigenstates of $\hat H$ in the flux sector with $e_3 =0,1$, respectively. In a theory where semiclassics is a good guide to the dynamics, these exact minimum energy eigenstates are expected to have substantial overlap with the above  $|e_3\rangle$ states,  built from the classical $| \pm \rangle$ vacua.
Thus, we expect that, as
$\beta \rightarrow \infty$, the  flux states matrix elements  \begin{eqnarray} \label{fluxzero}
 \langle e_3  | e^{- \beta \hat H} |e_3 \rangle \sim e^{- \beta E_{e_3}}~.
 \end{eqnarray}
 can be used to find the lowest energy in the corresponding  sector of Hilbert space.
Thus, to find $E_{e_3}$, we rewrite (\ref{fluxzero})  using (\ref{flux12})
 \begin{eqnarray}\label{flux1}
 \langle e_3  | e^{- \beta \hat H} |e_3 \rangle &=& {1\over 2}(\langle + | e^{- \beta \hat H} | + \rangle + \langle - | e^{- \beta \hat H} | - \rangle + (-)^{e_3} \langle + | e^{- \beta \hat H} | - \rangle + (-)^{e_3}  \langle - | e^{- \beta \hat H} | + \rangle ) \nonumber \\
 &=&  \langle + | e^{- \beta \hat H} | + \rangle  + (-)^{e_3} \langle + | e^{- \beta \hat H} \hat T_3 | + \rangle~, \end{eqnarray} 
 where on the second line we used $|-\rangle = \hat T_3 |+\rangle$,  that $\hat T_3$ commutes with the Hamiltonian, as well as its unitarity.
  
  Next, we note that the  $\langle + | e^{- \beta \hat H} | + \rangle$ matrix element  in (\ref{flux1}) receives contributions from fields with integer topological charge,  $Q_{top.} = n$, $n \in \Z$, since without twists in the time direction the topological charge on $\T^4$ is integer. In the semiclassical regime we are studying, the least suppressed contribution is the one of the perturbative sector ($n=0$). The $|n|>0$ contributions are suppressed by powers of at least $e^{- 2 S_0}$, $S_0 = 4 \pi^2/g^2$, because the minimum action in each topological sector is $2 S_0 |Q_{top.}| = {8 \pi^2 \over g^2} |Q_{top}|$. Thus, the leading semiclassical contribution of this matrix element will be $\langle + | e^{- \beta \hat H} | + \rangle \sim e^{- \beta E_{pert.}}$, where $E_{pert}$ is the perturbative vacuum energy in  the $|\pm\rangle$  degenerate vacua.
  
In contrast, the matrix element $\langle + | e^{- \beta \hat H} \hat T_3 | + \rangle$, receives contributions with half-integer topological charge, $Q_{top.} = n + {1 \over 2}$, $n\in \Z$, due to the twist by $\hat T_3$ (which carries half-integer winding number, recall (\ref{windingt3})). Based on power counting, it is clear that there are at least two lowest-order contributions with the same semiclassical suppression, those with $n=0$ and $n=-1$. Each is accompanied by a factor of $e^{\pm i {\theta \over 2}} e^{- S_0}$. Thus, the leading contribution to the second term in (\ref{flux1}) in the $\beta \rightarrow \infty$ limit will be that of an (anti-) instanton of topological charge $\pm 1/2$ and action $S_0$, located anywhere in the time interval $\beta$. For the purpose of our argument here, it suffices to simply assume that such localized objects with a time-translation zero mode exist. This is based on numerical evidence in the femtouniverse case and on calculations in the infinite-volume limit of dYM. 

Thus, combining (\ref{fluxzero}) and (\ref{flux1}), we obtain
\begin{eqnarray} \label{flux2}
e^{- \beta E_{e_3}} \sim e^{- \beta E_{pert}} + (-)^{e_3}  {\beta c \over L} e^{- S_0}  \cos {\theta \over 2} + \ldots
\end{eqnarray}
Here, our ignorance about the details of the instanton solutions, including their multiplicity, fluctuation determinants, etc., is encoded in the dimensionless constant $c$ which can have pre-exponential coupling dependence as well as depend on the ratios of the periods of $\T^3$. The sign of the 1-instanton term in (\ref{flux2}) was written in accordance with the understanding that semiclassical objects with positive fugacity give positive contributions to the partition function. 

The final step in our derivation is to exponentiate the single $|Q|=1/2$ instanton contributions of (\ref{flux2}) in the dilute gas approximation, to obtain
 \begin{eqnarray} \label{vacthetafinal}
&& e^{- \beta E_{e_3}}  \simeq    e^{- \beta \left[ E_{pert} - (-)^{e_3} {\tilde c \over L} e^{- S_0} \cos {\theta\over 2} + {\cal{O}}(e^{- 2 S_0)}\right]} \nonumber\\
&& ~\implies~
 E_{e_3} = E_{pert.} - {\tilde c \over L} e^{- S_0} \cos\left( {\theta \over 2} - \pi e_3\right), 
 \; e_3 = 0,1.  
\end{eqnarray}
 The upshot is  that the vacuum energies in the  $e_3 = 0,1$ electric flux sectors are split, due to nonperturbative effects, by an amount  ${2 \tilde c \over L} e^{- S_0} \cos {\theta\over 2}$ and that they remain degenerate at $\theta = \pi$, in accordance with the anomaly arguments. 
 
Clearly, in (\ref{vacthetafinal}) we have found a particular case of the $SU(N)$ formula (first written using non-semiclassical large-$N$ arguments \cite{Witten:1979vv,Witten:1980sp})
\begin{equation}\label{vacenergytheta}
E_{vac.}(\theta, e_3) - E_{pert.} = - {\tilde{c} \over L} e^{- S_0} \cos({\theta \over N} - {2 \pi e_3 \over N})~, ~ e_3 = 0, \ldots , N-1.
\end{equation}
We also note that an expression identical to (\ref{vacenergytheta}) for the vacuum energy has been obtained in dYM theory on $\R^3 \times \S^1_L$, with $e_3$ replaced by an index labeling the $N$ different extrema of the dYM potential for the dual photons on $\R^3 \times \S^1_L$. Here, the semiclassical objects contributing\footnote{The calculation using monopole-instantons on $\R^3 \times \S^1_L$ is in  \cite{Unsal:2008ch,Unsal:2012zj}.} are much better known than the ones in the femtouniverse (where they have been studied numerically, see  \cite{vanBaal:2000zc}). For the femtouniverse, the result for the $\theta$-dependence of the vacuum energy (\ref{vacenergytheta})  appears explicitly in  \cite{vanBaal:2000zc}, but also much earlier  in van Baal's thesis \cite{vanBaal:1984ra}---see Figure 3  in the unpublished Ch.~III,  attributed there to 't Hooft. 

In summary, our main point here is that (\ref{vacthetafinal},\ref{vacenergytheta}) is a consequence of the classical vacuum structure and the action of center symmetry, in dYM on $\T^2\times \S^1_L$ and in YM in the femtouniverse, both with $\T^2$ twist. Thus, eqn.~(\ref{vacenergytheta}) follows simply from symmetries and the applicability of semiclassics in these two limits, irrespective of our detailed knowledge of the corresponding instanton configurations. 

 \subsubsection{Future studies}
 \label{future}
 
The fact that (\ref{vacthetafinal})\footnote{And, we expect, more generally (\ref{vacenergytheta}), with the details left as  the subject of another set of future studies.} applies in two distinct semiclassical limits may be taken to suggest that the semiclassical configurations contributing in each case can be related to each other (see \cite{GarciaPerez:1999hs} for some related results in this regard). This expectation as well as the results of Section \ref{sec:stability} suggest several interesting directions of future studies:
\begin{enumerate}
\item A more explicit description of the configurations with fractional $Q_{top.}$, in either limit, would help to better understand the relation between dYM on  the finite $\R \times \T^2 \times \S^1_L$  and its infinite-$\T^2$ limit counterpart on $\R^3 \times \S^1_L$.
In ref.~\cite{Unsal:2020yeh}, it was suggested that in  dYM, in the limit of large but finite $\T^2$, in the vacua (\ref{genbackground}) with $W = \pm \pi$, the well-known self-dual monopole-instantons  in center-symmetric vacua  on $\R^3 \times \S^1_L$ could be used to construct (approximate) solutions of fractional topological charge on $\T^2 \times \S^1_L$, with flux through $\T^2$.
It would be interesting to explicitly construct such configurations, obeying the $\T^2$ boundary conditions (\ref{fields}) with appropriate transition functions in a conveniently chosen gauge. 

\item We suspect that the above is more than  an interesting mathematical exercise, as it may help elucidate some lingering issues with our eqn.~(\ref{vacthetafinal}). In particular, it is known  from studies of dYM on $\R^3 \times \S^1_L$ that, for $SU(2)$, there are two extrema of the dual-photon effective field theory at all values of $\theta$. Only one of them is a minimum of the energy functional, while the other one is a maximum. Our arguments concerning semiclassics and the $\theta$-dependence of the previous section were not detailed enough to offer insight into the stability of these states. The role of higher fluxes should also be better understood (see \cite{Banks:2014twn,Anber:2011gn,Unsal:2020yeh}). 

One's hope is that a more explicit construction and application of semiclassical ideas on $\T^2 \times \S^1_L$ would shed light both on these dynamical issues (and, more generally, address the question: what are the states responsible for the multi-branched structure of the $\theta$-vacuum?) and on the continuity between the finite and infinite volume limits.

\item The study of the semiclassical expansion in other theories, such as QCD(adj), with massive or massless fermions, where the configurations (\ref{genbackground}) were shown (in Section \ref{sec:stability} and Figure \ref{fig:theorycomp}) to be stable---and thus continuously connected to the known $\R^3 \times \S^1_L$ vacua of these theories---would also be of interest. 

Notably, as the plots on Figures \ref{fig:epscomp} and \ref{fig:theorycomp} show, the $\R^3\times \S^1_L$ vacua remain  stable, or metastable, at least until the $\T^2$ and $\S^1_L$ become of comparable size (e.g. for ${\epsilon} =0.2$), a region that overlaps with the semiclassical regime of the femtouniverse. 
Thus, it might be of interest investigate the transition, as one changes $\T^2$ from large to small, between the large-volume vacua (\ref{omegavacuumdym}), connecting to the known $\R^3 \times \S^1_L$ limit  to the small-volume ``femtouniverse'' vacua (\ref{femtogamma})  in   theories with semiclassical calculability.
\end{enumerate}
 
 \begin{figure}[h]
\includegraphics[width=0.8\textwidth]{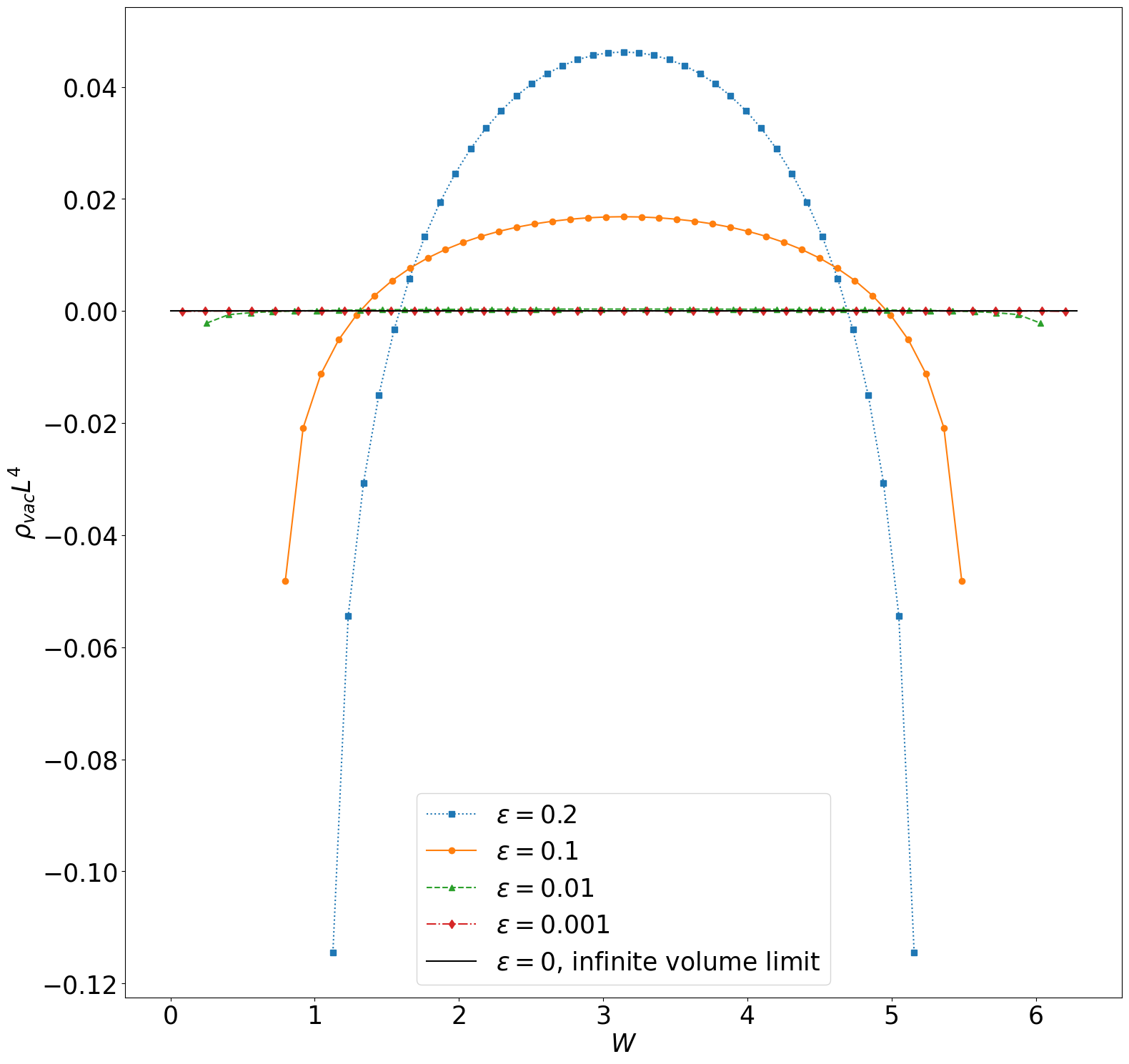}
\centering
\caption{The dimensionless vacuum energy as a function of the dimensionless parameter, $W$, for SYM. The numerical evaluation used $N_n=100000$ and $N_p=500$ as the upper limit for the $n$- and $p$-sums in Equations (\ref{fermiondoublesum}) and (\ref{bosondoublesum}). The upper limit for the Taylor series in Equation (\ref{tachyonseries}) was taken to be $N_m=500$. The numerical reliability breaks down near the boundary of the region without a tachyon, so the plot is only covers the values in the interval, $\left[\sqrt{\frac{2\pi \epsilon}{0.99}},2\pi - \sqrt{\frac{2\pi \epsilon}{0.99}}\right]$. The value diverges down to negative infinity as $W$ approaches the values $W=\sqrt{2\pi \epsilon}$ and $W=2\pi - \sqrt{2\pi \epsilon}$ making the $W=\pi$ vacuum not quite stable.\label{fig:symcomp}}
\end{figure}

\subsection{Semiclassics vs. supersymmetry: a puzzle  on $\R \times \T^2 \times \S^1_L$ vs. $\R^3 \times \S^1_L$?}
 
 Here we turn to some features of the dynamics of SYM on $\R \times \T^2 \times \S^1_L$ with a twist in $\T^2$ and, in particular, to its relation to the SYM dynamics on $\R^3 \times \S^1_L$. 
 
 To begin, we recall that  on $\R^3 \times \S^1_L$ SYM possesses a classical flat direction, given by the $\S^1_L$ holonomy, equivalently $A_3 = {W \over L} {\sigma^3 \over 2}$. Perturbative effects do not lift this flat direction, owing to supersymmetry. However, nonperturbative effects due to monopole-instantons and twisted monopole-instantons (and the related ``neutral bions'')  stabilize $|W|=\pi$, the unique, up to gauge identifications, center-symmetric vacuum on $\R^3 \times \S^1_L$. The semiclassical nonperturbative effects further lead to the appearance of two vacua (for $SU(2)$) breaking the discrete chiral symmetry $\Z_4 \rightarrow \Z_2$.\footnote{See e.g.~\cite{Poppitz:2021cxe} for a review and a full list of references. Here we only note that the $SU(2)$ semiclassical calculation in SYM on $\R^3 \times \S^1_L$ was performed in \cite{Davies:1999uw}.}
 
Now, we turn to our results. They allow us to plot the GPY potential for SYM, and on Figure \ref{fig:symcomp}  we show the potential  for $W$ for  different values of $\epsilon$. 
It is  seen, most clearly from the curves for $\epsilon =0.2, 0.1$, that the $\tr W_3=0$ value for the $\S^1_L$-holonomy, $W =   \pi$, is a local maximum, rather than a local minimum of the  potential (as in dYM or QCD(adj)).
This should, perhaps, not come as a surprise if one recalls that the backgrounds (\ref{omegavacuumdym}) carry  nonzero magnetic flux  (\ref{centerorderdym}) and  thus have nonzero vacuum energy (\ref{dymenergy})---hence they violate supersymmetry. 
As is also clear from the plot, the flatness of the $W$-direction is restored in the infinite-$\T^2$ limit. The negative mass squared of the holonomy near the center symmetric point is a quantity that depends on $\epsilon$ and goes to zero as $\epsilon \rightarrow 0$, in a manner that can be precisely determined from our expressions (although we have not done so). 

If, as suggested in \cite{Unsal:2020yeh},  the semiclassical expansion in the background (\ref{genbackground}), near $|W| = \pi$, can be understood via semiclassical objects of size $1/L$ known from the infinite-$\T^2$ limit, it is feasible that the same stabilization mechanism (neutral bions) operative on $\R^3 \times \S^1_L$ also stabilizes the center-symmetric point $|W| = \pi$---at least for sufficiently large $L^1 L^2$. This is because the effect of such localized neutral bions is not expected to scale with the volume of $\T^2$. Thus, despite the fact that neutral bions generate only an exponentially small (in the gauge coupling $g = g(1/L)$) positive mass squared around the center symmetric point, at sufficiently small $\epsilon$, but fixed $L$, the stabilizing order-$e^{- {\cal{O}}(1)/g^2}$ effect of the neutral bions can overcome the negative mass squared due to the instability. Thus, at a sufficiently large $\T^2$ one might envisage a semiclassically-stabilized vacuum at $W=\pi$, albeit one with a small nonzero  vacuum energy.\footnote{Barring a miraculous cancellation, at weak coupling, between the classical energy of the flux, an order $1/g^2$ effect, the Casimir energy of order $g^0$, and the bion effects, of order $e^{- {\cal{O}}(1)/g^2}$.}  This putative  nonsupersymmetric but metastable state would ``collide,'' as $\epsilon \rightarrow 0$, with the established supersymmetric ground state $W=\pi$ on $\R^3 \times \S^1_L$. Whether such a semiclassical scenario is realized remains to be seen. Clearly, addressing this issue requires answering  the questions already raised above, in Section \ref{future}.

Let us now turn to considerations of supersymmetry. The $\R \times \T^2 \times \S^1_L$ boundary conditions with $n_{12}=1$ are the same for bosons and fermions and thus preserve supersymmetry (in fact,   they were useful in the calculation of the Witten index \cite{Witten:1982df}). It is our flux background (\ref{omegavacuumdym}) that breaks supersymmety, inviting the nonsupersymmetric scenario of the previous paragraph. But $SU(2)$ SYM, with supersymmetric boundary conditions, should have supersymmetric vacua (at least 2, according to the index arguments) at any size   $\T^2$. The feature we find puzzling, or at least unusual, is the following. There are two---and only two \cite{Witten:1982df}---zero energy classical configurations on  $\R \times \T^2 \times \S^1_L$ with $n_{12}=1$. These are the configurations (\ref{femtogamma}) in $\Gamma$-gauge, or, equivalently,  (\ref{zerofluxomega})
in  $\Omega$-gauge, i.e. the femtouniverse classical ground states. These backgrounds have $\tr W_3 = \pm 2$ and, at large $\T^2$, do not allow for a semiclassical treatment irrespective of the $\S^1_L$ size: the spectrum of nonabelian gauge bosons has a gap of order $1/\sqrt{L^1 L^2} \rightarrow 0$, causing the coupling to  run  to large values; thus, there is no sign of abelianization and the ensuing applicability of weak coupling methods. The only feature one can take for granted is that whatever non-semiclassical dynamics determines the supersymmetric ground states on  
$\R \times \T^2 \times \S^1_L$ for large $\T^2$, as $\epsilon \rightarrow 0$, these supersymmetric ground states should land on the semiclassical point $W = \pi$ on the $\R^3 \times \S^1_L$ classical flat direction. 

 We stress that there is no inconsistency here. Our observation really says that, no matter how large $\T^2$ is taken, the imposition of a twist lifts the classical flat direction associated with the $\S^1_L$ holonomy. Thus, while we expect that there is a continuity of the supersymmetric vacua as $\epsilon$ is varied,  semiclassical methods---at least as we currently understand them---appear to only apply to the   study of the supersymmetric ground states at $\epsilon=0$. 
 
{\flushleft{In}} conclusion of this Section,   in view of the suggested continuity between gauge theories at finite and infinite $\T^2$ with twist, it is desirable to better understand the various issues raised.

\bigskip

{\flushleft{\bf Acknowledgements:}} EP is grateful to Antonio Gonz\' alez-Arroyo for many useful conversations during the ``Fluxtube-22'' Workshop at the KITP in the Winter of 2022.  Many results of this paper were reported at CAQCD-2022 at the University of Minnesota in May 2022, see talk by EP at \url{https://www.youtube.com/watch?v=FdypKO96x90&t=7s}.  Research at the KITP was supported in part by the National Science Foundation under Grant No. NSF PHY-1748958. The authors   also acknowledge support  by a Discovery Grant from NSERC. 

\bigskip

\appendix
\section{Constructing a smooth map between the $\Omega$- and $\Gamma$-gauges}
\label{appx:transform}

\subsection{$\Omega_{(k=0)}$ to $\Gamma$}
\label{appx:omega0}
Here, we explicitly construct  a smooth transformation that maps between the two gauges of interest to us. To simplify notation, we  set $L_1=L_2=L = 1$ (dimensions can be restored in the end of the day). It is clear from (\ref{transforms}) that a gauge transformation mapping (\ref{gammagauge}) to (\ref{omegagauge}) can be taken $x^3$-independent and should obey, for all $x^1,x^2 \in \R^2$ (see also (\ref{mapgauge1})):
\begin{eqnarray}
\label{gauge1}
g(x^1+1,x^2) &=& i \sigma_1 \;g(x_1, x_2),\\
g(x^1, x^2+1)&=& i \sigma_3 \; g(x_1, x_2)\; e^{- i \pi x^1 \sigma_3}.\nonumber
\end{eqnarray}
We now write the $SU(2)$ group element $g$ as
\begin{equation}\label{gab}
g = \left(\begin{array}{cc}a & b^* \cr -b & a^* \end{array}\right),\; \text{with} \; |a|^2 + |b|^2 = 1. 
\end{equation}
Then, the conditions (\ref{gauge1}) imply that $a$ and $b$ obey the $x^1$-periodicity conditions:
\begin{eqnarray}
\label{gauge1x}
a(x^1+1, x^2) &=& - i b(x^1,x^2), \nonumber \\
b(x^1+1, x^2) &=& - i a(x^1,x^2).
\end{eqnarray}
Likewise, in $x^2$, we must have that
\begin{eqnarray}
\label{gauge1y}
a(x^1, x^2+1) &=&  i a(x^1,x^2) \;e^{- i \pi x^1}, \nonumber \\
b(x^1, x^2+1) &=& - i b(x^1,x^2)\;e^{- i \pi x^1}.
\end{eqnarray}
It is now easy to see that (\ref{gauge1x}) imply that $a$ and $b$ are periodic functions of $x^1$ of period $4$ and can be Fourier expanded in $x^1$ (with Fourier coefficient which depend on $x^2$). Further, one finds that (\ref{gauge1x}) also demands that the $a$ and $b$ Fourier components in $x^1$ are related, and that, furthermore,  only  the odd $x^1$-Fourier components can be nonzero. Then, the second set of conditions (\ref{gauge1y}) is seen to relate all $x^1$-Fourier components to each other. This  allows one to express everything in terms of a single undetermined $\C$-valued function of $x^2$. 

The end result, as is easy to explicitly check,  is that the general solution of the conditions imposed on $a$ and $b$ by (\ref{gauge1x}, \ref{gauge1y}) are satisfied by the following expressions given in terms of a single  function $f(x^2)$ ($\R \rightarrow \C$):
\begin{eqnarray}
\label{gauge2x}
a(x^1,x^2) &=& e^{i {\pi x^1 \over 2}} \sum_{n \in \Z} e^{i \pi n (x^1 - {1 \over 2})} f(x^2+n),\nonumber \\
b(x^1,x^2) &=&- e^{i {\pi x^1  \over 2}}  \sum_{n \in \Z}  e^{i \pi n (x^1 + {1 \over 2})} f(x^2 + n) .
\end{eqnarray}
Finally, we impose the normalization condition $|a|^2 + |b|^2 = 1$, which implies that
\begin{equation}\label{gauge3}
\sum_{n \in\Z} e^{i 2 \pi k x^1} (-1)^k \sum_{m \in \Z} f^*(x^2 + m) f(x^2 + m + 2 k) = {1 \over 2}, ~ \text{for all} ~ x^1, x^2 \in \R^2.
\end{equation}
To see the consequences of (\ref{gauge3}), we denote $F_k(x^2) \equiv \sum_{m \in \Z} f^*(x^2 + m) f(x^2 + m + 2 k)$ and note that (\ref{gauge3}) implies that $F_k(x^2)$ is real, that $F_{k \ne 0}(x^2) = 0$, and that,  for all $x^2$, 
$F_{k=0}(x^2) = {1 \over 2}$.

We shall now construct an example of a real $f(x^2)$ obeying (\ref{gauge3}), smooth and defined for all $x^2 \in \R$. Thus, we will have found a gauge transformation $g(x^1,x^2)$ obeying (\ref{gauge1}) (such a function is not uniquely determined, reflecting the fact that there are gauge transformations that preserve the $\Gamma$-gauge or the $\Omega$-gauge). The idea we shall use is that the condition $F_{k \ne 0}(x^2) = 0$ for all $x^2$ can be automatically satisfied by having $f(x^2)$ be only nonzero for $0 \le x^2 \le 2$. The other condition (\ref{gauge3}), $F_{k=0}(x^2)=1/2$, can be obeyed by demanding, e.g. for real $f$, that $f^2$ has a symmetry w.r.t. reflections across $x^2=1$: $f^2(x^2+1) = 1/2 - f^2(x^2), 0 \le x^2 \le 1$.

  We begin by noting that one can construct, using a ``bump function'', an infinitely differentiable function $\tilde{f}(x)$ obeying
\begin{equation}\label{tildef}
\tilde{f}(x) = \left\{ \begin{array}{ccc} 0 & \text{for} &x \notin [0, 2], \cr
{1 \over 2} & \text{for}& x = 1, \cr
{1 \over 2} - \tilde{f}(1+x)& {\text{for}} & x \in [0,1]~. \end{array} \right.
\end{equation} The details of the construction, along with a plot of $\tilde f(x)$, are shown on Figure~\ref{fig1} (in Section \ref{sec:gauges1}).
In terms of $\tilde f$, the function $f(x^2)$ that determines the transformation between the $\Omega$-gauge and $\Gamma$-gauge is simply $f(x^2) = \sqrt{\tilde{f}(x^2)}$, which is also infinitely differentiable.  
As already mentioned, the vanishing of $f$ outside the $[0,2]$ interval guarantees that $F_{k \ne 0}$ vanishes identically, while the relation $f^2(x) + f^2(1+x) = 1/2$ guarantees that the $g$, 	expressed in terms of  $f(x^2)$ and $a, b$, via (\ref{gauge2x}) and (\ref{gab}) is an $SU(2)$ group element. 

To summarize, for any unit square of the $x^1, x^2$ plane, the expression of $g(x^1,x^2)$ in terms of $f(x^2) = \sqrt{\tilde f(x^2)}$ only contains two terms. For example, for $0 \le x^2 \le 1$, only the $n=0$ and $n=1$ terms in (\ref{gauge3}) are nonzero and we have that 
\begin{eqnarray}\label{gunitsquare}
 g(x^1, 0 \le x^2 \le 1) &&=\\
 && \left(\begin{array}{cc} e^{i {\pi x^1 \over 2}}[f(x^2) -i e^{i \pi x^1} f(x^2+1)] &  -e^{-i {\pi x^1 \over 2}} [f(x^2) - ie^{-i \pi x^1} f(x^2 + 1)] \cr     e^{i {\pi x^1 \over 2}} [f(x^2) + ie^{i \pi x^1} f(x^2 + 1)]&e^{-i {\pi x^1 \over 2}}[f(x^2) +i e^{-i \pi x^1} f(x^2+1)]  \end{array} \right)~. \nonumber
\end{eqnarray}
 For future use, let us now compute various quantities of interest that involve $g(x^1, x^2)$. We shall do so using the $x^{1,2}$-plane  strip $0 \le x^2 \le 1$, i.e. the form of eqn.~(\ref{gunitsquare}).
 We obtain the antihermitean traceless matrices $g^{-1} d g$:
 \begin{eqnarray} \label{gderivatives}
g^{-1} \partial_1 g& =& - i \sigma^1\; 2\pi  f(x^2) f(1+x^2) \sin 2 \pi x^1 + i \sigma^2\; 2 \pi  f(x^2) f(1+x^2) \cos 2 \pi x^1\\
&& + i \sigma^3\; { \pi  \over 2}\left( 3 - 4 f^2(x^2)\right),   \nonumber \\
g^{-1} \partial_2 g&=&    (i \sigma^1   \cos 2\pi x^1 + i \sigma^2   \sin 2 \pi x^1) 2\left[f(x^2) f'(1+x^2)- f'(x^2) f(1+x^2)\right] , 
 \end{eqnarray}
 as well as the hermitean $g \sigma^3 g^{-1}$:
 \begin{eqnarray}
 \label{sigma3trfd}
 &&g \sigma^3 g^{-1}   = \\
&&\qquad \sigma^1\; 2\left[ f^2(x^2) - f^2(1+x^2)\right] + \sigma^2\; 4 f(x^2) f(1+x^2) \cos \pi x^1 +  \sigma^3\; 4 f(x^2) f(1+ x^2) \sin \pi x^1.\nonumber
 \end{eqnarray}

\subsection{$\Omega_{(k\neq0)}$ to $\Omega_{(k=0)}$}
\label{omegaktoomegazero}

Here, we construct an explicit smooth gauge transformation between the $\Omega_{(k)}$ gauges with $k=0$ and with $k\neq0$. As before, we set $L_1=L_2=L=1$ to simplify notation. It is clear from (\ref{transforms}) that a transformation mapping (\ref{omegagauge}) to (\ref{omegakgauge}) can be taken $x^3$-independent and for all $x^1,x^2\in \R^2$:
\begin{equation}
\begin{split}
g(x^1+1,x^2) = & g(x^1,x^2) \\
g(x^1,x^2+1) = & e^{i\pi x^1 \sigma^3} g(x^1,x^2) e^{-i\pi(2k+1)x^1\sigma^3}~.
\end{split}
\end{equation}
We can write $g\in \text{SU}(2)$ as\footnote{Note that the definition of $a,b$ here slightly differs from (\ref{gab}).}
\begin{equation}
g = \begin{pmatrix}a & -b \\ b^* & a^* \end{pmatrix}~.
\end{equation}
Now, from the periodicity conditions of $g$ we can work out that $a$ and $b$ are $x^1$-periodic and obey
\begin{equation}
\begin{split}
a(x^1,x^2+1) =  & e^{-i2\pi k x^1} a(x^1,x^2)\\
b(x^1,x^2+1) = & e^{i2\pi (k+1)x^1} b(x^1,x^2)~.
\end{split}
\end{equation}
Hence, if we write 
\begin{equation}
\begin{split}
a(x^1,x^2) = & \sum_{k_1\in\Z} e^{i2\pi k_1x^1} a_{k_1}(x^2)\\
b(x^1,x^2) = & \sum_{k_1\in\Z} e^{i2\pi k_1x^1} b_{k_1}(x^2)~,
\end{split}
\end{equation}
then the boundary conditions become
\begin{equation}
\begin{split}
a_{k_1}(x^2+1) = & a_{k_1+k}(x^2)\\
b_{k_1}(x^2+1) = & b_{k_1-k-1}(x^2)~.
\end{split}
\end{equation}
Because of this, the $a_{k_1}$ functions are determined by $k$ independent functions, whereas the $b_{k_1}$ are determined by $k+1$ functions via the relations,
\begin{equation}
\begin{split}
a_{nk+m}(x^2) = & a_m(x^2+n)\\
b_{p(k+1)+q}(x^2) = & b_q(x^2-p)~,
\end{split}
\end{equation}
where $m \in \{0,1,\ldots,k-1\}$ and $q\in\{0,1,\ldots,k\}$.

This gives us a lot of freedom in finding solutions, but we are only interested in finding one simple example. To avoid over-complication, we suppose out of all $k$ equations for $a_m$, only the $m=0$ equation is non-zero. Similarly, we only consider $b_0$. Solutions of this sort will take the form
\begin{equation}
\begin{split}
a = & \sum_{n\in\Z} e^{i2\pi n k x^1} a_0(x^2+n)\\
b = &  \sum_{p\in\Z} e^{i2\pi p (k+1) x^1} b_0(x^2-p)~.
\end{split}
\end{equation}
To ensure $g\in SU(2)$, we must pick $a$ and $b$ such that 
\begin{equation}
\begin{split}
\label{unitarycondition}
1 = & a^*a + b^*b \\
= & \sum_{n,n'\in\Z} e^{i2\pi (n-n') k x^1}a_0^*(x^2+n) a_0(x^2+n')\\
& + \sum_{p,p'\in\Z} e^{i2\pi (p-p') (k+1) x^1}b_0^*(x^2-p) b_0(x^2-p')~.
\end{split}
\end{equation}

To construct a solution, consider defining a bump function, $h$, with the properties:
\begin{itemize}
\item $h(x)$ has support only on $x\in[0,1]$
\item for all $x\in[0,1/2]$, $h(x) + h(x+1/2) = 1$
\end{itemize}
Such bump functions were shown to be possible to construct in Figure \ref{fig1}.
Set $a_0(x^2) = h(x^2)$ and $b_0(x^2) = h(x^2 + 1/2)$. To see that this will define a solution, take an arbitrary $x^2\in\R$. Then the first property of $h$ guarantees that first sum in (\ref{unitarycondition}) is only non-zero for the term $n=n'=-\lfloor x^2\rfloor$, so we get 
\begin{equation}
 \sum_{n,n'\in\Z} e^{i2\pi (n-n') k x^1}a_0^*(x^2+n) a_0(x^2+n') = h^2\left(x^2-\lfloor x^2\rfloor\right).
\end{equation}
Similarly for the second sum at most one term is non-zero. If $x^2 - \lfloor x^2 \rfloor \in [0,1/2)$, then the contributing term is $p=p'=\lfloor x^2 \rfloor$, giving 
\begin{equation}
\sum_{p,p'\in\Z} e^{i2\pi (p-p') (k+1) x^1}b_0^*(x^2-p) b_0(x^2-p') = h^2\left(x^2-\lfloor x^2\rfloor + 1/2\right).
\end{equation}
Since $x^2 - \lfloor x^2 \rfloor \in [0,1/2)$, the two sums add up to $h^2\left(x^2-\lfloor x^2\rfloor\right) + h^2\left(x^2-\lfloor x^2\rfloor + 1/2\right) = 1$, by the second property of $h$, which agrees with (\ref{unitarycondition}).

If $x^2 - \lfloor x^2 \rfloor \in (1/2,1]$, then the contributing term is $p=p'=\lfloor x^2 \rfloor+1$, giving
\begin{equation}
\sum_{p,p'\in\Z} e^{i2\pi (p-p') (k+1) x^1}b_0^*(x^2-p) b_0(x^2-p') = h^2\left(x^2-\lfloor x^2\rfloor - 1/2\right).
\end{equation}
Since $x^2 - \lfloor x^2 \rfloor \in (1/2,1]$, $x^2 - \lfloor x^2 \rfloor - 1/2 \in (0,1/2]$ and the two sums add up to 
$$h^2\left(\left(x^2-\lfloor x^2\rfloor - 1/2\right)+1/2\right) + h^2\left(x^2-\lfloor x^2\rfloor - 1/2\right) = 1,$$ by the second property of $h$, which agrees with (\ref{unitarycondition}).

Finally, if $x^2- \lfloor x^2 \rfloor = 1/2$, then there are no non-zero terms of the second sum (since $h$ must vanish on all integers). However, in this case, the second property of $h$ guarantees that $h^2\left(x^2-\lfloor x^2\rfloor\right) = h^2(1/2) = 1$, so the total is still 1 as needed for (\ref{unitarycondition}).

Hence, using the bump function $h$, we can construct a smooth gauge transformation that maps between the $\Omega_{(k\neq 0)}$ and $\Omega_{(k=0)}$ gauges.

\section{Determination of spectra}
\label{appx:spectra}

In this Section, we describe in detail the calculation leading to equations (\ref{bosonspectrum}) and (\ref{fermionspectrum}), as well as the corresponding degeneracies. 

\subsection{The boundary conditions}
In the following sections we will make use of the boundary conditions for adjoint fields. As the fluctuations of the gauge fields around the classical background (\ref{genbackground}) obey homogeneous boundary conditions (eqn.~(\ref{fields}) without the non-homogeneous term present in the $\Omega$-gauge), the boundary conditions are identical for the boson and fermion modes. Thus, in this section we just use a generic adjoint field $\phi$ in place of the gauge boson or fermion fields. 

We use the $\Omega$-gauge from (\ref{omegagauge}), so the boundary conditions are
\begin{eqnarray}
\label{omegabc}
\phi(x^1+L_1,x^2,x^3) & = & \phi(x^1,x^2,x^3)\nonumber\\
\phi(x^1,x^2+L_2,x^3) & = &  e^{i\pi \frac{x^1}{L_1}\sigma^3} \phi(x^1,x^2,x^3)e^{-i\pi \frac{x^1}{L_1}\sigma^3} \\
\phi(x^1,x^2,x^3+L) & = & \phi(x^1,x^2,x^3)~.\nonumber 
\end{eqnarray} 
The boundary conditions around the $x^2$ direction behave differently for the three colour components of the adjoint field. The Cartan component commutes with the factor $e^{i\pi x^1 \sigma^3 / L_1}$, so it is $L_2$-periodic in the $x^2$ direction. Hence we can write the Cartan component as a typical Fourier series:
\begin{equation}
\label{3field}
\phi^3(x^1,x^2,x^3) = \sum_{k_1,k_2,k_3 \in \Z} e^{i2\pi k_1\frac{x^1}{L_1}} e^{i2\pi k_2\frac{x^2}{L_2}} e^{i2\pi k_3\frac{x^3}{L}} \phi^3_{k_1,k_2,k_3},
\end{equation} 
where the $\phi^3_{k_1,k_2,k_3}$ are constants. 
For the non-Cartan components, it is easiest to split the components into $+,-$ components instead of $1,2$ components. While this is a standard practice, we give the explicit formulae here to be clear about our conventions:
\begin{eqnarray}
\sigma^+ &=& \frac12 \left(\sigma^1 + i \sigma^2 \right) = \begin{pmatrix} 0 & 1 \\ 0 & 0 \end{pmatrix}\nonumber\\
\sigma^- &=& \frac12 \left(\sigma^1 - i \sigma^2 \right)= \begin{pmatrix} 0 & 0 \\ 1 & 0 \end{pmatrix}
\end{eqnarray}
and
\begin{eqnarray}\label{br4}
\phi &=& \phi^1 \frac{\sigma^1}{2} + \phi^2 \frac{\sigma^2}{2} + \phi^3 \frac{\sigma^3}{2} =  \phi^+\sigma^+ + \phi^-\sigma^- + \phi^3\frac{\sigma^3}{2}~.
\end{eqnarray}
Combined, this leaves the relation $\phi^\pm  =  \frac12 \left(\phi^1 \mp i\phi^2\right)$. In these components, the $x^2$ boundary condition becomes
\begin{equation}
\phi^\pm(x^1,x^2+L_2,x^3) = e^{\pm i 2\pi \frac{x^1}{L_1}} \phi^\pm(x^1,x^2,x^3)~.
\end{equation} 
Therefore, we can write these components as 
\begin{equation}
\phi^\pm(x^1,x^2,x^3) = \sum_{k_1,k_3\in\Z} e^{i2\pi k_1\frac{x^1}{L_1}} e^{i2\pi k_3\frac{x^3}{L}} \phi^\pm_{k_1,k_3} (x^2)~,
\end{equation}
with the condition
\begin{equation}
\phi^\pm_{k_1,k_3} (x^2 + L_2) = e^{\pm i 2\pi \frac{x^1}{L_1}}\phi^\pm_{k_1,k_3} (x^2)~.
\end{equation}
By rearranging terms and comparing Fourier components, we find that 
\begin{equation}
\label{k1relations}
\phi^\pm_{k_1\pm 1,k_3} (x^2 + L_2) = \phi^\pm_{k_1,k_3} (x^2)~.
\end{equation}
By induction for all $k_1\in\Z$, we can write everything in terms of the $k_1=0$ functions:
\begin{equation}
\phi^\pm_{k_1,k_3}(x^2) = \phi^\pm_{0,k_3}(x^2\mp k_1L_2).
\end{equation}
For the sake of reducing the number of indices, we will leave the $k_1=0$ index off in further equations. Therefore, we can write the $\pm$ components of $\phi$ as 
\begin{equation}
\label{pmfield}
\phi^\pm(x^1,x^2,x^3) = \sum_{k_1,k_3\in\Z} e^{i2\pi k_1\frac{x^1}{L_1}} e^{i2\pi k_3\frac{x^3}{L}} \phi^\pm_{k_3} (x^2\mp k_1L_2)~.
\end{equation}
With these boundary conditions, we are ready to find the allowed energy levels.

\subsection{The boson spectrum}
For the boson spectrum, we start by expanding our gauge field, $A$, into a background part, $A^\Omega$, and a dynamical part, $a$: $A = A^\Omega + a$. We then expand the $FF$ terms of the Lagrangian (\ref{action}) into terms quadratic in the dynamical field $a$:
\begin{eqnarray}
&&\left.F_{\mu\nu}^i F^{i,\mu\nu}\right|_{a^2} = 2\varepsilon^{ijk}F^{i,\mu\nu}\left[A^\Omega\right] a_\mu^j a_\nu^k\\
&&= \left(\partial_\mu a^i_\nu - \partial_\nu a^i_\mu -\varepsilon^{ijk}\left(a^j_\mu A^{\Omega,k}_\nu - a^j_\nu A^{\Omega,k}_\mu\right) \right)   \left(\partial^\mu a^{i,\nu} - \partial^\nu a^{i,\mu} -\varepsilon^{ijk}\left(a^{j,\mu} A^{\Omega,k,\nu} - a^{j,\nu} A^{\Omega,k,\mu}\right) \right)~.\nonumber
\end{eqnarray}
Here we use $i,j,k = 1,2,3$ to denote color indices and $g_{\mu\nu} = {\rm diag}(+,-,-,-)$ and $F[A^\Omega]$ is the field strength tensor of the background field (\ref{genbackground}). Expanding the colour index sums, we find, in terms of the components defined in (\ref{br4})
\begin{equation}
\begin{split}
\left.F_{\mu\nu}^i F^{i,\mu\nu}\right|_{a^2} = & -i\frac{16\pi}{L_1L_2} \left(a^-_1a^+_2 - a^+_1a^-_2\right) + 4 \left( D^+_\mu a^+_\nu - D^+_\nu a^+_\mu\right)\left(D^{-\mu} a^{-\nu} - D^{-\nu} a^{-\mu} \right)\nonumber\\
& + \left(\partial_\mu a^3_\nu - \partial_\nu a_\mu^3\right)\left(\partial^\mu a^{3,\nu} - \partial^\nu a^{3,\mu}\right)~,
\end{split}
\end{equation}
where $D^\pm_\mu  = \left(\partial_\mu \pm i A^{\Omega,3}_\mu\right)$. From this description, we can see that the $a^3$ field is a free abelian gauge field. Hence, its spectrum is identical to that of a  photon in a box, with periodic boundary conditions, and does not depend on $W$. This will not make any significant contribution to the potential, so we drop it moving forward and focus only on the $a^\pm$ fields. 

Next we find the Euler-Lagrange equations of motion from these quadratic terms:
\begin{equation}
D^{\pm,\mu} \left(D^\pm_\mu a^\pm_\nu - D^\pm_\nu a^\pm_\mu \right) \pm i\frac{2\pi}{L_1L_2}\left(g_{\nu1}a^\pm_2 - g_{\nu2}a^\pm_1\right) = 0
\end{equation}
This is a set of four equations corresponding to $\nu = 0,1,2,3$. To find the allowed energy levels, we solve for time independent solutions of the form:
\begin{equation}
a^\pm(x^0,x^1,x^2,x^3) = e^{-i E x^0} a^\pm(x^1,x^2,x^3)~,
\end{equation}
with energy $E$. This effectively makes the substitution $\partial_0\rightarrow-i E$ in our equations of motion. Plugging this into the $\nu=0$ equation and substituting the known background field, we obtain the version of Gauss' Law for this system:
\begin{equation}
\label{nu0eom}
\left(\partial_1 \pm i \frac{\alpha_1}{L_1}  \mp i\frac{2\pi x^2}{L_1L_2} \right)a^\pm_1 + \left(\partial_2\pm i \frac{\alpha_2}{L_2}\right)a^\pm_2 + \left(\partial_3\pm i\frac{W}{L}\right)a^\pm_3 = 0~.
\end{equation}
Applying a similar procedure and employing Gauss' Law for simplification, we find the other three equations motion can be written as 
\begin{equation}\label{b15}
\begin{split}
0 = & \left[E^2 + \left(\partial_1 \pm i \frac{\alpha_1}{L_1} \mp i \frac{2\pi x^2}{L_1L_2}\right)^2 + \left(\partial_2\pm i \frac{\alpha_2}{L_2}\right)^2 + \left(\partial_3\pm i \frac{W}{L}\right)^2\right]a^\pm_1 \pm i \frac{4\pi}{L_1L_2} a^\pm_2\\
0 = & \left[E^2 + \left(\partial_1 \pm i \frac{\alpha_1}{L_1} \mp i \frac{2\pi x^2}{L_1L_2}\right)^2 + \left(\partial_2\pm i \frac{\alpha_2}{L_2}\right)^2 + \left(\partial_3\pm i \frac{W}{L}\right)^2\right]a^\pm_2 \mp i \frac{4\pi}{L_1L_2} a^\pm_1\\
0 = & \left[E^2 + \left(\partial_1 \pm i \frac{\alpha_1}{L_1} \mp i \frac{2\pi x^2}{L_1L_2}\right)^2 + \left(\partial_2\pm i \frac{\alpha_2}{L_2}\right)^2 + \left(\partial_3\pm i \frac{W}{L}\right)^2\right]a^\pm_3~.
\end{split}
\end{equation}
The change of basis given by $a^\pm_1 \rightarrow a^\pm_+ + a^\pm_-$ and $a^\pm_2 \rightarrow i\left(a^\pm_+ - a^\pm_-\right)$, results in three independent differential equations:
\begin{equation}\label{b16}
\begin{split}
0 = & \left[E^2 + \left(\partial_1 \pm i \frac{\alpha_1}{L_1} \mp i \frac{2\pi x^2}{L_1L_2}\right)^2 + \left(\partial_2\pm i \frac{\alpha_2}{L_2}\right)^2 + \left(\partial_3\pm i \frac{W}{L}\right)^2 \mp i \frac{4\pi}{L_1L_2} \right]a^\pm_+\\
0 = & \left[E^2 + \left(\partial_1 \pm i \frac{\alpha_1}{L_1} \mp i \frac{2\pi x^2}{L_1L_2}\right)^2 + \left(\partial_2\pm i \frac{\alpha_2}{L_2}\right)^2 + \left(\partial_3\pm i \frac{W}{L}\right)^2 \pm i \frac{4\pi}{L_1L_2} \right] a^\pm_-\\
0 = & \left[E^2 + \left(\partial_1 \pm i \frac{\alpha_1}{L_1} \mp i \frac{2\pi x^2}{L_1L_2}\right)^2 + \left(\partial_2\pm i \frac{\alpha_2}{L_2}\right)^2 + \left(\partial_3\pm i \frac{W}{L}\right)^2\right]a^\pm_3~.
\end{split}
\end{equation}
Including the forms of the fields given in (\ref{3field}) and (\ref{pmfield}), we can further simplify these equations into three ODEs:
\begin{equation}\label{b17}
\begin{split}
0 = & \left[E^2 - \left(\frac{2\pi k_1}{L_1} \pm \frac{\alpha_1}{L_1} \mp  \frac{2\pi x^2}{L_1L_2}\right)^2 + \left(\partial_2\pm i \frac{\alpha_2}{L_2}\right)^2 - \left(\frac{2\pi k_3}{L} \pm \frac{W}{L}\right)^2 \mp i \frac{4\pi}{L_1L_2} \right]a^\pm_{k_3,+}\\
0 = & \left[E^2 - \left(\frac{2\pi k_1}{L_1} \pm \frac{\alpha_1}{L_1} \mp  \frac{2\pi x^2}{L_1L_2}\right)^2 + \left(\partial_2\pm i \frac{\alpha_2}{L_2}\right)^2 - \left(\frac{2\pi k_3}{L} \pm \frac{W}{L}\right)^2 \pm i \frac{4\pi}{L_1L_2} \right]a^\pm_{k_3,-}\\
0 = & \left[E^2 - \left(\frac{2\pi k_1}{L_1} \pm \frac{\alpha_1}{L_1} \mp  \frac{2\pi x^2}{L_1L_2}\right)^2 + \left(\partial_2\pm i \frac{\alpha_2}{L_2}\right)^2 - \left(\frac{2\pi k_3}{L} \pm \frac{W}{L}\right)^2\right]a^\pm_{k_3,3}~.
\end{split}
\end{equation}
Now, we can take some simplifying transformations. We factor out an $x^2$ dependent phase to remove the $\alpha_2$ dependence from the equation, and we shift the $x^2$ coordinate to remove the $\alpha_1$ and $k_1$ dependence. To combine our equations into a single expression, we also introduce the symbol
\begin{equation}\label{es}
s_i = \begin{cases}
1 & i = + \\
-1 & i = - \\
0 & i=3
\end{cases}~.
\end{equation}
These simplifications make (\ref{b17}) become
\begin{equation}
\label{SHO}
\frac12\left(E^2 - \left(\frac{2\pi k_3}{L}\pm \frac{W}{L}\right)^2 \mp s_i \frac{4\pi}{L_1L_2}\right) a^\pm_{k_3,i} =  \left(-\frac12 \partial_2^2 + \frac12 \left(\frac{2\pi}{L_1L_2}\right)^2 \left(x^2\right)^2\right) a^\pm_{k_3,i}~.
\end{equation}
These equations take the form of the Schr{\"o}dinger equation for a simple harmonic oscillator, so we can easily find the allowed energy levels and corresponding solutions. Since the eigenvalues of the operator on the r.h.s. in (\ref{SHO}) are ${2 \pi \over L_1 L_2}(n+{1 \over 2})$, we find that the allowed energies are
\begin{equation}\label{levels1}
E = \sqrt{\frac{2\pi}{L_1L_2}\left(2(n\pm s_i) + 1\right) + \frac{1}{L^2}\left(2\pi k_3 \pm W\right)^2}~,
\end{equation}
where $n=0,1,2,...$ and $s_i = 0, \pm 1$ as per (\ref{es}). Here, either the real or imaginary part (see discussion below) of 
$a^\pm_{k_3,i}$ is equal to the $n$\textsuperscript{th} harmonic oscillator solution, which we denote $\phi_n$.\footnote{That only $n \ge 0$, i.e. solutions normalizable on the entire real line $\R$ spanned by $x^2$, are permissible follows from the $\T^2 \times \S^1$ normalizability of the modes (\ref{pmfield}).} From (\ref{levels1}), after a simplifying relabeling $n\pm s_i \rightarrow n$, we see that all energies are of the form 
\begin{equation}\label{levels2}
E_{k_3,n} = \sqrt{\frac{2\pi}{L_1L_2}\left(2n+ 1\right) + \frac{1}{L^2}\left(2\pi k_3 \pm W\right)^2}~,~ {\rm{with}}~ n=-1,0,1,2,\ldots.
\end{equation}

Now we discuss the degeneracies of each of these levels. The energy levels are consistent with any linear combination of $a^\pm_{k3,i}$ solutions with the same energy, but not all possible linear combinations are consistent with the equation of motion, Equation (\ref{nu0eom}). To simplify these considerations, we note that the reality of the gauge field requires $\left(a^+_{k_3,+}\right)^* = a^-_{-k_3,-}$, $\left(a^+_{k_3,-}\right)^* = a^-_{-k_3,+}$, and $\left(a^+_{k_3,3}\right)^* = a^-_{-k_3,3}$, where we recall that between (\ref{b15}) and (\ref{b16}), we introduced $a_+^{\pm} = (a_1^\pm - i a_2^\pm)/2$ and $a_-^{\pm} = (a_1^\pm + i a_2^\pm)/2$. Hence, the $a^+$ solution is not independent of the $a^-$ solution, and we need only consider the equation of motion on the $a^+$ modes. 

It is important to be careful: the fact that the $a^+$ modes determine the $a^-$ does not mean that each solution for $a^+$ contributes one physical mode. The reality of the gauge field means we should be counting real degrees of freedom. Hence, as already alluded to after (\ref{levels1}), the real and imaginary parts of $a^+$ both constitute independent modes, and so the degeneracy associated with each allowed complex solution to Equation (\ref{nu0eom}) is 2. 

Rewriting (\ref{nu0eom}) in terms of $a_\pm$ and using the same simplifications as used for (\ref{SHO}), we find the identity
\begin{equation}
\begin{split}
\label{Gauss}
&\left(\frac{2\pi}{L_1L_2}x^2 - \partial_2\right) a^+_+  + \left(\frac{2\pi}{L_1L_2}x^2 + \partial_2\right)a^+_- - \frac{1}{L} \left(2\pi k_3 + W\right)a^+_3 = 0 \\
\Rightarrow ~ & \sqrt{\frac{4\pi}{L_1L_2}} \hat{A}^\dag a^+_+ + \sqrt{\frac{4\pi}{L_1L_2}} \hat{A} a^+_-    - \frac{1}{L} \left(2\pi k_3 + W\right)a^+_3 = 0~.
\end{split}
\end{equation}
Here in the second line we have defined the differential operators, $\hat{A}\equiv \sqrt{\frac{\pi}{L_1L_2}}\left(x^2 + \frac{L_1L_2}{2\pi}\partial_2\right)$ and $\hat{A}^\dag\equiv \sqrt{\frac{\pi}{L_1L_2}}\left(x^2 - \frac{L_1L_2}{2\pi}\partial_2\right)$. These are, respectively, the annihilation and creation operators for the simple harmonic oscillator defined in Equation (\ref{SHO}).

We can now start counting degeneracies. First consider the $n=-1$ case. This case can only be achieved by a solution of the form (for every value of $k_3$): $a^+_- = \phi_0$ (corresponding to $n=0$, $i=-$ in (\ref{levels1})). If this mode satisfies (\ref{Gauss}), it is allowed and will be doubly degenerate. Plugging it in, we find
\begin{equation}
 \sqrt{\frac{4\pi}{L_1L_2}}\hat{A} \phi_0 = 0~.
\end{equation}
This equation is true because the annihilation operator annihilates the zeroth level solution. Thus, the $n=-1$ energy levels are doubly degenerate. 

Next, consider the $n=0$ case. Here there are two potential modes: $a^+_- = \phi_1$ and $a^+_3 = \phi_0$. A general mode is a linear combination of these: $a^+_+ = 0$,  $a^+_- = \beta_- \phi_1$, and $a^+_3 = \beta_3 \phi_0$. Plugging this into equation (\ref{Gauss}) gives
\begin{equation}
\begin{split}
0 = & \sqrt{\frac{4\pi}{L_1L_2}} \hat{A} \beta_- \phi_1 - \frac{1}{L} \left(2\pi k_3 + W\right)\beta_3 \phi_0 = \left(\sqrt{\frac{4\pi}{L_1L_2}} \beta_- - \frac{1}{L} \left(2\pi k_3 + W\right)\beta_3\right) \phi_0 ~.
\end{split}
\end{equation}
Here we find a linear equation relating $\beta_-$ and $\beta_3$, which allows only one complex solution. Hence the $n=0$ energy levels are also doubly degenerate.

Lastly, consider the cases with $n\geq 1$. Here the $a^+_+$ field can turn on and our general modes take the form: $a^+_+ = \beta_+ \phi_{n-1}$, $a^+_- = \beta_- \phi_{n+1}$, and $a^+_3 = \beta_3 \phi_{n}$. Plugging this into equation (\ref{Gauss}) gives
\begin{equation}
\begin{split}
0 = & \sqrt{\frac{4\pi}{L_1L_2}} \hat{A}^\dag \beta_+ \phi_{n-1} +\sqrt{\frac{4\pi}{L_1L_2}} \hat{A} \beta_- \phi_{n+1} - \frac{1}{L} \left(2\pi k_3 + W\right)\beta_3 \phi_n\\
= & \left(\sqrt{\frac{4\pi n}{L_1L_2}}\beta_+ + \sqrt{\frac{4\pi(n+1)}{L_1L_2}} \beta_- - \frac{1}{L} \left(2\pi k_3 + W\right)\beta_3\right) \phi_n ~.
\end{split}
\end{equation}
Here we get a single linear equation relating the coefficients $\beta_+$, $\beta_-$, and $\beta_3$. This will always admit two linear independent complex solutions, so each of these energy levels will be fourfold degenerate. 

Thus, (\ref{levels2}) and the above discussion give us the boson spectrum and degeneracies stated after (\ref{bosonspectrum}). In Section \ref{sec:calculating}, we explain how we use the results for the energy levels and their degeneracies to compute the $\S^1_L$ holonomy ($W$) one-loop potential. Finally, we note that as an important check on the energy levels and degeneracies obtained in this Section, in Section \ref{sec:infinite} we take the infinite volume limit ($L_1 L_2 \rightarrow \infty$) to obtain precisely the well-known $\R^3 \times \S^1$ GPY potential. 

\subsection{The fermion spectrum}
In this section, we calculate the spectrum of the Weyl fermions with Majorana mass $M$. To start, we consider the fermion Lagrangian (\ref{fermionlagrangian}) with just the background gauge field, ignoring the interaction terms with the dynamical gauge boson. From there, we can find the Euler-Lagrange equations:
\begin{equation}
\begin{split}
\bar{\sigma}^{\mu,\dot{\alpha}\alpha} \partial_\mu \psi^3_\alpha & - iM \varepsilon^{\dot{\beta}\dot{\alpha}} \left(\psi^3\right)^\dag_{\dot{\beta}} = 0\\
\bar{\sigma}^{\mu,\dot{\alpha}\alpha} \partial_\mu \psi^+_\alpha & +iA^{\Omega,3}_\mu \bar{\sigma}^{\mu,\dot{\alpha}\alpha} \psi^+_\alpha - iM \varepsilon^{\dot{\beta}\dot{\alpha}} \left(\psi^-\right)^\dag_{\dot{\beta}}=0\\
\bar{\sigma}^{\mu,\dot{\alpha}\alpha} \partial_\mu \psi^-_\alpha & -iA^{\Omega,3}_\mu \bar{\sigma}^{\mu,\dot{\alpha}\alpha} \psi^-_\alpha - iM \varepsilon^{\dot{\beta}\dot{\alpha}} \left(\psi^+\right)^\dag_{\dot{\beta}}=0~.
\end{split}
\end{equation}
Here we can see, as in the case of the gauge bosons, that the Cartan component will not contribute to the potential for $W$, so we ignore it. For the other two equations, we can rewrite these into a single matrix equation:\footnote{Recall that we are using the definition, $D^\pm = \partial_\mu \pm i A^{\Omega,3}_\mu$}
\begin{equation}
\label{matrixeom}
\begin{pmatrix}D^+_\mu \bar{\sigma}^\mu & \bar{\sigma}^2 M\\ 
\left(\bar{\sigma}^2 M\right)^* & \left(D^+_\mu \bar{\sigma}^\mu \right)^*\end{pmatrix}
\begin{pmatrix}\psi^+ \\ \left(\psi^-\right)^* \end{pmatrix} = 0~.
\end{equation}
Now, we can use the top row of the matrix to solve for $\left(\psi^-\right)^*$, then we can plug it into the bottom row to get:
\begin{equation}
\left(\sigma^2 M + \frac{1}{M} D^+_\nu \left(\bar{\sigma}^\nu\right)^* \sigma^2 D^+_\mu \bar{\sigma}^\mu\right) \psi^+ = 0~.
\end{equation}
Multiplying this by $\sigma^2M$ and using the identity $\sigma^2 \left(\bar{\sigma}^\nu\right)^* \sigma^2 = \sigma^\mu$, we find
\begin{equation}
\left(M^2 +  D^+_\nu  \sigma^\nu D^+_\mu \bar{\sigma}^\mu\right) \psi^+ = 0~.
\end{equation}
Following similar steps for the $\psi^-$, we find the equation
\begin{equation}
\left(M^2 +  D^-_\nu  \sigma^\nu D^-_\mu \bar{\sigma}^\mu\right) \psi^- = 0~.
\end{equation}
These equations can be further simplified using the commutation relations between the $D^\pm_\mu$. They all commute except for the one combination: $\left[D^\pm_1,D^\pm_2\right] = \pm i\frac{2\pi}{L_1L_2}$. Using this we find the matrices are actually diagonal and we get the four uncoupled equations,
\begin{equation}
\begin{split}
\left(\left(D^+_0\right)^2 - \left(D^+_1\right)^2 - \left(D^+_2\right)^2 - \left(D^+_3\right)^2 + M^2 + \frac{2\pi}{L_1L_2}\right)\psi^+_1 & = 0 \\
\left(\left(D^+_0\right)^2 - \left(D^+_1\right)^2 - \left(D^+_2\right)^2 - \left(D^+_3\right)^2 + M^2 - \frac{2\pi}{L_1L_2}\right)\psi^+_2 & = 0 \\
\left(\left(D^-_0\right)^2 - \left(D^-_1\right)^2 - \left(D^-_2\right)^2 - \left(D^-_3\right)^2 + M^2 - \frac{2\pi}{L_1L_2}\right)\psi^-_1 & = 0 \\
\left(\left(D^-_0\right)^2 - \left(D^-_1\right)^2 - \left(D^-_2\right)^2 - \left(D^-_3\right)^2 + M^2 + \frac{2\pi}{L_1L_2}\right)\psi^-_2 & = 0 ~.
\end{split}
\end{equation}
From here, we apply the same steps as in the boson case. We look for states of the form
\begin{equation}
\psi^\pm = e^{-iEx^0} \sum_{k_1,k_3\in\Z} e^{i2\pi k_1\frac{x^1}{L_1}} e^{i2\pi k_3\frac{x^3}{L}} \psi^\pm_{k_3} \left(x^2\mp k_1L_2\right)~,
\end{equation}
and shift coordinates and phases to eliminate extraneous variables. This leaves the equations:
\begin{equation}
\begin{split}
\left(-\partial_2^2 + \left(\frac{2\pi}{L_1L_2}\right)^2 \left(x^2\right)^2\right) \psi^+_1 & = \left[E^2 - M^2 - \frac{2\pi}{L_1L_2} - \frac{1}{L}\left(2\pi k_3 + W_3\right)^2\right] \psi^+_1 \\
\left(-\partial_2^2 + \left(\frac{2\pi}{L_1L_2}\right)^2 \left(x^2\right)^2\right) \psi^+_2 & = \left[E^2 - M^2 + \frac{2\pi}{L_1L_2} - \frac{1}{L}\left(2\pi k_3 + W_3\right)^2\right] \psi^+_1 \\
\left(-\partial_2^2 + \left(\frac{2\pi}{L_1L_2}\right)^2 \left(x^2\right)^2\right) \psi^-_1 & = \left[E^2 - M^2 + \frac{2\pi}{L_1L_2} - \frac{1}{L}\left(2\pi k_3 - W_3\right)^2\right] \psi^+_1 \\
\left(-\partial_2^2 + \left(\frac{2\pi}{L_1L_2}\right)^2 \left(x^2\right)^2\right) \psi^-_2 & = \left[E^2 - M^2 - \frac{2\pi}{L_1L_2} - \frac{1}{L}\left(2\pi k_3 - W_3\right)^2\right] \psi^+_1 ~.
\end{split}
\end{equation}
Hence the solutions are simple harmonic oscillator solutions, $\phi_n$, with corresponding energies
\begin{equation}
E = \sqrt{\frac{2\pi}{L_1L_2}(2n+1) \pm \frac{2\pi}{L_1L_2} + \frac{1}{L^2}\left(2\pi k_3 \mp W\right)^2 + M^2}~.
\end{equation}
These energies can all be written in a form similar to (\ref{levels2}):
\begin{equation}\label{levels3}
E_{k_3,n} = \sqrt{\frac{4\pi}{L_1L_2}n  + \frac{1}{L^2}\left(2\pi k_3 + W\right)^2 + M^2}~,~ \text{with} ~n=0,1,2,\ldots.
\end{equation}

For determining the degeneracies, similarly to the boson case, we plug the solutions back into the equation of motion, Equation (\ref{matrixeom}). To compare to the boson case we look for real degrees of freedom, hence each complex solution contributes 2 physical modes. In terms of the creation and annihilation operators, $\hat{A}$ and $\hat{A}^\dag$ introduced after (\ref{Gauss}), Equation (\ref{matrixeom}) can be rewritten as 
\begin{equation}
\label{degeneracymatrix}
\begin{pmatrix}
E + \frac{1}{L}\left(2\pi k_3+W\right) & -\sqrt{\frac{4\pi}{L_1L_2}} \hat{A} & 0 & -M \\
-\sqrt{\frac{4\pi}{L_1L_2}} \hat{A}^\dag & E - \frac{1}{L}\left(2\pi k_3+W\right) & M & 0 \\
0 & M & E + \frac{1}{L}\left(2\pi k_3+W\right) & -\sqrt{\frac{4\pi}{L_1L_2}} \hat{A}^\dag \\
-M & 0 & -\sqrt{\frac{4\pi}{L_1L_2}} \hat{A} & E - \frac{1}{L}\left(2\pi k_3+W\right)
\end{pmatrix}
\begin{pmatrix}
\psi^+_1 \\
\psi^+_2 \\
\left(\psi^-_1\right)^* \\
\left(\psi^-_1\right)^*
\end{pmatrix} = 0~.
\end{equation}

We first  consider the case $n=0$. This can only happen with modes of the form $\psi^+_1 = 0$, $\psi^+_2 = \beta^+_2 \phi_0$, $\psi^-_1 = \beta^-_1 \phi_0$, and $\psi^-_2 = 0$. Plugging this into (\ref{degeneracymatrix}), we find
\begin{equation}
\begin{pmatrix}
0 \\
\left(E_{k_3,0} - \frac{2\pi k_3}{L} - \frac{W}{L}\right) \beta^+_2 + M \beta^-_1 \\
M \beta^+_2 + \left(E_{k_3,0} + \frac{2\pi k_3}{L} + \frac{W}{L}\right) \beta^-_1 \\
0
\end{pmatrix}=0~.
\end{equation}
This can be rearranged into the matrix equation
\begin{equation}
\begin{pmatrix}
\left(E_{k_3,0} - \frac{2\pi k_3}{L} - \frac{W}{L}\right) & M \\
M & \left(E_{k_3,0} + \frac{2\pi k_3}{L} + \frac{W}{L}\right)
\end{pmatrix}
\begin{pmatrix}
\beta^+_2 \\
\beta^-_1
\end{pmatrix}=0~.
\end{equation}
This has only one linearly independent solution, so the $n=0$ modes are doubly degenerate.

For the case $n\geq1$, the modes take the form $\psi^+_1 = \beta^+_1 \phi_{n-1}$, $\psi^+_2 = \beta^+_2 \phi_{n}$, $\psi^-_1 = \beta^-_1 \phi_{n}$, and $\psi^-_2 = \beta^-_2 \phi_{n-1}$. Plugging this into (\ref{degeneracymatrix}), we find
\begin{equation}
\begin{pmatrix}
\left(\left(E_{k_3,n} + \frac{2\pi k_3}{L} + \frac{W}{L}\right)\beta^+_1 - \sqrt{\frac{4\pi}{L_1L_2}}\beta^+_2 -M\beta^-_2 \right) \phi_{n-1} \\
\left( - \sqrt{\frac{4\pi}{L_1L_2}}\beta^+_1 + \left(E_{k_3,n} - \frac{2\pi k_3}{L} - \frac{W}{L}\right)\beta^+_2 +M\beta^-_1\right) \phi_n \\ 
\left(M\beta^+_2. + \left(E_{k_3,n} + \frac{2\pi k_3}{L} + \frac{W}{L}\right)\beta^-_1 - \sqrt{\frac{4\pi}{L_1L_2}}\beta^-_2\right) \phi_n \\
\left(-M\beta^+_1 - \sqrt{\frac{4\pi}{L_1L_2}}\beta^-_1 + \left(E_{k_3,n} - \frac{2\pi k_3}{L} - \frac{W}{L}\right)\beta^-_2\right) \phi_{n-1}
\end{pmatrix}=0~.
\end{equation}
This gives the matrix equation
\begin{equation}
\begin{pmatrix}
E_{k_3,n} + \frac{1}{L}\left(2\pi k_3+W\right) & -\sqrt{\frac{4\pi}{L_1L_2}} & 0 & -M \\
-\sqrt{\frac{4\pi}{L_1L_2}} & E_{k_3,n} - \frac{1}{L}\left(2\pi k_3+W\right) & M & 0 \\
0 & M & E_{k_3,n} + \frac{1}{L}\left(2\pi k_3+W\right) & -\sqrt{\frac{4\pi}{L_1L_2}}  \\
-M & 0 & -\sqrt{\frac{4\pi}{L_1L_2}}  & E_{k_3,n} - \frac{1}{L}\left(2\pi k_3+W\right)
\end{pmatrix}
\begin{pmatrix}
\beta^+_1 \\
\beta^+_2 \\
\beta^-_1 \\
\beta^-_2
\end{pmatrix}=0~.
\end{equation}
There are two linearly independent solutions to this matrix, hence, these modes are fourfold degenerate.

Thus, the fermion energy levels (\ref{levels3}) along with the degeneracies given above constitute the non-Cartan spectrum of massive or massless fermions given in (\ref{fermionspectrum}).

\subsection{Higher flux ($k>1$)}
In the previous Sections, we discussed the spectra of the $\Omega$-gauge, but here we discuss the modifications to the above arguments when we change to the $\Omega_{(k)}$-gauge (\ref{omegakgauge}) where the simplest abelian  background  (\ref{kflux})  corresponds to higher values of the magnetic flux through $\T^2$
\begin{equation}
\label{Omegakbackground}
A^{W, \Omega_k}(x) = \left(\left(- {2 \pi (2k+1) x^2 \over L^1 L^2} + {\alpha_1 \over L^1}\right) dx^1 + {\alpha_2 \over L^2} dx^2   + {W \over  L} dx^3\right) {\sigma_3 \over 2}~. 
\end{equation}
There are only two differences between this and the $k=0$ case. Firstly, the boundary conditions are different. This difference is apparent in the relations (\ref{k1relations}), which become
\begin{equation}
\phi^\pm_{k_1\pm (2k+1),k_3} (x^2 + L_2) = \phi^\pm_{k_1,k_3} (x^2)~.
\end{equation}
Hence, we cannot write the $x^2$ dependent coefficients in terms of just the $k_1=0$ function. Instead, there are now $\left|2k+1\right|$ independent functions. If we follow each of these through the calculation, we find that each of these functions solves the same set of differential equations, and hence produce identical spectra. Thus, the degeneracy of every energy level and the overall vacuum energy density are multiplied by $\left|2k+1\right|$. This overall factor multiplies the overall $\frac{1}{L_1L_2}$ from dividing by the volume of space.

Secondly, the factor of $2k+1$ in the background will show up with all the factors of $\frac{1}{L_1L_2}$ throughout the calculation. Thus, the ``frequency'' of the simple harmonic oscillator will be $\frac{2\pi}{L_1L_2}\left|2k+1\right|$. This leads to the energy levels:
\begin{equation}
E_{boson} = \sqrt{\frac{2\pi\left|2k+1\right|}{L_1L_2}\left(2n+ 1\right) + \frac{1}{L^2}\left(2\pi k_3 \pm W\right)^2}
\end{equation}
and
\begin{equation}
E_{fermion} = \sqrt{\frac{4\pi\left|2k+1\right|}{L_1L_2}n  + \frac{1}{L^2}\left(2\pi k_3 + W\right)^2 + M^2}~.
\end{equation}
Note that if $2k+1 < 0$, this will rearrange which modes correspond to the $n=0$ and $n=-1$ energy levels, but will not change the final results.

Hence, the only change to the vacuum energy density is introducing a factor of $\left|2k+1\right|$ everywhere there is a factor of $\frac{1}{L_1L_2}$. This means we can find the $k\neq 0$ potential by making the replacement 
\begin{equation}
\epsilon \rightarrow \epsilon_k \equiv \left|2k+1\right|\epsilon
\end{equation}
in the $k=0$ potential. 

   \bibliography{fluxsemiclassics.bib}
 
  \bibliographystyle{JHEP}

\end{document}